\documentclass[sn-nature]{sn-jnl}

\usepackage{graphicx}%
\usepackage{multirow}%
\usepackage{amsmath,amssymb,amsfonts}%
\usepackage{amsthm}%
\usepackage{mathrsfs}%
\usepackage[title]{appendix}%
\usepackage{xcolor}%
\usepackage{textcomp}%
\usepackage{manyfoot}%
\usepackage{booktabs}%
\usepackage{algorithm}%
\usepackage{algorithmicx}%
\usepackage{algpseudocode}%
\usepackage{listings}%
\usepackage{pdfpages}%

\newcommand{\araa}{Annu. Rev. Astron. Astrophys.}   
\newcommand{\aj}{Astron. J.}   
\newcommand{\apj}{Astrophys. J.}   
\newcommand{\apjl}{Astrophys. J. Lett.}   
\newcommand{\apjs}{Astrophys. J. Suppl. Ser.}   
\newcommand{\aap}{Astron. Astrophys.}   
\newcommand{\aaps}{Astron. Astrophys. Suppl.}   
\newcommand{\mnras}{Mon. Not. R. Astron. Soc.}   
\newcommand{\nat}{Nature} 
\newcommand{\pasj}{Publ. Astron. Soc. Jpn}   
\newcommand{\pasp}{Publ. Astron. Soc. Pac.}   

\newcommand{\Ha}{H$\alpha$}
\newcommand{\Hb}{H$\beta$}

\newcommand{\OIII}{[O\,\textsc{iii}]}

\newcommand{\NII}{[N\,\textsc{ii}]}

\newcommand{\FeII}{Fe\,\textsc{ii}}
\newcommand{\sersic}{S\'{e}rsic}
\newcommand{\msersic}{$m_{\rm S\acute{e}rsic}$}

\unnumbered

\begin{document}

\title[AGNs and Their Host Galaxies]{Evolutionary Paths of Active Galactic Nuclei and Their Host Galaxies}


\author*[1,2,3]{\fnm{Ming-Yang} \sur{Zhuang}}\email{mingyangzhuang@pku.edu.cn}

\author[1,2]{\fnm{Luis C.} \sur{Ho}}

\affil[1]{\orgdiv{Kavli Institute for Astronomy and Astrophysics}, \orgname{Peking University}, \orgaddress{\city{Beijing}, \postcode{100871}, \country{China}}}

\affil[2]{\orgdiv{Department of Astronomy}, \orgname{School of Physics, Peking University}, \orgaddress{\city{Beijing}, \postcode{100871}, \country{China}}}

\affil[3]{\orgdiv{Department of Astronomy}, \orgname{University of Illinois Urbana-Champaign}, \orgaddress{\city{Urbana}, \postcode{61801}, \state{IL}, \country{USA}}}


\abstract{
The tight correlations between the masses of supermassive black holes (BHs) and the properties of their host galaxies suggest that BHs coevolve with galaxies. However, what is the link between BH mass ($M_{\rm BH}$) and the properties of the host galaxies of active galactic nuclei (AGNs) in the nearby Universe? We measure stellar masses ($M_*$), colors, and structural properties for  $\sim11,500$ $z\leq0.35$ broad-line AGNs, nearly 40 times larger than that in any previous work. We find that early-type and late-type AGNs follow a similar $M_{\rm BH}-M_*$ relation. The position of AGNs on the $M_{\rm BH}-M_*$ plane is connected with the properties of star formation and BH accretion. Our results unveil the evolutionary paths of galaxies on the $M_{\rm BH}-M_*$ plane: objects above the relation tend to evolve more horizontally with substantial $M_*$ growth; objects on the relation move along the local relation; and objects below the relation migrate more vertically with substantial $M_{\rm BH}$ growth. These trajectories suggest that radiative-mode feedback cannot quench the growth of BHs and their host galaxies for AGNs that lie below the relation, while kinetic-mode feedback hardly suppress long-term star formation for AGNs situated above the relation. This work provides important constraints for numerical simulations and offers a framework for studying the cosmic coevolution of supermassive BHs and their host galaxies. }


\maketitle
\newpage

\section{Main}
The discovery of tight correlations between the masses of supermassive BHs and the properties of their host galaxies suggests that BHs and galaxies coevolve\cite{Kormendy&Ho2013ARA&A}. No consensus has been reached, however, as to how the BH-galaxy relations arose or how they evolved over time. Popular scenarios ascribe a critical role for energy feedback from the AGN in regulating the growth of the BH and its host galaxy\cite{Di_Matteo+2005Natur, Weinberger+2018MNRAS}, but yet some contend that the scaling relations merely reflect the natural consequences of the hierarchical assembly of BHs and their host galaxies via merging events\cite{Peng2007ApJ}.

Considerations of when and where the BH-galaxy scaling relations were first established and how they evolved over time face the immediate practical difficulty that direct $M_{\rm BH}$ measurements currently are limited only to the nearest galaxies whose BH sphere-of-influence can be resolved\cite{Kormendy&Ho2013ARA&A, Greene+2020ARA&A}. To avoid the contamination and disturbances from the active nucleus, direct $M_{\rm BH}$ measurements also deliberately and nearly universally have avoided galaxies with substantial nuclear activity. 
Consequently, the BH-galaxy scaling relations have been established securely only in the most local galaxies, ones whose central BH is essentially quiescent and whose overall star formation activity is also largely muted. This severe limitation can be bypassed by exploiting indirect methods of estimating $M_{\rm BH}$ using the properties of the broad-line region, either through reverberation mapping experiments or empirical estimators of $M_{\rm BH}$ based on single-epoch spectroscopy calibrated against reverberation-mapped AGNs\cite{Vestergaard&Peterson2006ApJ, Wang+2009ApJ}. Securing reliable estimates of $M_{\rm BH}$ is only the first step. Obtaining robust host galaxy parameters with which to compare $M_{\rm BH}$ presents an equally, if not more, daunting challenge.

Definitive conclusions on the nature of the $M_{\rm BH}$-host scaling relations of AGNs, including whether and how they evolve with redshift, remain elusive\cite{Peng+2006bApJ, Alexander+2008AJ, Schulze&Wisotzki2014MNRAS, Neeleman+2021ApJ}. The study of cosmic evolution of $M_{\rm BH}$-host scaling relations using high-$z$ AGNs relies critically on our understanding of the relation of their low-redshift counterparts. Do nearby AGNs differ with respect to local inactive galaxies? Do AGN hosts with different morphologies or bulge types behave the same? 
Progress has been stymied by the small samples (typically a few tens, at most a few hundred) and the systematics of host measurements. To anchor our understanding of the local BH-host galaxy scaling relations, we urgently need a large sample of nearby AGNs that covers a wide parameter space and has reliable host measurements, one that can serve as the definitive benchmark for evolutionary studies.

We derive the structural and photometric properties of the host galaxies of 11,474 $z\leq0.35$ type~1 AGNs, by performing simultaneous, multi-band decomposition of Pan-STARRS1 $grizy_{\rm P1}$ images that considers the wavelength-dependent variation of galaxy structure (see Methods). This sample allows us to revisit this topic with a more comprehensive, less biased population of AGNs that spans more than 4 orders of magnitude in AGN bolometric luminosity ($L_{\rm bol}= 10^{41.9}-10^{46.4} \,{\rm erg\, s^{-1}}$), almost 5 orders of magnitudes in BH mass ($M_{\rm BH}=10^{5.1}-10^{10.0}\, M_{\odot}$), and a much broader distribution of Eddington ratios ($\lambda_{\rm E} =  10^{-3.5}-10^{0.7}$).
\\
\newpage

\begin{figure}[t!]
\centering
\includegraphics[width=0.7\textwidth]{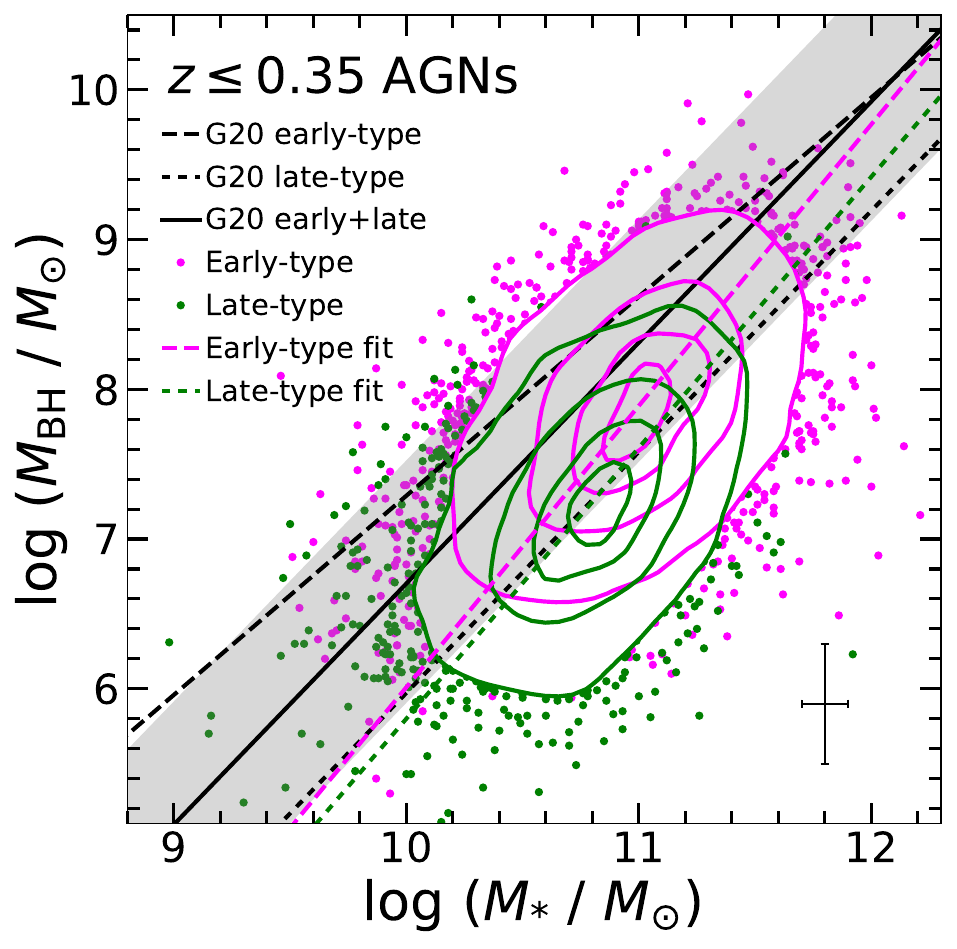}
\caption{The relation between $M_{\rm BH}$ and $M_*$ for $z\leq 0.35$ type~1 AGNs. Objects with late-type, disk-dominated and early-type, bulge-dominated host morphologies are shown in green and magenta, respectively. Best-fit relations from G20\cite{Greene+2020ARA&A} for local early-type galaxies (black dashed line), late-type galaxies (black dotted line), and both types combined (black solid line; intrinsic scatter of 0.81~dex indicated as shaded gray region). We plot the best-fit relation for the early-type AGNs as a magenta dashed line and for the late-type AGNs as green dotted line. Data are presented as median values and typical uncertainty is shown in the lower-right corner. Contours indicate the distribution of 10\%, 30\%, 60\%, and 90\% of the entire sample, respectively. Dots are individual objects located outside the 90\% contour.}
\label{fig1}
\end{figure}

\noindent \textbf{\large The $M_{\rm BH}-M_*$ Relation of AGNs in the Nearby Universe} 

We consider the $M_{\rm BH}-M_*$ relations of Greene et al. (2020; hereinafter G20\cite{Greene+2020ARA&A}) in the local Universe as the reference framework. G20 use dynamical $M_{\rm BH}$ measurements of local, inactive galaxies spanning more than 6 orders of magnitude in $M_{\rm BH}$ and a wide range of morphological types. They provide separate relations for early-type and late-type galaxies, and for both galaxy types combined.

AGNs generally follow the local relations (Fig.~1). To better facilitate detailed quantitative comparison with G20, we present linear regression fits to our AGNs. A fit to the early-type, bulge-dominated AGNs using \texttt{linmix}\cite{Kelly2007ApJ} (the independent and dependent variables are switched when performing the fits in Equations~(\ref{eq1}--\ref{eq4}), to better trace the orientation of the contours) gives 
\begin{equation} \label{eq1}
{\log\, (M_{\rm BH}}/M_{\odot}) = (1.88 \pm 0.05) \log\, (M_*/10^{11} M_{\odot}) + (7.89 \pm 0.40),
\end{equation}

\noindent with a total scatter of 0.64~dex. For the late-type, disk-dominated AGNs,
\begin{equation} \label{eq2}
{\log\, (M_{\rm BH}}/M_{\odot}) = (1.81 \pm 0.06) \log\, (M_*/10^{11} M_{\odot}) + (7.61 \pm 0.50),
\end{equation}

\noindent
with a total scatter of 0.59~dex. AGNs with late-type hosts follow a relation close to that of late-type inactive galaxies, with a slightly steeper slope. Although the BHs in early-type AGNs are on average $\sim0.6$~dex systematically more massive than late-type AGNs ($10^{7.8} M_{\odot}$ versus $10^{7.2} M_{\odot}$), they are still significantly less massive ($-0.7$~dex at $M_*=10^{11} M_{\odot}$) than those in early-type, bulge-dominated inactive galaxies, which are predominantly quiescent galaxies with very weak star formation and nuclear activity. Interestingly, the relations of early-type and late-type AGNs are quite similar. At $M_* = 10^{11} M_{\odot}$, the difference in $M_{\rm BH}$ is merely 0.28~dex, smaller than the typical uncertainty of the single-epoch virial $M_{\rm BH}$ estimates (0.4~dex).

The classical bulges of inactive galaxies, whose formation physics likely parallels that of ellipticals, follow the same relation as ellipticals, while pseudo bulges, which share properties akin to those of galactic disks, do not correlate strongly with $M_{\rm BH}$ and tend to be offset below the relation of classical bulges\cite{Kormendy&Ho2013ARA&A}. In terms of total stellar mass, late-type galaxies exhibit $\sim1.2$~dex lower $M_{\rm BH}$ than early-type galaxies (G20). For active galaxies, however, definitive conclusions on whether they follow the relations of inactive galaxies have remained elusive. While some\cite{Bennert+2021ApJ} contend that local AGNs follow a similar correlation between $M_{\rm BH}$ and bulge stellar mass as inactive galaxies, with no difference between classical and pseudo bulges, others\cite{Reines&Volonteri2015ApJ, Kim&Ho2019ApJ} maintain that nearby AGNs in general lie offset below the scaling relation of normal classical bulges and ellipticals, regardless of whether the bulge of the AGN host belongs to the pseudo or classical variety. \\

\begin{figure*}[t!]
\centering
\includegraphics[width=\textwidth]{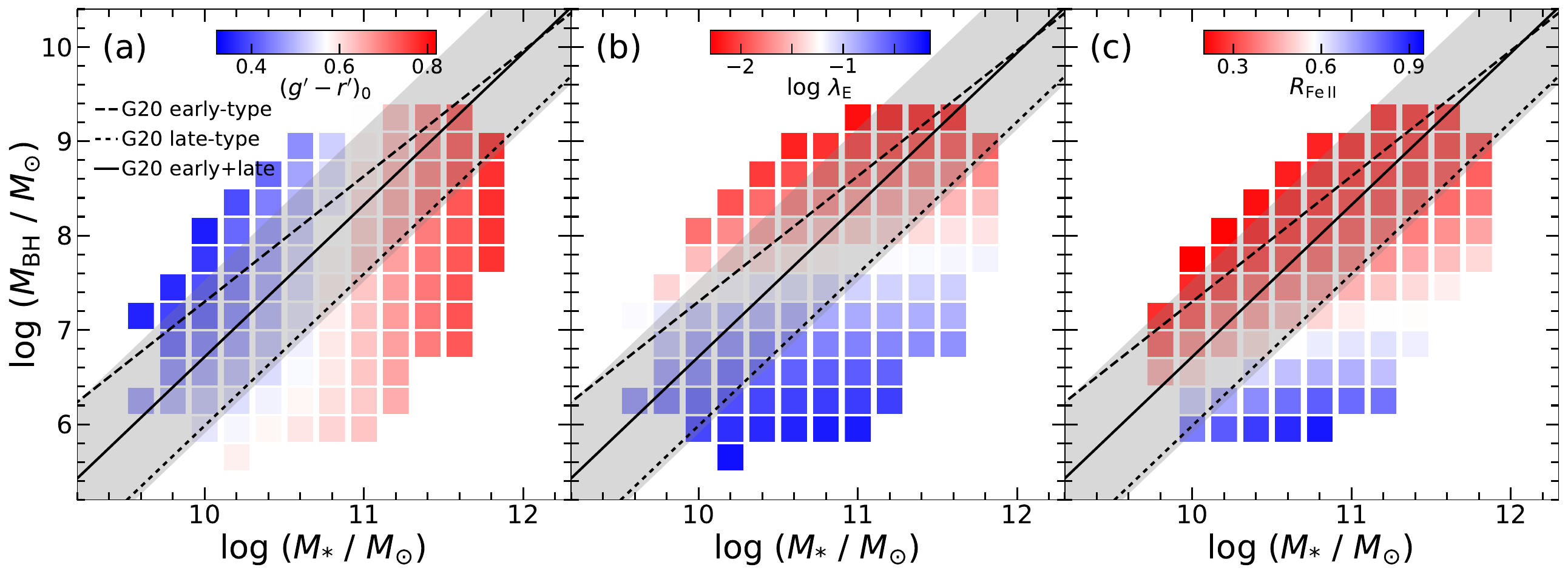}
\caption{The relation between $M_{\rm BH}$ and $M_*$ for $z\leq 0.35$ AGNs. Colors represent (a) rest-frame optical color $(g^{\prime}-r^{\prime})_0$, (b) Eddington ratio ($\lambda_{\rm E}$), and (c) relative \FeII\ strength. Best-fit relations from G20 for local early-type galaxies (black dashed line), late-type galaxies (black dotted line), and both types combined (black solid line; intrinsic scatter of 0.81~dex indicated as shaded gray region). Squares indicate median values of objects within a box of size 0.2~dex in $M_*$ and 0.3~dex in $M_{\rm BH}$ and source number $\geq5$, after performing locally weighted regression smoothing using the {\tt Python} package \texttt{LOESS}\cite{Cappellari+2013MNRAS}. The above trends persist regardless of whether the data are smoothed (Extended Fig.~2) or the choice of $M_{\rm BH}$ estimator (Extended Fig.~3).}
\label{fig2}
\end{figure*}

\begin{figure*}[t!]
\centering
\includegraphics[width=\textwidth]{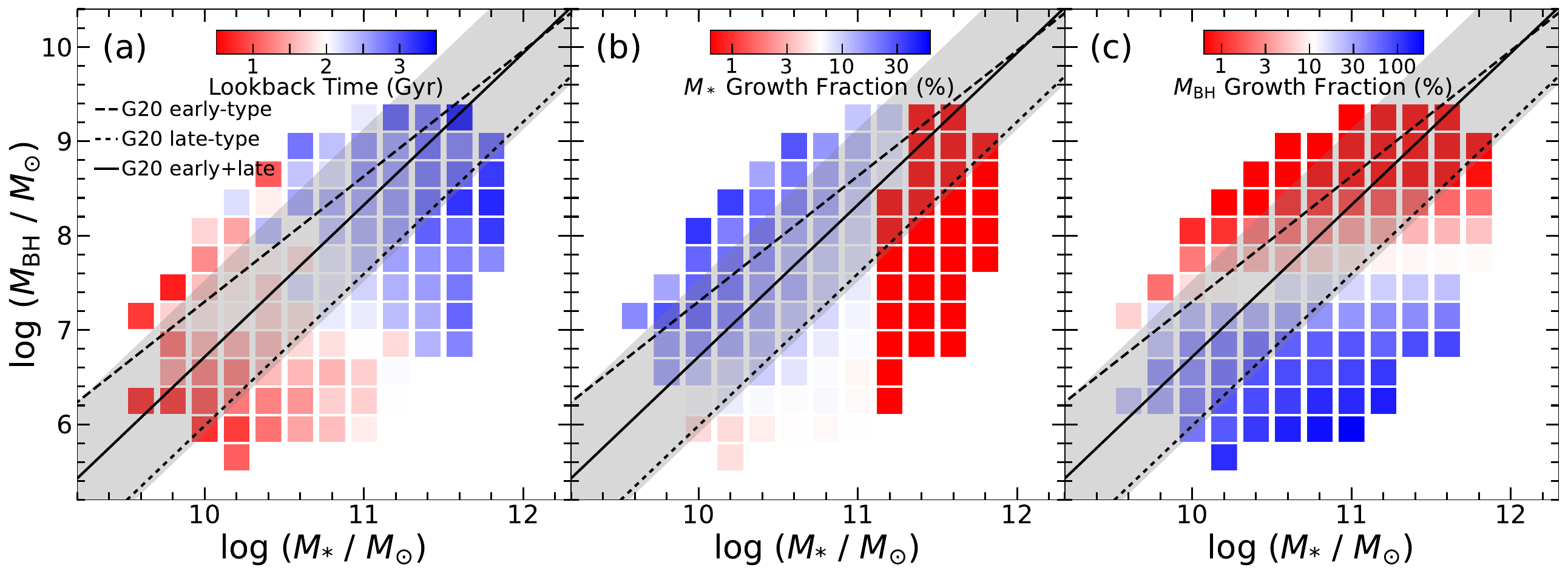}
\caption{The relation between $M_{\rm BH}$ and $M_*$ for $z\leq 0.35$ AGNs. Colors represent (a) the median lookback time, (b) growth fraction of $M_*$, and (c) growth fraction of $M_{\rm BH}$. The growth fraction is defined as $\Delta M/M$ at $z=0$. The host galaxies are assumed to be able to sustain star formation at the current rate. $M_{\rm BH}$-dependent duty cycle of the BH accretion is adopted to estimate the mass gain of BHs (see Methods). }
\label{fig3}
\end{figure*}

\noindent \textbf{\large Connections with Star Formation and BH Accretion Properties}

The large sample size and reliable image decomposition of this study allow us to investigate, systematically for the first time, the star formation and BH accretion properties of AGNs distributed across the $M_{\rm BH}-M_*$ plane. In light of the similarity between early-type and late-type AGNs on this plane (Fig.~1), we do not separate them in the following analysis. First, we confirm that active galaxies, as in inactive galaxies in the color$-M_*$ diagram (see Extended Fig.~1), exhibit a clear positive correlation between rest-frame $(g^{\prime}-r^{\prime})_0$ color and $M_*$, with redder colors found in more massive hosts (Fig.~2a). We also see a clear correlation between $\lambda_{\rm E}$ and $M_{\rm BH}$, such that more massive BHs have lower $\lambda_{\rm E}$ (Fig.~2b), in line with trends established in nearby galaxies\cite{Ho2008ARA&A}  that arise mostly from sample selection bias in magnitude-limited surveys. Intriguingly, besides the aforementioned expected trends, we find that $(g^{\prime}-r^{\prime})_0$ decreases systematically (color becomes bluer) toward higher $M_{\rm BH}$ at fixed $M_*$, while $\lambda_{\rm E}$ increases significantly toward higher $M_*$ at fixed $M_{\rm BH}$. As $\lambda_{\rm E}$ is derived from $M_{\rm BH}$, it suffers from the large uncertainty ($\sim0.4$~dex) associated with single-epoch estimates of $M_{\rm BH}$ and thus could be driven potentially by unknown systematics. To mitigate against this concern, we use an independent, empirical indicator of Eddington ratio, $R_{\rm \FeII}$, defined as the ratio between the equivalent width of the broad optical \FeII\ emission (integrated over 4434--4684~\AA) to that of broad \Hb\cite{Boroson&Green1992ApJS}. The surrogate indicator $R_{\rm \FeII}$ also increases toward higher $M_*$ at fixed $M_{\rm BH}$ (Fig.~2c), confirming the robustness of the trends observed based on $\lambda_{\rm E}$ directly. The above trends also persist regardless of whether the data are smoothed (Extended Fig.~2) or the choice of $M_{\rm BH}$ estimator (Extended Fig.~3).

Taking into account their relative positions with respect to the local $M_{\rm BH}-M_*$ relation, our results show that objects lying below the local relation tend to have higher $\lambda_{\rm E}$, while objects lying above the local relation have characteristically higher specific star formation rate (sSFR). This supports previous works based on much smaller samples\cite{Urrutia+2012ApJ, Sun+2015ApJ, Kim&Ho2019ApJ, Zhao+2021ApJ}. Within this context, it is noteworthy that the class of AGNs known as narrow-line Seyfert~1s (NLS1s), which are suspected of experiencing high accretion rates, also appears to lie offset below the BH-galaxy scaling relations of inactive galaxies\cite{Mathur+2012ApJ}, although these claims are controversial\cite{Greene&Ho2006ApJ}.

To understand the cumulative effect of star formation and BH accretion after the AGN population has evolved to $z=0$, we estimate the amount of mass gain of the BH and of the stars in the host galaxy (see Methods). Fig.~3 redisplays the $M_{\rm BH}-M_*$ relation, now color-coded by (a) the lookback time and the relative growth fraction ($\Delta M/M$) till $z=0$ of (b) $M_*$ and (c) $M_{\rm BH}$. The period available for growth ranges from 0.5 to 3.5~Gyr. We find that the stellar masses of AGNs generally will experience limited growth (fraction $\lesssim10\%$), except for objects located on or above the G20 relation of early-type galaxies, which may grow up to $35\%$. Objects lying on or below the G20 relation of late-type galaxies can grow their BHs substantially, by as much as $200\%$ for the most extreme outliers. 

Combining the growth trajectories of $M_*$ and $M_{\rm BH}$, we schematically outline in Fig.~4 the predicted evolutionary paths for sources located in different positions across the $M_{\rm BH}-M_*$ plane. Three clear trends emerge: (1) objects above the local relation experience significant $M_*$ growth and migrate mostly horizontally; (2) objects already on the local relation move along it by growing both $M_{\rm BH}$ and $M_*$ in lockstep for less massive objects, or remain on the relation with little change in either $M_{\rm BH}$ or $M_*$ for massive objects; (3) objects below the local relation enhance $M_{\rm BH}$ substantially and mainly ascend vertically toward the relation. Furthermore, objects that depart the most from the local relation have a greater tendency to evolve toward it. Our results imply that the location of an AGN on the $M_{\rm BH}-M_*$ plane is closely linked to the star formation history of its host galaxy on the one hand, and the accretion history of the BH on the other hand.

\begin{figure}[t!]
\centering
\includegraphics[width=0.7\textwidth]{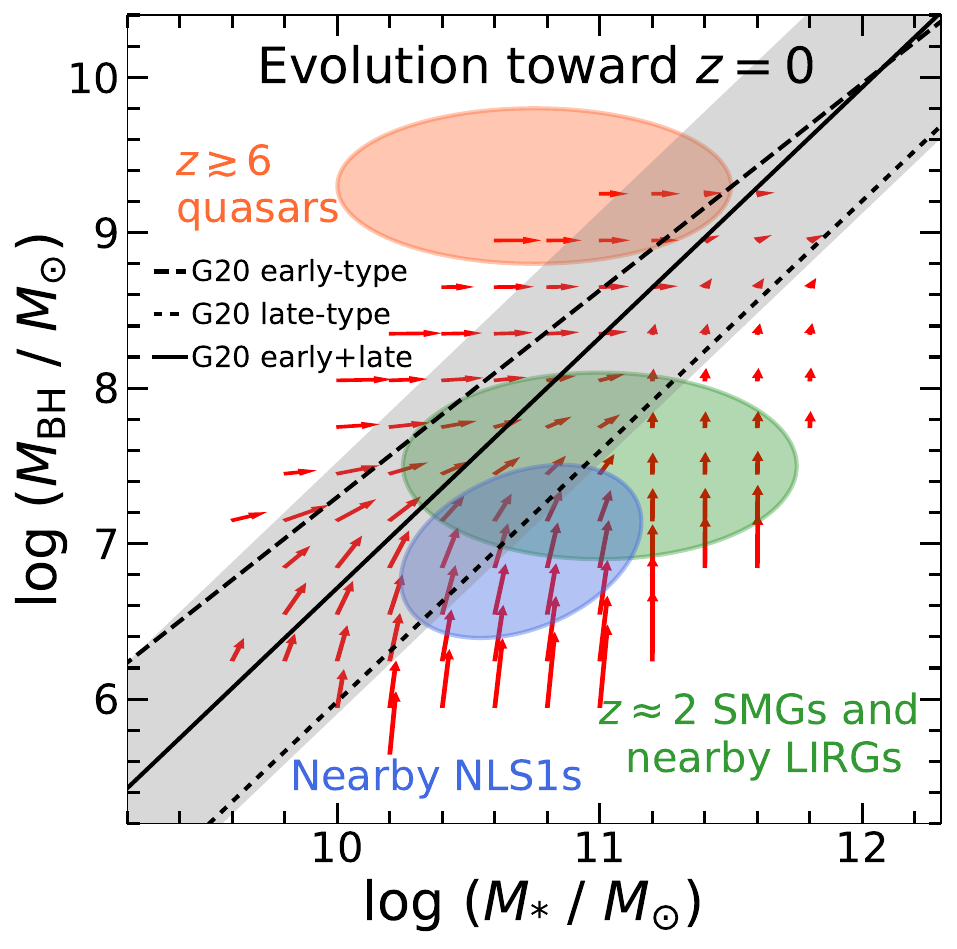}
\caption{Evolutionary paths toward $z=0$ of objects on the $M_{\rm BH}-M_*$ plane. Evolutionary paths are indicated by the direction and length of each arrow. The red, green, and blue ellipses represent typical locations of $z\gtrsim6$ quasars\cite{Decarli+2018ApJ, Neeleman+2021ApJ}, $z\approx2$ submillimeter galaxies, SMGs and nearby luminous infrared galaxies, LIRGs\cite{Borys+2005ApJ, Alexander+2008AJ, Farrah+2022MNRAS}, and nearby narrow-line Seyfert 1s, NLS1s\cite{Mathur+2012ApJ, Zhao+2021ApJ}. Best-fit relations from G20 are shown for local early-type galaxies (black dashed line), late-type galaxies (black dotted line), and both types combined (black solid line; intrinsic scatter of 0.81~dex indicated as shaded gray region).}
\label{fig4}
\end{figure}

\begin{figure}[t!]
\centering
\includegraphics[width=0.7\textwidth]{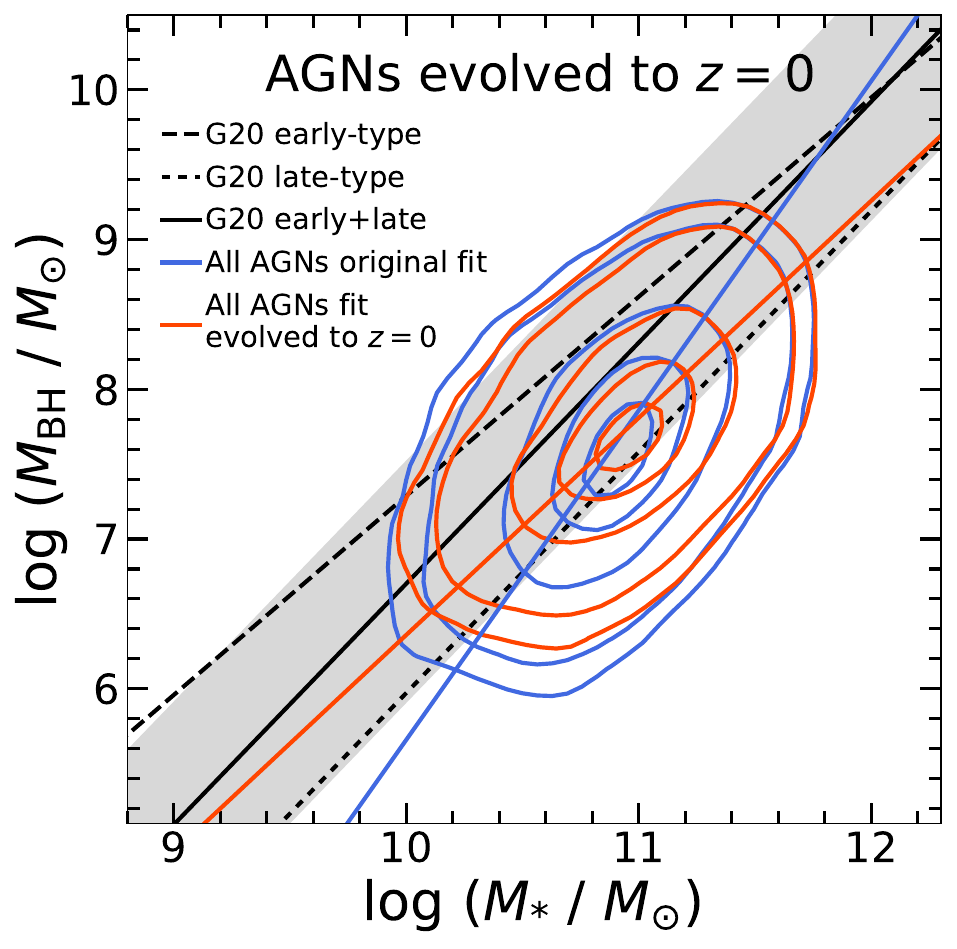}
\caption{Comparison of the current distribution of low-$z$ AGNs (blue) and the expected distribution (red) when they have evolved to $z=0$. Contours indicate the distribution of 10\%, 30\%, 60\%, 90\%, and 95\% of the entire sample, respectively. The blue solid line gives the fit to AGNs of all morphological types, and the red solid line is the fit to the expected counterparts of the AGNs at $z=0$. Best-fit relations from G20 for local early-type galaxies (black dashed line), late-type galaxies (black dotted line), and both types combined (black solid line; intrinsic scatter of 0.81~dex indicated as shaded gray region).}
\label{fig5}
\end{figure}

To visualize the relative positional shift of AGNs toward $z=0$ on the $M_{\rm BH}-M_*$ plane, we compare their distributions in Fig.~5, together with the fits to the entire sample of AGNs (regardless of morphology) before and after applying the evolutionary correction. The original distribution (total scatter 0.72~dex)
\begin{equation} \label{eq3}
\log\, (M_{\rm BH}/M_{\odot}) = (2.20 \pm  0.04) \log\, (M_*/10^{11} M_{\odot}) +  (7.86 \pm 0.30)
\end{equation}

\noindent
evolves to $z=0$ to become (total scatter 0.51~dex)
\begin{equation} \label{eq4}
\log\, (M_{\rm BH}/M_{\odot}) = (1.45 \pm 0.03) \log\, (M_*/10^{11} M_{\odot}) + (7.81 \pm 0.21).
\end{equation}
\noindent The fit to the original distribution yields a rather steep slope of 2.20, likely a consequence of mixing hosts of early and late types, which have slightly different normalizations. After applying the evolutionary correction, the slope of 1.45 lies between the values of inactive early-type ($1.33\pm0.12$) and late-type ($1.61\pm0.24$) galaxies\cite{Greene+2020ARA&A}.

Intriguingly, the zero point for the AGNs remains offset below that of inactive galaxies, by 0.5~dex at $M_*=10^{11} M_{\odot}$. This difference is likely contributed by many factors, such as sample size ($\sim11,500$ in this work versus $\sim120$ in G20), potential selection biases, and normalization and relative fractions of different morphological types. The upper envelope of the distribution of objects after applying the evolution correction overlaps with early-type inactive galaxies at $z=0$. The relatively small fraction of these objects suggests that the BHs in the majority of the local inactive ellipticals and classical bulges that define the local relation have not been active since $z=0.35$. As both cosmic star formation history and BH accretion history peak at $z\approx2$\cite{Kormendy&Ho2013ARA&A, Madau&Dickinson2014ARA&A}, these massive objects likely experienced most of their mass growth at high redshifts. \\

\noindent \textbf{\large Implications for the Coevolution of Supermassive BHs and their Host Galaxies} 

Our results reveal clear indications that the level of BH accretion and star formation of a galaxy is related to its position on the $M_{\rm BH}-M_*$ plane. We illustrate these trends graphically by overlaying in Fig.~4 the typical locations of objects that are found to exhibit various degrees of departures from the canonical locus of inactive galaxies. We do not consider the effects of mergers\cite{Graham&Sahu2023}, which play a minor role in triggering AGN activity in the nearby Universe\cite{Steinborn+2018MNRAS}. 

Why are there undermassive BHs, and how do they grow? Objects with undermassive BHs on the $M_{\rm BH}-M_*$ plane invariably have late-type, disk-dominated morphologies (Fig.~1). In light of their relatively blue colors (Extended Fig.~4), it is likely that they contain substantial amounts of cold gas. Low-redshift AGNs similar to those studied here typically have $\sim 10^9\,M_\odot$ of molecular gas\cite{Zhuang&Ho2020ApJ}, which maintains healthy levels of star formation\cite{Zhuang&Ho2022ApJ}. In this context, a parallel may be drawn with another class of actively star-forming, massive galaxies notable for having BHs undermassive with respect to their stellar masses, namely the nearby luminous infrared galaxies (LIRGs)\cite{Farrah+2022MNRAS} and their distant counterparts, the submillimeter galaxies (SMGs) at $z \approx 2-4$\cite{Borys+2005ApJ, Alexander+2008AJ}. These, too, possess a significant stellar disk component\cite{Veilleux+2006ApJ, Chen+2022ApJ}, and they are naturally rich in dust and gas\cite{Greve+2005, Shangguan+2019}. The comparison can be further extended to the NLS1s, which also share some similarities insofar as their BHs appear to be undermassive\cite{Mathur+2012ApJ, Zhao+2021ApJ} and their host galaxy morphologies are conspicuously late-type\cite{Orban+2011MNRAS}. Statistical studies of the cold interstellar medium of low-redshift quasars, among them members that qualify as NLS1s, detect abundant cold gas that sustains vigorous ongoing star formation\cite{Shangguan+2020ApJ, Xie+2021ApJ}. In the gas-rich environments that typify these classes of sources, gas inflow leads to star formation and BH accretion. Although both processes draw from a common fuel reservoir, the two are not tightly synchronized because they occur on different spatial and temporal scales. Star formation can take place over a wide range of spatial scales, so long as the gas is gravitationally unstable, while BH accretion occurs only in the innermost region of a galaxy after gas has lost sufficient angular momentum. Star formation proceeds on a local dynamical timescale, which depending on the location within a galaxy, on the order of $10^8$~yr\cite{Kennicutt1998ApJ}, while AGN lifetimes are $\sim1-2$ orders of magnitude shorter\cite{Martini2004}. Hence, star formation outpaces BH growth, resulting in more rapid increase of $M_*$ compared to $M_{\rm BH}$. Of course, other factors can also stunt the growth of the BH\cite{Habouzit+2017MNRAS, Catmabacak+2022MNRAS}, such as strong supernova feedback or insufficient early BH merger events.

Recent studies of the host galaxies of luminous AGNs conclude that, far from being gas-deficient, they have at least as much molecular gas as normal star-forming galaxies of the same stellar mass\cite{Koss+2021ApJS}. Closer inspection of the subset with adequate follow-up observations shows the gas distribution to be compact\cite{Molina+2021ApJ}, and the star formation efficiency to be elevated\cite{Shangguan+2020ApJ}. The relatively constant molecular gas fraction and especially the correlation between molecular gas mass and gas fraction with high $\lambda_{\rm E}$\cite{Koss+2021ApJS} support the idea that the gas supply to the BH, at fixed $M_{\rm BH}$, increases with increasing $M_*$. This would trigger a higher BH accretion rate and thus larger mass growth along the horizontal direction toward larger $M_*$, as observed (Figures~2b and 3c). Super-Eddington accretion associated with the late-stage mergers of LIRGs can grow undermassive BHs by up to an order of magnitude\cite{Farrah+2022MNRAS}, as found in simulations\cite{Weinberger+2018MNRAS}, and restore them to the local $M_{\rm BH}-M_*$ relation of early-type galaxies. The predicted evolutionary path of systems with undermassive BHs is qualitatively consistent with recent simulations of BH growth\cite{Habouzit+2017MNRAS, Catmabacak+2022MNRAS}, wherein at early times, when $M_*$ is low, the growth of the BH lags that of the stars because of supernova feedback, while later on the BH catches up its host after the gas reservoir has become stabilized at higher $M_*$. This resonates with the $M_*$-dependent ratio of $\dot{M}_{\rm BH}$ to SFR in X-ray--selected AGNs, for which $\dot{M}_{\rm BH}$/SFR increases toward higher $M_*$\cite{Aird+2019MNRAS}. 

At the opposite extreme, we draw attention to the subset of AGNs that lies clearly above the $M_{\rm BH}-M_*$ relation of inactive early-type galaxies. Although such objects account for $\lesssim10\%$ of our sample, they may be just the tip of an iceberg representing the most extreme of the galaxy population. These overmassive BHs are reminiscent of $z \gtrsim 6$ quasars\cite{Decarli+2018ApJ, Neeleman+2021ApJ}, which, in view of the limited time ($< 1$~Gyr) available for their rapid formation and growth, pose a challenge for possible formation scenarios\cite{Inayoshi+2020ARA&A}. Although the requirement is less stringent for our low-$z$ AGNs, they may share similar growth channels. Moreover, at least some of these objects may be the remnants of less extreme high-$z$ quasars. 

Studies of local massive elliptical galaxies---the direct descendants of the most massive, distant quasars---have established that they experienced substantial size growth compared to their precursor red nuggets\cite{Daddi+2005ApJ} at $z \approx 2$, through the accumulation of an extended outer envelope\cite{van_Dokkum+2010ApJ}. In the popular two-phase formation scenario of early-type galaxies\cite{Oser+2010ApJ}, the transformation of red nuggets to today's massive ellipticals proceeded largely through a series of minor (but see \cite{Davari+2017}), dry mergers. The relatively blue colors and hence high sSFR of the AGNs with overmassive BHs (Fig.~2a), which necessarily must gain stellar mass in order to converge to the $M_{\rm BH}-M_*$ relation of inactive early-type galaxies (Fig.~3b) suggest that their growth evidently was not completely dissipationless and involved some level of star formation.

The growth trajectories for the majority of the AGNs in our sample follow the local $M_{\rm BH}-M_*$ relation. Applying an evolutionary correction affects the bulk of the objects rather modestly, with significant changes occurring only at the outskirts of the distribution (Fig.~5). The mutual dependence of $L_{\rm bol}$ and SFR on the molecular gas reservoir\cite{Zhuang+2021ApJ} may account for the lockstep growth of $M_{\rm BH}$ and $M_*$. 

Our results are based on nearby AGNs. Abundant gas and frequent mergers at higher redshifts should lead to rapid evolution with more pronounced growth on the $M_{\rm BH}-M_*$ plane. For the obscured type~2 AGNs, it is difficult to directly measure BH mass for a large sample to investigate their $M_{\rm BH}-M_*$ relation and whether they behave similarly to the type~1 AGNs. However, based on 14 Seyfert~2 AGNs with near-infrared broad-line detections (see Methods), we find that they are located on the local relation and follow the same trends as their unobscured counterparts (Extended Fig.~5). This similarity suggests a common evolutionary scenario for both type~1 and type~2 AGNs.\\

\noindent \textbf{\large Implications for AGN Feedback}

AGNs with blue host galaxies have relative slowly accreting BHs, whereas AGNs with rapidly accreting BHs tend to be hosted by redder galaxies (Fig.~2). Are we witnessing the effects of negative AGN feedback? Radiative-mode feedback has been an appealing mechanism proposed for quenching star formation in galaxies\cite{Di_Matteo+2005Natur}, although linking it to observable consequences has proven not straight-forward\cite{Ward+2022MNRAS}. If negative AGN feedback operates effectively for the general population of active galaxies as viewed through their position on the $M_{\rm BH}-M_*$ plane, we expect $(g^{\prime}-r^{\prime})_0$ to correlate with $\lambda_{\rm E}$: more highly accreting BHs, having more powerful AGNs, inhibit star formation and turn the hosts red. This basic expectation is not borne out by our sample. Indeed, the exact opposite is seen (Extended Fig.~6). This disfavors the proposition that strong, instantaneous, negative feedback from accreting BHs quenches star formation, at least in nearby galaxies, in concert with the conclusions of previous studies of AGN hosts based on measurements of gas content, SFR, and star formation efficiency\cite{Shangguan+2020ApJ, Zhuang&Ho2020ApJ, Koss+2021ApJS, Xie+2021ApJ, Zhuang&Ho2022ApJ}. 

What about the AGNs with undermassive BHs with high $\lambda_{\rm E}$ and red $(g^{\prime}-r^{\prime})_0$ colors (Fig.~2)? At first glance, these traits appear to implicate negative AGN feedback. We argue, however, that AGN feedback only minimally affects the interstellar medium in the host galaxies of this population of objects. The consequences of AGN winds on the host galaxy depend on the properties of the shocks that they impart on the interstellar medium\cite{King&Pounds2015ARA&A}. If the wind cools efficiently on a timescale shorter than its flow time, most of its kinetic energy is lost. As momentum is conserved, a momentum-driven outflow injects only $\sim10\%$ of the binding energy of the bulge to the interstellar medium, which is hardly capable of clearing the surrounding gas and preventing accretion. It has been demonstrated that an outflow is momentum-driven when $M_{\rm BH}$ is smaller than a critical mass\cite{King&Pounds2015ARA&A}, which scales with the fourth power of the velocity dispersion (or, equivalently, mass) of the bulge, very close to the observed local relation\cite{Tremaine+2002ApJ}. If we assume a bulge-to-total ratio of 0.2, as typical of late-type AGN host galaxies\cite{Zhao+2021ApJ} and nearby inactive disk galaxies\cite{Gao+2019}, most objects with a predicted increase of $M_{\rm BH}$ larger than $20\%$ still lie below the $M_{\rm BH}-M_{\rm bulge}$ relation of early-type galaxies. This implies that $M_{\rm BH}$ is smaller than the critical mass and that the outflows are momentum-driven, rendering them incapable of shutting down BH growth or curtailing star formation (Extended Fig.~7). This is consistent with the results from a recent detailed analysis of the properties of the X-ray outflows in I~Zwicky~1, a well-studied NLS1 that is located far below the local relation\cite{Ding_YZ+2022ApJ}.

When $M_{\rm BH}$ reaches the critical mass, as is the case for AGNs with overmassive BHs above the early-type relation, the outflow becomes energy-driven, and the adiabatically expanding bubble from kinetic feedback efficiently expels the inflowing gas and suppresses further BH growth. However, our analysis suggests that $M_*$ increases significantly even after substantial gain in $M_{\rm BH}$ (Fig.~3b). This implies that although AGN feedback can suppress the growth of the BH, it fails to prevent long-term star formation on galactic scales, seemingly at odds with the conclusions of recent cosmological simulations\cite{Weinberger+2018MNRAS, Dave+2019MNRAS}.

\newpage
\section{Methods}

\subsection{Low-$z$ Type~1 AGN Sample}

We analyze 14,574 type~1 AGNs with $z\leq 0.35$ selected by Liu et al. (2019)\cite{Liu+2019ApJS} from the seventh data release (DR7)\cite{Abazajian+2009ApJS} of the Sloan Digital Sky Survey (SDSS)\cite{York+2000AJ} to have broad \Ha\ emission. The redshift cut ensures that \Ha\ lies within the SDSS spectral coverage. This constitutes the largest low-$z$ type~1 AGN sample studied in terms of their host galaxy properties. It surpasses other similar samples by a factor of $\sim3$ in size, and it covers a wider range of AGN luminosities owing to the selection based on the detection of broad \Ha\ instead of \Hb, as well as the lower threshold on line width used to define broad emission lines\cite{Shen+2011ApJS, Oh+2015ApJS, Lyke+2020ApJS, Rakshit+2020ApJS}. The emission lines are measured after subtracting a model for the continuum, which includes a broken power law for the AGN accretion disk, a Balmer continuum, if present, an optical \FeII\ pseudo-continuum, and the host galaxy starlight represented by stellar population templates\cite{Liu+2019ApJS}.
Liu et al.'s catalog provides basic properties such as the AGN continuum luminosity at 5100~\AA\ ($L_{5100}$), the fluxes, line widths, and equivalent widths of strong permitted and forbidden lines, as well the strength of the broad \FeII\ emission blend at 4434--4684~\AA. We estimate the AGN bolometric luminosity ($L_{\rm bol}$) from $L_{5100}$ with a bolometric correction of $L_{\rm bol}=9.8L_{5100}$\cite{McLure&Dunlop2004MNRAS}.

We obtain BH masses ($M_{\rm BH}$) from the full width at half-maximum (FWHM) of broad \Hb\ and $L_{5100}$, using the virial mass estimator based on single-epoch spectra\cite{Ho&Kim2015ApJ}. This estimator takes into account different calibration coefficients for pseudo and classical bulges, which have virial factors of $f=3.2$ and 6.3, respectively\cite{Ho&Kim2014}. We adopt the calibration for pseudo bulges for hosts with reliable late-type, disk-dominated morphology, and we use the zero point for classical bulges for hosts that are robustly classified as early-type, bulge-dominated systems. Hosts without reliable morphologies get assigned the zero point for both bulge types combined. When broad \Hb\ is absent, we convert the FWHM of broad \Ha\ to that of broad \Hb\ using the empirical relation between the two line widths\cite{Greene&Ho2005ApJ}. The derived BH masses are on average  $0.17\pm0.26$~dex lower for late-type hosts and $0.23\pm0.28$~dex higher for early-type hosts compared to those derived from broad \Ha\ alone with $f=4.47$\cite{Woo+2015ApJ}. Switching to the \Ha-based $M_{\rm BH}$ estimator does not qualitatively affect our main conclusions. Extended Fig.~3 shows a different version of Fig.~2 without performing smoothing and with $M_{\rm BH}$ estimated from broad \Ha.

Recent reverberation mapping observations of nearby type~1 AGNs suggest that the scatter in the relation between broad-line region radius and luminosity depends systematically on Eddington ratio\cite{Du+2016ApJ, Du&Wang2019ApJ, Dalla_Bonta+2020ApJ} (but see counter argument\cite{Fonseca2020}), $\lambda_{\rm E} \equiv L_{\rm bol}/L_{\rm E}$, where $L_{\rm E}=1.26 \times 10^{38}\,(M_{\rm BH}/M_\odot)$ is the Eddington luminosity. If this dependence on Eddington ratio is not taken into account, the virial BH masses estimated from single-epoch spectra, which rely on the radius-luminosity relation, will be underestimated for AGNs with low $\lambda_{\rm E}$ and overestimated for those with high $\lambda_{\rm E}$. This effect might impact our results, in view of the variation of Eddington ratio with morphological type observed in our sample. To mitigate this potential systematic effect, we conservatively assign an uncertainty of 0.4~dex to the $M_{\rm BH}$ estimates.

The final sample after excluding objects with unreliable measurements on host properties ($\sim78\%$ of the original) consists of 11,474 moderately luminous to powerful AGNs, which are characterized by $L_{\rm bol} = 10^{41.9}-10^{46.4}$ erg~s$^{-1}$ (median $10^{44.6}$ erg~s$^{-1}$), $M_{\rm BH} = 10^{5.1}-10^{10.0}\, M_{\odot}$ (median $10^{7.6} M_{\odot}$), and $\lambda_{\rm E} = 10^{-3.5}-10^{0.7}$ (median $10^{-1.2}$). It still well represents the parent sample (Supplementary Fig.~1), with a major difference being the significant reduction (by $\sim 50\%$) of the number of objects with $L_{\rm bol}\gtrsim 10^{45.4}$ erg~s$^{-1}$. This sample is nearly 40 times larger than that in any previous work in the nearby ($z\leq0.35$) Universe\cite{Treu+2007ApJ, Bennert+2010ApJ, Falomo+2014MNRAS, Matsuoka+2014ApJ, Reines&Volonteri2015ApJ, Sun+2015ApJ, Bentz&Manne-Nicholas2018ApJ, Kim&Ho2019ApJ, Ishino+2020PASJ, Bennert+2021ApJ, Li_Junyao+2021aApJ, Li_Jennifer+2021ApJ, Son+2022JKAS}.

\subsection{Low-$z$ Inactive Sample}

To investigate systematically the AGN host galaxies and the potential biases that can be introduced from the presence of a bright nucleus, it is vital to select a sample of inactive galaxies in the same redshift range that covers a similar parameter space in host galaxy properties as the AGN sample. We start from the SDSS data release 16 spectroscopic catalog (\texttt{specObj} table; \url{https://data.sdss.org/sas/dr16/sdss/spectro/redux/specObj-dr16.fits}) and limit to galaxies within a radius of 2.5$^{\prime}$ of each AGN. This is the minimum size of the image cutout of the AGNs, one that yields similar image properties, in terms of depth and point-spread function (PSF), for both the galaxy and AGN sample. We exclude duplicate observations of the same objects and spectroscopically identified AGNs by omitting objects with \texttt{CLASS=QSO}, \texttt{SUBCLASS=AGN} or \texttt{AGN BROADLINE}. The above \texttt{SUBCLASS} criteria remove objects classified as (broad-line) Seyferts and low-ionization nuclear emission-line regions (LINERs)\cite{Heckman1980, Ho2008ARA&A} based on the \OIII/\Hb\ versus \NII/\Ha\ line-intensity ratio diagnostic diagram\cite{Baldwin+1981PASP}, to leave a clean sample of inactive galaxies free of potential effects of the central AGN on the galaxy properties. After selecting objects with $0.01\leq z \leq 0.35$ and \texttt{ZWARNING=0}, we obtain 12,236 inactive galaxies, among which 11,576 are retained after excluding 538 objects without image coverage (located in a different image tile) and 122 objects affected by image masks. As we are mainly concerned about the reliability of photometric and structural properties of the AGN host galaxies, the final inactive galaxy sample consists of 11,017 objects ($95.2\%$) with robust (measurement/error $> 3$) magnitudes in all five bands, half-light radii, and \sersic\ indices, as well as successful spectral analysis from spectral energy distribution (SED) fitting. The inactive sample, apart from the absence of an active nucleus, ideally should have exactly the same distribution of galaxy properties as the sample of active galaxies. This is difficult to realize in practice, but the above-constructed inactive galaxy sample suffices for our primary goal of probing the typical photometric parameter space spanned by the AGN host galaxy sample.

\subsection{Reduction of the Pan-STARRS1 Data}

We use the coadded five-band ($grizy_{\rm P1}$) images (pixel size 0.25$^{\prime \prime}$) from Pan-STARRS1 (PS1)\cite{Chambers+2016arXiv}. We retrieve the stack images, weight images, and mask images, as well as the source catalog that includes coordinates, PSF magnitudes ($m_{\rm PSF}$), and Kron\cite{Kron1980} magnitudes ($m_{\rm Kron}$) from the second data release (DR2)\cite{Flewelling+2020ApJS} of PS1. Each image cutout is centered on the position of the AGN with an initial size of $5^{\prime} \times 5^{\prime}$, and the size increases in steps of 1$^{\prime}$ according to the number of stars available for PSF construction. We exclude 92 objects that lack an image in at least one band, that are affected by close bright stars and foreground galaxies, or that have at least 50\% of the pixels masked in the central 1 arcsec$^2$. For the remaining 14,482 AGNs, we subtract a two-dimensional background model and generate segmentation maps and source catalogs using \texttt{SExtractor}\cite{1996A&AS..117..393B}. It has been demonstrated that PSF models generated by stacking nearby stars perform best in describing the light profiles and recovering the magnitudes of point-like sources by comparing the behavior of different codes for PSF construction, finding no significant spatial variation within the $\sim$30 arcmin$^2$ field-of-view of the stacked PS1 images\cite{Zhuang&Ho2022ApJ}. Following their procedure, we use \texttt{SWarp}\cite{Bertin+2002ASPC} to construct PSF models by stacking isolated point-like sources in the image cutout, which are selected by requiring $m_{\rm PSF}<21$~mag and $m_{\rm PSF} - m_{\rm Kron}<0.05$~mag in the $i_{\rm P1}$ band\cite{Farrow+2014MNRAS}. The number of applicable stars is determined by gradually increasing the size of the image cutout so that the magnitudes of the PSF models are brighter than $m_{\rm PSF}$ of the AGN from the PS1 catalog (conservative overestimation) in all five bands, to ensure enough signal-to-noise ratio to model the light profile of the nucleus. 

We summarize the image properties of our sample in Supplementary Fig.~2 and Supplementary Table~1, which gives the statistics of the FWHM of the PSF obtained from fitting a circular, two-dimensional Moffat\cite{Moffat1969} function, the $3\sigma$ surface brightness depth, and the $5\sigma$ point source depth, where $\sigma$ is the standard deviation of 100 flux measurements made using a circular aperture of diameter 15$^{\prime \prime}$ for extended sources and 2 times the PSF FWHM for point sources, randomly placed in the source-masked background of the image.

\subsection{Two-dimensional Simultaneous Multiwavelength Fits} 

The morphology and structure of a galaxy can change significantly with wavelength as a result of variations in stellar population, metallicity gradient, or dust attenuation. Using \texttt{GALFITM}\cite{Haussler+2013MNRAS, Vika+2013MNRAS}, it has been demonstrated that both the \sersic\cite{Sersic1968} index ($n$) and half-light radius ($R_e$) of a galaxy vary smoothly from the ultraviolet to the near-infrared, with increasing $n$ and decreasing $R_e$ toward longer wavelengths\cite{Kelvin+2012MNRAS}. \texttt{GALFITM} is a multiwavelength version of \texttt{GALFIT}\cite{Peng+2002AJ, Peng+2010AJ}, which includes all the functions of the original code and models the wavelength dependence of parameters using a series of Chebyshev polynomials. By simultaneously fitting multiple images at different wavelengths, \texttt{GALFITM} can improve the accuracy, reliability, and signal-to-noise ratio of the derived parameters\cite{Haussler+2013MNRAS, Haussler+2022A&A}. 

We use \texttt{GALFITM} to model the unresolved nucleus with the PSF and the underlying host galaxy with a single \sersic\ component. The choice of wavelength dependence of model parameters follows that of previous work\cite{Zhuang&Ho2022ApJ}. We keep the ellipticity and position angle constant across all bands, but allow the magnitudes of the PSF and \sersic\ components to vary with wavelength as a fourth-order polynomial, while $n$ and $R_e$ change linearly with wavelength. The positions of the PSF and \sersic\ components are tied together, under the assumption that the supermassive BH is located at the center of the galaxy. 

We also perform a set of decompositions with the positions of the PSF and \sersic\ components untied, allowing their central coordinates to vary by up to 10 pixels. Consistent results are obtained in the $i_{\rm P1}$ band for \msersic, $R_e$, and $n$, with no systematic differences in the median and a scatter of $\lesssim0.06$ mag, 0.03 dex, and 0.07 dex, respectively. We caution that leaving too much freedom to the central positions of the two components may lead to artificial offsets in hosts with fainter surface brightness\cite{Zhuang&Shen2023}. The segmentation map and source properties returned by \texttt{SExtractor} are used to generate the source mask and to set up the initial guesses for the model parameters and fitting region. We use the weight maps from PS1 DR2, which properly account for the uncertainty of the image using information from the image stacking and data reduction. All the sources inside the fitting region are masked, except for the AGN and possible blended companions identified by \texttt{SExtractor} with \texttt{DEBLEND\_NTHRESH=32} and \texttt{DEBLEND\_MINCONT=0.005}, which, if present, are fitted simultaneously. The fitting region is set to be the larger value between 50 pixels and $10$ times \texttt{A\_IMAGE}, which is the length of the semi-major axis of the isophotal elliptical shape of the source. If a companion is present, the source extent and the separation between the AGN and the companion are summed up to determine the final fitting region. We model the companion with a PSF function if its \texttt{CLASS\_STAR > 0.9}, and as a \sersic\ function otherwise. We use $m_{\rm PSF}$ from the PS1 catalog and \msersic\,$=-2.5\, \log\,(10^{-m_{\rm Kron}/2.5} - 10^{-m_{\rm PSF}/2.5})$ as initial guesses of the magnitudes for the PSF and \sersic\ components, respectively. For the structural parameters of the \sersic\ model, we take values from the output of \texttt{SExtractor} for $R_e$, axis ratio, and position angle, but we restrict the initial guess of $n$ to 2. We constrain $R_e$ to vary within $0.5-500$ pixels and $n$ within values of $0.3-8$. 

The simultaneous, multi-object, multiwavelength decomposition is illustrated in Supplementary Fig.~3 for a late-type host galaxy, and in Supplementary Fig.~4 for an early-type system. In Supplementary Fig.~3, the prominent emission from the nucleus is well extracted and clearly visible in the surface brightness radial profiles, despite its small contribution (as low as 3\%) to the total flux. The hole in the central region, evident in the data $-$ model and residuals images, indicates that the nucleus is oversubtracted. As discussed in the following subsection, this is more severe in objects with lower AGN-to-host contrast but does not significantly affect the measured galaxy properties. Although substructures from the bulge, bar, and spiral arms can leave noticeable residuals, a single-component \sersic\ model can recover the total magnitude of the galaxy as robustly as more detailed (e.g., bulge+disk) models\cite{Casura+2022MNRAS, Haussler+2022A&A}. 

Successful fits were obtained for a total of 12,903 objects. We cannot derive reliable magnitudes (magnitude error $>0.36$~mag, equivalent to flux/flux error $<3$) for 1,579 sources, which are mainly powerful ($L_{\rm bol}\approx 10^{45.0}$~erg~s$^{-1}$) AGNs at higher redshift ($z\approx0.3$).

\subsection{Verification of Results}

Although a single-component \sersic\ model can deliver a reliable total magnitude for the galaxy, the situation is more complicated with the presence of a prominent active nucleus. The structure and color of a galaxy can significantly affect the magnitude of the nucleus, and the bright nucleus can also bias the measurements of the galaxy parameters. Moreover, the formal errors returned by fitting codes always underestimate the true uncertainties, regardless of the goodness of the fit\cite{Haussler+2007ApJS, Gao2017}. We perform the most comprehensive input-output tests to date to derive realistic uncertainties and study the potential systematics introduced by the presence of a bright nucleus. We produce two sets of mock AGN images: one from model galaxies generated from the best-fit parameters, and the other from actual observations of low-$z$ inactive galaxies with structural and photometric properties carefully matched to each AGN. The second approach is often employed in the literature\cite{Fan+2014ApJ, Li_Junyao+2021aApJ}, but only using control samples matched in redshift and stellar mass, neglecting the effect of color and morphology of the AGN host galaxies.

\subsubsection{Mock AGNs from Idealized Galaxies}

We quantify the uncertainty of each model parameter by constructing idealized mock AGN images using the best-fit parameters of each object while mimicking the conditions of its imaging properties. Three sources of noise are applied independently to the data: (1) Poisson noise associated with the source, (2) Gaussian noise of the individual background pixels, and (3) Gaussian noise of the constant background after background subtraction prior to the decomposition. The last source of noise is related to the extent of the source and is measured as the standard deviation of 100 ellipses randomly distributed in the source-masked, background-dominated regions (the background constitutes more than $95\%$ of all pixels), with semi-major axis 3 times $R_e$ and axis ratio and position angle fixed to the best-fit values of the AGN host; these apertures enclose $96\%$ and $79\%$ of the total light for \sersic\ profiles with $n=1$ and $n=4$, respectively. We generate 100 realizations of mock AGNs for each object in each band, and we repeat the decomposition using \texttt{GALFITM} with the same configuration file and constraints as for the real observations. The final measurement and its uncertainty are taken as the median value and standard deviation of the 100 results. We exclude 580/12,903 ($4.5\%$) objects for the following reasons: (1) they have unreliable magnitudes for both the PSF and \sersic\ components; (2) the fitting results for the magnitudes, $R_e$, and $n$ of the observed and mock AGNs differ by more than 3 times the uncertainty of the mock AGNs; and (3) the magnitude of the \sersic\ component is unreliable in at least two bands for either the real observations or the mock AGNs, which would render the SED analysis highly uncertain. The uncertainties derived from the idealized model AGNs are added in quadrature to the final error budget of each parameter. The same procedure is adopted also for the low-$z$ inactive galaxy sample, which resulted in 11,479 objects.

\subsubsection{Mock AGNs from Real Galaxies}

Model mismatch due to the presence of galactic substructures, such as the bulge, bar, ring, lens, or spiral arms, even if they are not well resolved at relatively higher redshifts, can lead to significant systematic parameter bias from our adoption of a simplified two-component (PSF + \sersic) model. It is of fundamental importance to understand how the final parameters of each AGN host galaxy may be affected by this possible source of systematic bias. To accomplish this goal, we generate mock AGNs by manually adding an artificial point source to the center of each real galaxy selected from the low-$z$ inactive sample. We then study the derived parameters before and after adding the point source. Although we do not consider explicitly galaxy substructures, we note that $n$ does correlate with galaxy concentration and bulge-to-total ratio\cite{Simard+2011ApJS}.

Galaxies that are similar to each AGN are selected based on the error-weighted difference (${\rm \mid AGN-gal \mid}/\sqrt{\rm error(AGN)^2 + error(gal)^2}$) of four parameters: observed $(g-i)_{\rm P1}$ color, $R_e$, $n$, and axis ratio. The error of axis ratio is not used but instead the largest difference between the galaxy and the AGN cannot exceed 0.1. If the error is too small, we adopt a fiducial value of 0.1 mag for color and 10\% for $R_e$ and $n$. An integer factor of $2-5$ is multiplied to the error to increase the number of inactive galaxies if it is fewer than 20. We exclude 33 AGNs that do not have enough inactive galaxies even after applying an error boost factor of 5. If more than 20 objects satisfy our criteria, we choose the top 20 with the smallest magnitude difference between the AGN and the inactive galaxy in the $i_{\rm P1}$ band. Supplementary Fig.~5 illustrates our procedure for the target AGN SDSS~J011448.68$-$002946.1 (already shown in Supplementary Fig.~3), in which we compare its $i_{\rm P1}$-band image to those of a collection of 20 inactive galaxies that have similar structural properties and color. In total, we use 245,800 inactive galaxies (each galaxy can be used multiple times) to generate mock images for 12,290 AGNs. The error boost factor is $\leq3$ for $\sim95\%$ of the sources. The distributions of differences between AGNs and inactive galaxies in observed-frame $(g-i)_{\rm P1}$ color, $R_e$, and $n$ confirm the similarity between the real AGNs and their mocks (Supplementary Fig.~6). The median differences cluster near 0 with a small dispersion; the tails of the distribution arise from a tiny fraction of sources with relatively large parameter errors. 

To generate the corresponding mock active galaxies for a given AGN target, we add a point source to the center of each of the 20 inactive galaxies. The magnitude of the point source in each band is calculated from $L_{\rm bol}$ and the composite quasar SED\cite{2006ApJS..166..470Richards+}, independent of the robustness and reliability of $m_{\rm PSF}$ from the image decomposition. To ensure a consistent AGN-to-host contrast in each band, these magnitudes are further subject to corrections according to the magnitude difference between the host galaxy of the AGN target and each inactive galaxy.

We perform simultaneous multiwavelength decomposition of the mock AGNs following the same procedure as employed for the real AGNs. Supplementary Fig.~7 shows the general trend of the difference ($\rm \Delta=AGN-galaxy$) between the results measured in the mock AGNs and those measured in the inactive galaxies used to generate them, as a function of \msersic, $R_e$, $n$, and $m_{\rm AGN}-$\msersic, which indicates the AGN-to-host contrast. We find that the $R_e$ of AGN host galaxies can be measured reliably with no systematic bias ($\Delta \log R_e\approx 0.01$~dex; scatter $\sim0.03$~dex in all five bands) and with little dependence on the properties of the galaxy or relative AGN brightness. By contrast, \msersic\ and $n$ exhibit small systematic biases ($\Delta$\msersic$=0.07$~mag; $\Delta \log\, n=-0.11$~dex) and slightly larger scatter compared to $R_e$ (90\% of the sources are located within a dispersion of $\sim0.25$~mag for $\Delta$\msersic\ and $\sim0.2$~dex for $\Delta \log\,n$). The scatter of $\Delta$\msersic\ and $\Delta \log\,n$ are affected by \msersic\ and the AGN-to-host contrast, with larger scatter toward fainter host magnitude and relatively smaller scatter in redder, more star-dominated bands. We also find a tendency for $n$ to be underestimated in galaxies with larger $R_e$. Among the parameters investigated here, $m_{\rm AGN}$ performs least favorably. The AGN brightness is systematically overestimated when the nucleus is much fainter than its host galaxy, in the most extreme cases by as much as $\sim 3$~mag when the AGN is only 1\% as bright as the host. We apply these object-tailored corrections to every AGN and combine them with the results from previous subsection to derive the final value and uncertainty of each parameter.

\subsection{Comparison to Decomposition Results for Two-component \sersic\ Models}

To investigate in detail the effect on host magnitude when neglecting substructures, we select all late-type ($n<2$) objects with sufficient resolution (PSF FWHM $<2$ kpc in all five bands; resolution better than 1 kpc in radius) to detect potential central substructures, such as a bulge. This gives us 581 AGNs, $\sim 5\%$ of the parent sample. We perform a new set of image decomposition for these objects, following exactly the same procedure used for the parent sample but using, instead, a two-component \sersic\ model, where for the first component the \sersic\ index set to $n = 1$ to mimic emission from a disk.  The position of both components are tied together, and $R_e$ for the second component is constrained to be smaller than that of the first. 
 
We find that in the $i_{\rm P1}$ band the total host magnitude (sum of the fluxes of both \sersic\ components) from the double-component fit is consistent with that from the single-component fit, with a median difference of $-0.05$~mag ($5\%$; fluxes from the double-\sersic\ fit are brighter) and a scatter of 0.16~mag ($15\%$). This indicates that the single-component can recover the total magnitude of the galaxy reliably. Moreover, the median difference is in excellent agreement with the median correction of $-0.06$ mag applied to \msersic\ for the single-component decomposition, as estimated from our mock AGNs based on real inactive galaxies. Nevertheless, systematic biases may still be present even for the two-\sersic\ models. It is much harder to follow our procedure of generating mock AGNs with similar morphologies to study systematic biases, because of the lack of a sufficient amount of real inactive galaxies with similar bulge and disk parameters. Assuming negligible systematic bias in host magnitude from the double-\sersic\ component decomposition, we estimate the stellar masses of host galaxies following the procedures and configurations to be introduced below. We find consistent results: the median difference in stellar mass is $\sim0.05 \pm 0.09$~dex, which is smaller than the uncertainty of individual objects ($\sim 0.10$ dex). 

We conclude that adding an extra host component has only a minor effect on the magnitudes and stellar masses of a small fraction of AGN host galaxies in our sample. For simplicity, throughout this work we adopt the results (\msersic, $R_e$, and $n$) from the single-component \sersic\ models.

\subsection{Stellar Masses of AGN Host Galaxies}

We derive the stellar masses of the host galaxies from the AGN-removed photometry, using version 2022.0 of the SED-fitting code \texttt{CIGALE}\cite{2019A&A...622A.103Boquien+, Yang+2022ApJ}. The flux densities of the five PS1 bands are first corrected for Galactic foreground extinction using the {\tt Python} package \texttt{dustmaps}\cite{2018JOSS....3..695Green}, using a Milky Way dust map\cite{1998ApJ...500..525Schlegel+} and converting $E(B-V)$ to extinction in the PS1 bands using $R_V=3.1$\cite{2011ApJ...737..103Schlafly&Finkbeiner}. An additional $10\%$ uncertainty is added in quadrature to the flux densities to account for the uncertainty in absolute flux calibration and other potential unknown factors. 
In light of the limited available photometric coverage of our SEDs, we adopt a parametric ``delayed'' star formation history combined with single stellar population models\cite{BC03}, a nebular emission-line model\cite{Inoue2011MNRAS}, and a dust attenuation law (\texttt{dustatt modified starburst}) based on the starburst attenuation curve\cite{Calzetti+2000ApJ}. Albeit simplistic, parametric star formation histories have been demonstrated to yield robust stellar masses with $\lesssim10\%$ systematic bias and $<20\%$ scatter, even when only optical photometry is available\cite{Mitchell+2013MNRAS, Ciesla+2015A&A}. The models and fitting parameters for \texttt{CIGALE} are given in Supplementary Table~2. We scale the stellar masses by a factor of 1.08\cite{Madau&Dickinson2014ARA&A} to convert the initial mass function from Chabrier\cite{Chabrier2003PASP} assumed by \texttt{CIGALE} to that of Kroupa\cite{Kroupa2001MNRAS} adopted in this paper. We also test the reliability of the stellar masses directly derived from the observed integrated (AGN$+$host) SED, using SED-fitting as described in the following section. In general, the stellar masses derived from integrating the SEDs can be overestimated by as much as $\sim0.5$ dex for AGNs of the strength typically found in our sample.

To select the most reliable SED measurements, we use the difference between the estimate of $M_*$ from the best-fitting template ($M_{*, \rm{best}}$) and the mean of the probability density function distribution of the likelihood of all templates ($M_{*, \rm{bayes}}$) and $\chi^2_{\nu}$. We apply a strict cut to keep only objects with the most secure SEDs, by excluding 816 ($6.6\%$) sources with $\mid \log \frac{M_{*, \rm{bayes}}}{M_{*, \rm{best}}}\mid>0.3$ and $\chi^2_{\nu} > 1$. For the remaining 11,474 AGNs, we obtain the rest-frame SDSS $g^{\prime}-r^{\prime}$ color, $(g^{\prime}-r^{\prime})_0$, and adopt $M_{*, \rm{bayes}}$ and its associated uncertainty as the final estimate of stellar mass. The median uncertainty of $M_*$ is $\sim0.10$ dex ($23\%$), which is larger than the uncertainties of the input flux densities ($\sim11\%$). This suggests that the uncertainty of $M_*$ is dominated by the degeneracy of the templates. 

G20\cite{Greene+2020ARA&A} assume a ``diet Salpeter'' initial mass function by using mass-to-light ratios from Ref.\cite{Bell+2003ApJS} to derive $M_*$. To match the initial mass function adopted in this work, we scale down their stellar masses by a factor of 1.06. 

\subsection{Stellar masses of AGNs derived from AGN$+$host SEDs}

It is common to derive the stellar masses of AGN host galaxies by fitting their total (AGN$+$host), integrated SEDs. To test the reliability of this method, we use the stellar masses and SEDs of the inactive galaxy sample to generate mock SEDs for the AGNs by adding a nuclear flux to each band, following the same procedure used to generate the mock AGN images. We consider three typical AGN strengths ($L_{\rm bol} = 10^{43}, 10^{44}$, and $10^{45}\, \rm erg~s^{-1}$) and limit the inactive sample to objects with $M_* = 10^{9.5}-10^{12}\, M_\odot$ to cover a parameter space similar to that of our actual AGN sample. We then fit their SEDs using exactly the same configuration of \texttt{CIGALE} models, but with an additional AGN component generated using \texttt{SKIRTOR}\cite{Stalevski+2012MNRAS, Stalevski+2016MNRAS}. We set the inclination angle to four discrete values ($i = 0^{\circ}$, $10^{\circ}$, $20^{\circ}$, $30^{\circ}$) and allow the AGN fraction (defined in the wavelength range 0.3--1.0 $\mu$m) to freely vary ($f_{\rm AGN}=0.001-0.9$), while fixing all other parameters to their default values, since they are related to the properties of the AGN torus and are not relevant here. 

Supplementary Fig.~8 compares the stellar masses derived from the mock AGNs and those measured in the inactive galaxies used to generate them (the ground truth). We find that the differences between the two sets of measurements and their scatter increase with increasing AGN $L_{\rm bol}$ (first column), likely a consequence of the increasing degeneracy between the AGN and host galaxy toward larger AGN-to-host contrast and bluer host galaxy color. As a result, \texttt{CIGALE} tends to overestimate $f_{\rm AGN}$ toward bluer host galaxies (vertical trend). We also find a turnover (``V''-shape) of the trend in relatively red host galaxies: $\Delta M_*$ first decreases and then increases with increasing AGN-to-host contrast. This can be understood as underestimation of $f_{\rm AGN}$ at low AGN-to-host contrast and overestimation of $f_{\rm AGN}$ toward high AGN-to-host contrast. As mass-to-light ratio rises for galaxies with redder colors\cite{Bell+2003ApJS}, stellar mass would be overestimated if $f_{\rm AGN}$ is overestimated, and vice versa. 

We also fit the AGN$+$host SEDs of the AGN sample using measurements of $m_{\rm Kron}$ from the PS1 DR2 catalog. Supplementary Fig.~9 compares stellar masses from fitting the AGN+host SEDs (SED) and those from SED-fitting to AGN contribution-subtracted host SEDs from image decomposition (IMG). Consistent with results of mock AGNs, stellar mass difference ($\Delta=$ SED$-$IMG) tends to be more positive toward smaller $M_*$, higher AGN-to-host contrast and bluer host color (Supplementary Figures~9a and 9b), but with larger scatter. The larger scatter is partly due to the difference in fluxes between model fitting and aperture photometry, as evident from a negative trend of $\Delta M_*$ with $n$. At larger $n$, a higher fraction of light is extended, with fluxes enclosed within an aperture of radius of 3 times $R_e$ contribute to 96\% and 79\% of the total fluxes for \sersic\ profiles with $n=1$ and $n=4$, respectively. This potential missing flux issue, together with the different aperture sizes for different bands, renders fluxes measured from aperture photometry less reliable and less self-consistent. Supplementary Fig.~9d indicates that after controlling for the effect of $n$, the scatter of $\Delta M_*$ is relatively small at low AGN-to-host contrast (consistent to the middle panel in Supplementary Fig.~8a), while the scatter increases as a combined effect of higher AGN-to-host contrast and large host color range (consistent to the middle panels in Supplementary Figures~8b and 8c). These results demonstrate the biases and uncertainties in using SED-fitting to derive stellar mass of type~1 AGNs using AGN$+$host SEDs.

\subsection{Color$-$Stellar Mass Diagram and Estimation of Specific Star Formation Rate}

Broad-band optical colors encode valuable information about galaxies. For example, the SDSS $g^{\prime}-r^{\prime}$ color is an indicator of star formation properties and is widely used to classify galaxies into red (quiescent) and blue (star-forming) sequences\cite{Baldry+2004ApJ, Blanton&Moustakas2009ARA&A}. Using the rest-frame SDSS $g^{\prime}-r^{\prime}$ color from SED fitting, we compare the distributions of active and inactive galaxies on the color-stellar mass diagram. Extended Fig.~1 splits the samples into two redshift bins, one local ($z < 0.15$) and the other slightly more distant ($0.15 \leq z \leq 0.35$). Note that the contours of the inactive galaxies are only used to establish the typical range of parameter space instead of giving a quantitative comparison with their active counterparts, since other properties (e.g., morphology or \sersic\ index) correlate with color\cite{Blanton+2003ApJ}, and inactive galaxies are biased toward luminous red galaxies at higher redshift. Nevertheless, we find that the AGNs in the two redshift bins span a wide range of color, residing in both blue and red galaxies. AGNs are hosted mostly in galaxies with stellar masses $M_* \gtrsim 10^{10}\, M_{\odot}$, which mainly reflects the selection effect that luminous AGNs are biased toward more massive BHs and hence galaxies of larger stellar masses\cite{Magorrian+1998AJ, Kormendy&Ho2013ARA&A}. Lower mass BHs, which live in less massive galaxies, are less luminous and thus would be missed from a flux-limited sample such as ours. Notably, $\sim40\%$ of the AGNs in the lower redshift bin and $\sim53\%$ of the ones in the higher redshift bin have colors bluer than $(g^{\prime}-r^{\prime})_0 \approx 0.6$~mag, which indicates the presence of significant star formation activity alongside BH accretion\cite{Trump+2013ApJ}. 

To further obtain a rough estimate of the SFR of the AGN host galaxies, we investigate the correlation between $(g^{\prime}-r^{\prime})_0$ color and specific SFR (${\rm sSFR \equiv SFR}/M_*$) in inactive galaxies. We retrieve the SFRs of our inactive galaxy sample from the second version of the GALEX-SDSS-WISE Legacy Catalog\cite{Salim+2016ApJS, Salim+2018ApJ}, which were derived from SED fitting of broad-band photometry from the ultraviolet, optical, and mid-infrared. Supplementary Fig.~10 shows that the local galaxy population delineates a sharp ``elbow'' feature\cite{Weinmann+2006MNRAS, Huang+2012ApJ}, such that $(g^{\prime}-r^{\prime})_0$ color tightly traces sSFR for colors $(g^{\prime}-r^{\prime})_0\lesssim 0.65$~mag, while $(g^{\prime}-r^{\prime})_0$ becomes almost insensitive to sSFR at $(g^{\prime}-r^{\prime})_0\gtrsim 0.65$~mag, likely due to the degeneracy between age, metallicity, and dust extinction\cite{Brinchmann+2004MNRAS}. We take advantage of this strong empirical trend and use the $(g^{\prime}-r^{\prime})_0$ color as a proxy for sSFR, as this color is consistently covered by the PS1 filters across the entire redshift range of interest for our study. A fit to the star-forming inactive galaxies, which we define by $(g^{\prime}-r^{\prime})_0<0.65$~mag and ${\rm sSFR} >10^{-11}$~yr$^{-1}$, with the linear regression code \texttt{linmix}\cite{Kelly2007ApJ} gives 
\begin{equation} \label{eq5}
\log\, ({\rm sSFR}/{\rm yr}^{-1}) = -(2.98 \pm 0.06) (g^{\prime}-r^{\prime})_0 - (8.61 \pm 0.03),
\end{equation}

\noindent
which has a total scatter of 0.26~dex. For $(g^{\prime}-r^{\prime})_0>0.65$~mag, we adopt a conservative estimate of ${\rm sSFR} =10^{-12}~{\rm yr}^{-1}$.

\subsection{Morphologies of the Host Galaxies} 

The \sersic\ index $n$, a flexible, effective representation of a galaxy's light distribution, is often used to classify the morphologies of both active and inactive galaxies into early-type, bulge-dominated (large $n$) and late-type, disk-dominated (small $n$) systems\cite{Ravindranath+2004ApJ, Gabor+2009ApJ}. However, as the image resolution can affect the measurement of $n$ even at low redshifts\cite{Dewsnap+2022arXiv}, it is necessary to calibrate the \sersic\ index with independent, more accurate morphology classifications. Comparison of detailed morphological classification of galaxies imaged with the Hubble Space Telescope\cite{Zhuang&Ho2022ApJ} reveals that for PS1 images of galaxies at $z \approx 0.3$ the criterion $n=2$ suffices to separate bulge-dominated from disk-dominated objects. For more nearby objects, we crossmatch the inactive galaxy sample with the Galaxy Zoo~2 catalog\cite{Willett+2013MNRAS, Hart+2016MNRAS} to obtain more detailed morphology. Galaxy Zoo 2 improves upon the original Galaxy Zoo classification by measuring more detailed morphological features (e.g., bulge, bar, spiral arms) in a subset of the brightest and largest galaxies. In total we obtain reliable morphological classification for 2,259 massive ($M_*\geq 10^{10}\, M_{\odot}$) galaxies with $0.01\lesssim z\lesssim0.25$, including 933 ellipticals, 478 Sa--Sb types (bulge-dominated disks), 686 Sc--Sd types (disk-dominated disks), and 162 edge-on disks. S0 galaxies are more commonly classified as ellipticals, but with a nonnegligible possibility as disk galaxies\cite{Willett+2013MNRAS}. Supplementary Fig.~11 shows the distributions of $i_{\rm P1}$-band \sersic\ indices for galaxies of different morphological types. Ellipticals and Sa--Sb galaxies tend to have a high S\'ersic index broadly clustered at $n \approx 4-5$, while Sc--Sd galaxies peak at $n \approx 1-2$, although all types have significant tails toward lower and higher values of $n$. To quantify the effectiveness with which galaxy morphology can be classified crudely using the \sersic\ index, we calculate the number ratio (the higher the better) of Sc--Sd galaxies over galaxies of all other types (i.e., ellipticals, Sa--Sb, and edge-on disks). We find that $n\leq2$ performs better compared to $n\leq2.5$, with a number ratio of 2.17 compared to 1.76. This is consistent with studies of galaxies at slightly higher average redshift ($z \approx 0.3$), which conclude that $n=2$ provides a good separation between disk-dominated (conservatively Sc--Sd) and bulge-dominated galaxies\cite{Zhuang&Ho2022ApJ}. Moreover, the high frequency of pseudo bulges in Sc--Sd galaxies---at least $\sim60\%$ and as high as $100\%$, depending on galaxy sample\cite{Kormendy&Kennicutt2004ARA&A, Gao+2020ApJS}---justifies our decision to use the bulge type-dependent calibration\cite{Ho&Kim2015ApJ} to estimate $M_{\rm BH}$. 

Applying the $n=2$ criterion to the AGN sample yields 6,672 ($58\%$) early-type, bulge-dominated galaxies, 4,116 ($36\%$) late-type, disk-dominated galaxies, and 686 ($6\%$) galaxies without reliable morphological classification owing to the large uncertainty in $n$ (measurement/uncertainty $<3$). In general, early-type hosts are $\sim0.2$~dex more massive and dominate over the late-type hosts across the entire redshift range (Supplementary Fig.~12). The number ratio between early-type and late-type hosts slightly decreases toward higher redshift, from 2.1 at $z<0.1$ to 1.5 at $z\geq0.1$. Closer inspection of the colors of the two types reveals that they span a similar color range from star-forming to quiescent, with preference for bluer $(g^{\prime}-r^{\prime})_0$ color for late types compared to early types across the entire range of stellar masses (Extended Fig.~4), consistent with the results of nearby inactive galaxies\cite{Blanton&Moustakas2009ARA&A}. A number of recent studies have also reported that low-redshift, luminous AGNs have a significant fraction of late-type host galaxies\cite{Kim+2017ApJS, Kim+2021ApJS, Zhao+2019ApJ, Li_Junyao+2021aApJ}, highlighting the importance of internal secular processes in regulating AGN activity\cite{Kormendy&Kennicutt2004ARA&A, Martin-Navarro+2022MNRAS}.

\subsection{Mass Gain of the BHs and Their Host Galaxies to $z=0$}
We obtain the lookback time using the median redshift, estimate sSFR using Equation~(\ref{eq5}), and convert bolometric luminosity to BH accretion rate following

\begin{equation} \label{eq6}
\dot{M}_{\rm BH}=0.15\left(\frac{0.1}{\epsilon}\right) \left(\frac{L_{\rm bol}}{10^{45}\ {\rm erg\, s^{-1}}}\right)\ M_{\odot}\, {\rm yr}^{-1},
\end{equation}

\noindent
with a radiative efficiency $\epsilon = 0.1$\cite{Frank+1992apa..book}. Since the sSFR and $\dot{M}_{\rm BH}$ are median quantities computed over a large number of objects, we assume that star formation can sustain the current sSFR during the entire lookback time. We need to consider the duty cycle of the BH accretion, which can be defined as the fraction of BHs that are actively accreting above a certain Eddington ratio with respect to all BHs, including inactive ones. The duty cycle depends on redshift and BH mass\cite{Aversa+2015ApJ, Delvecchio+2020ApJ, Ananna+2022ApJS}, with generally higher values toward higher redshift and smaller BH mass. High Eddington ratio leads to lower duty cycle, but BHs of lower mass experience a higher duty cycle. We adopt the relation between duty cycle of AGNs with $\lambda_{\rm E}>0.01$ and BH mass\cite{Ananna+2022ApJS}. The BH mass coverage of the adopted relation spans the range $M_{\rm BH} \approx 10^7 - 10^{9.6}\, M_{\odot}$, over which the duty cycle decreases from $\sim0.1$ to $6\times10^{-4}$. For $M_{\rm BH} < 10^7\, M_{\odot}$, we simply fix the duty cycle to 0.1.

\subsection{$M_{\rm BH}-M_*$ Relation in Local Type~2 AGNs}

It is difficult to obtain BH mass measurements in type~2 AGNs, which lack broad emission lines in optical spectra. Recent observations\cite{Lamperti+2017MNRAS, Ricci+2022ApJS} have reported detection of near-infrared broad lines in Seyfert 2s, which can be used to derive BH masses in these systems and offer us the opportunity to compare their location and evolution on the $M_{\rm BH}-M_*$ plane. We crossmatch the type~2 AGNs presented in Ref.\cite{Ricci+2022ApJS} with the PS1 DR2 catalog and Dark Energy Survey (DES) Data Release 2 catalog\cite{Abbott+2021ApJS} to obtain optical+infrared SEDs. In the end, 14 $z<0.1$ type~2 AGNs have suitable photometries in five bands $grizy_{\rm P1}$ ($grizY_{\rm DES}$). Assuming that AGN contamination to the optical and near-infrared continuum is negligible in type 2 AGNs, we obtain their stellar masses, $(g^{\prime}-r^{\prime})_0$ color, and sSFR following the same procedure as used for the host galaxies of type~1 AGNs. Their BH masses are derived with a virial factor of 6.3, same as our choice for early-type galaxies. We find that local type~2 AGNs generally follow the same relation as inactive galaxies and type~1 AGNs, with the caveat that the current sample is severely limited  by small-number statistics (Extended Fig.~5). Their evolution toward $z=0$, as indicated with red arrows, is more limited because of the low redshift ($z<0.1$), but it is still similar to that of type~1 AGNs (Fig.~4). 

\subsection{Cosmology}
Throughout the paper, we adopt a cosmology with $H_0=70$ km s$^{-1}$ Mpc$^{-1}$, $\Omega_{\rm m}=0.3$, and $\Omega_{\Lambda}=0.7$.

\backmatter

\bmhead{Data Availability}

Pan-STARRS1 images are available from the Pan-STARRS1 data archive \url{https://outerspace.stsci.edu/display/PANSTARRS/Pan-STARRS1+data+archive+home+page}. The basic properties of the final AGN sample, PS1 flux densities, and the derived physical properties of the host galaxies are available from https://doi.org/10.12149/101278\citep{dataproduct}. 

\bmhead{Code Availability}
\texttt{Astropy}, \texttt{dustmaps}, \texttt{LOESS}, \texttt{Matplotlib}, \texttt{Numpy}, and \texttt{Scipy} are available from python package index (PyPI) \url{https://pypi.org/}.  \texttt{CIGALE} is available from \url{https://cigale.lam.fr/}. \texttt{GALFITM} is available from \url{https://www.nottingham.ac.uk/astronomy/megamorph/}. \texttt{linmix} is available from \url{https://linmix.readthedocs.io/en/latest/}. \texttt{SExtractor} and \texttt{SWarp} are available from \url{https://www.astromatic.net/software/}. 

\bmhead{Acknowledgments}
This work was supported by the National Key R\&D Program of China (2022YFF0503401), the National Science Foundation of China (11721303, 11991052, 12011540375, 12233001) and the China Manned Space Project (CMS-CSST-2021-A04, CMS-CSST-2021-A06). 

\bmhead{Author Contributions Statement}
M.-Y. Z. and L. C. H. developed and discussed the idea. M.-Y. Z. reduced the data, performed the analyses, and wrote the manuscript. L. C. H. revised the manuscript. 

\bmhead{Competing Interests Statement}
The authors declare that they have no competing financial interests.

%

\bmhead{Extended Data}
\addtocounter{figure}{-5}
\begin{figure*}[t!]
\renewcommand{\figurename}{Extended Fig.}
\centering
\includegraphics[width=0.7\textwidth]{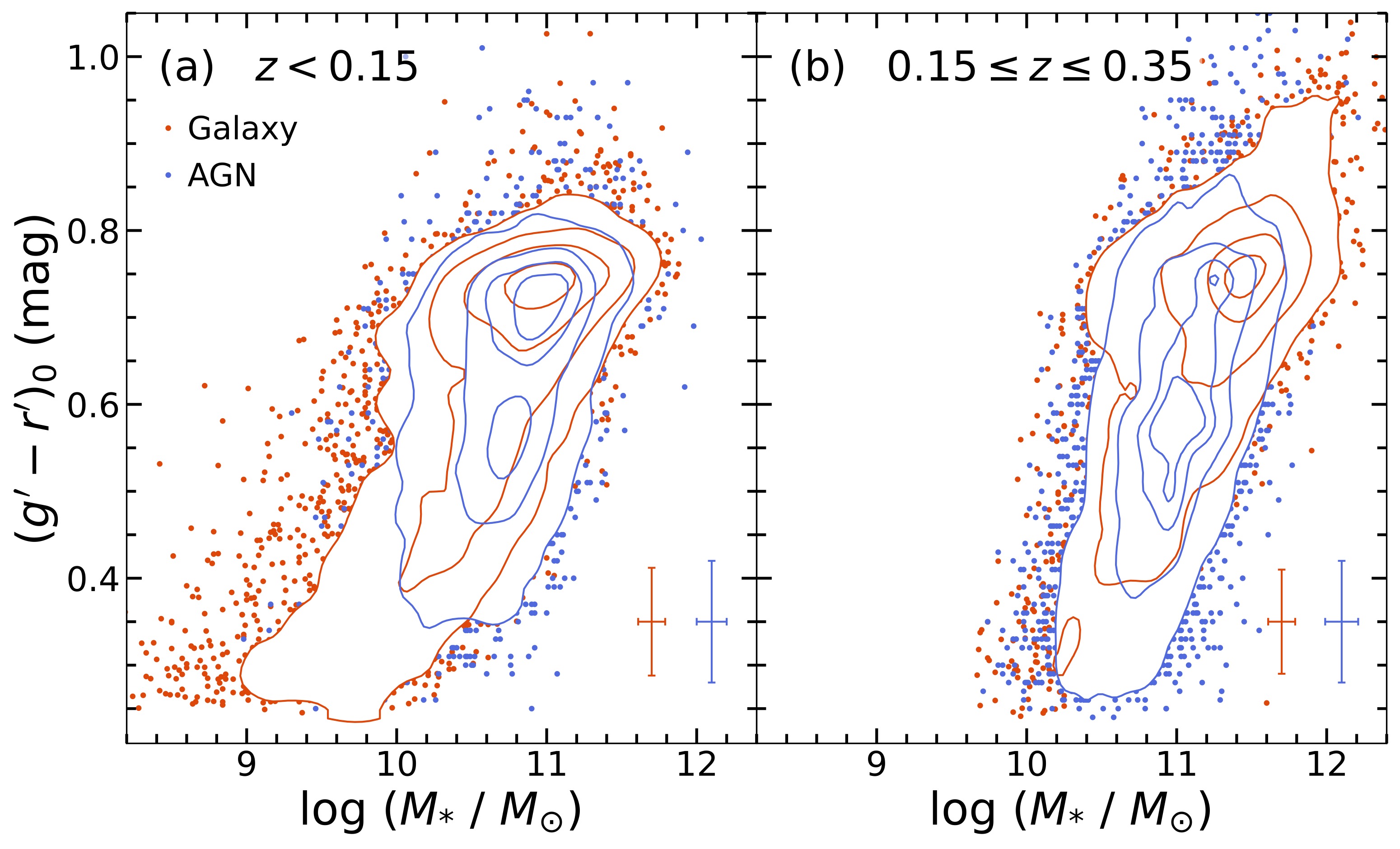}
\caption{Rest-frame SDSS $g^{\prime}-r^{\prime}$ color [$(g^{\prime}-r^{\prime})_0$] versus stellar mass ($M_*$) for the inactive galaxy sample (red) and AGN sample (blue). Objects with $z<0.15$ are shown in panel (a), and objects with $0.15\leq z \leq 0.35$ are shown in panel (b). Contours indicate the distribution of 10\%, 30\%, 60\%, and 90\% of the entire sample, respectively. Dots are individual objects located outside the 90\% contour. The lower-right corner of each panel shows the typical uncertainties.}
\end{figure*}

\begin{figure*}[t!]
\renewcommand{\figurename}{Extended Fig.}
\centering
\includegraphics[width=\textwidth]{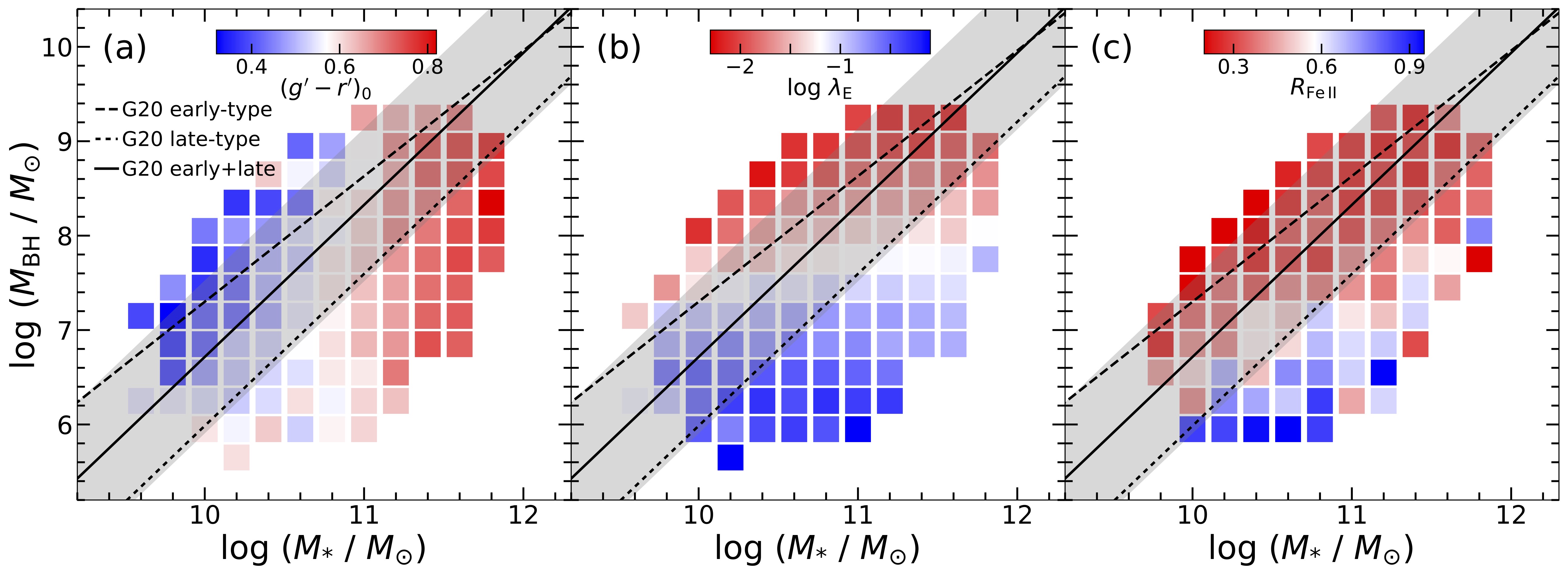}
\caption{The relation between $M_{\rm BH}$ and $M_*$ for $z\leq 0.35$ type~1 AGNs without smoothing. Colors represent (a) rest-frame optical color $(g^{\prime}-r^{\prime})_0$, (b) Eddington ratio ($\lambda_{\rm E}$), and (c) relative Fe\,\textsc{ii} strength. Best-fit relations from G20 for local early-type galaxies (black dashed line), late-type galaxies (black dotted line), and both types combined (black solid line; intrinsic scatter of 0.81~dex indicated as shaded gray region). This figure shows the version of Figure~2 without smoothing to demonstrate that the trends are still present in the original data.}
\end{figure*}

\begin{figure*}[t!]
\renewcommand{\figurename}{Extended Fig.}
\centering
\includegraphics[width=\textwidth]{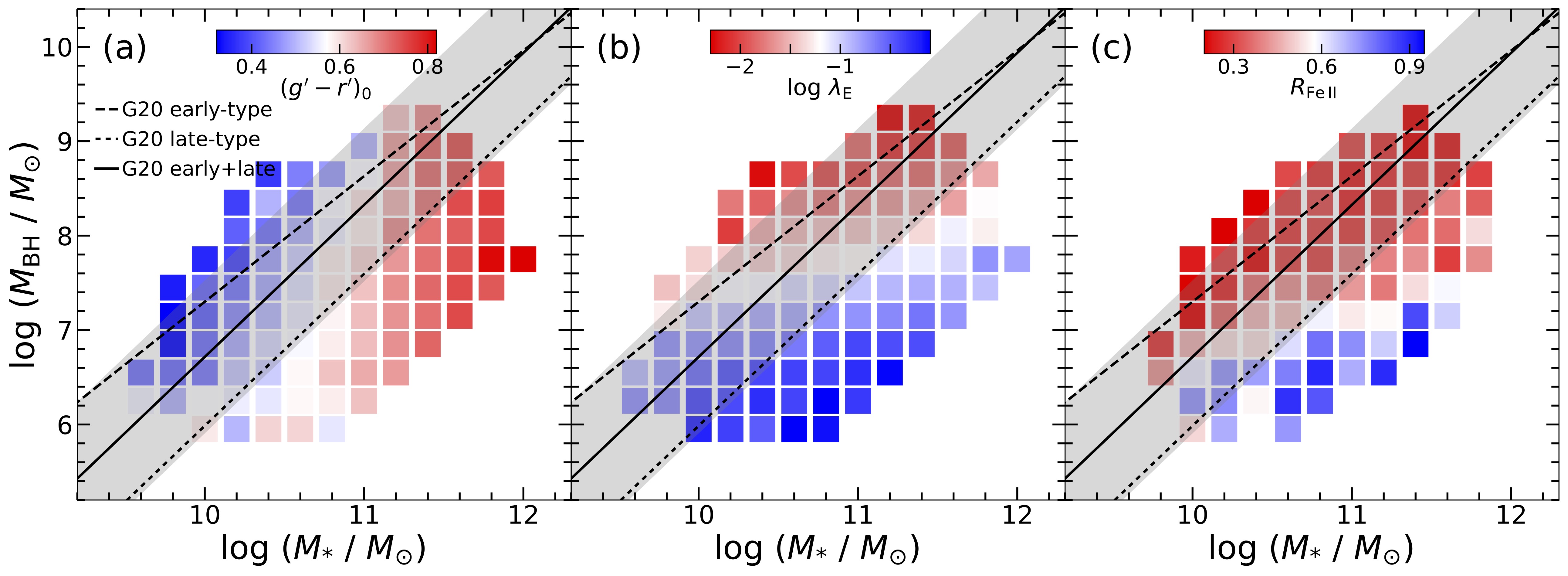}
\caption{The relation between $M_{\rm BH}$ and $M_*$ for $z\leq 0.35$ type~1 AGNs without smoothing and with $M_{\rm BH}$ derived from broad H$\alpha$ emission with a virial factor $f=4.47$. Colors represent (a) rest-frame optical color $(g^{\prime}-r^{\prime})_0$, (b) Eddington ratio ($\lambda_{\rm E}$), and (c) relative Fe\,\textsc{ii} strength. Best-fit relations from G20 for local early-type galaxies (black dashed line), late-type galaxies (black dotted line), and both types combined (black solid line; intrinsic scatter of 0.81~dex indicated as shaded gray region). This figure demonstrates that switching to the H$\alpha$-based $M_{\rm BH}$ estimator does not qualitatively affect our main conclusions.}
\end{figure*}

\begin{figure*}[t!]
\renewcommand{\figurename}{Extended Fig.}
\centering
\includegraphics[width=0.7\textwidth]{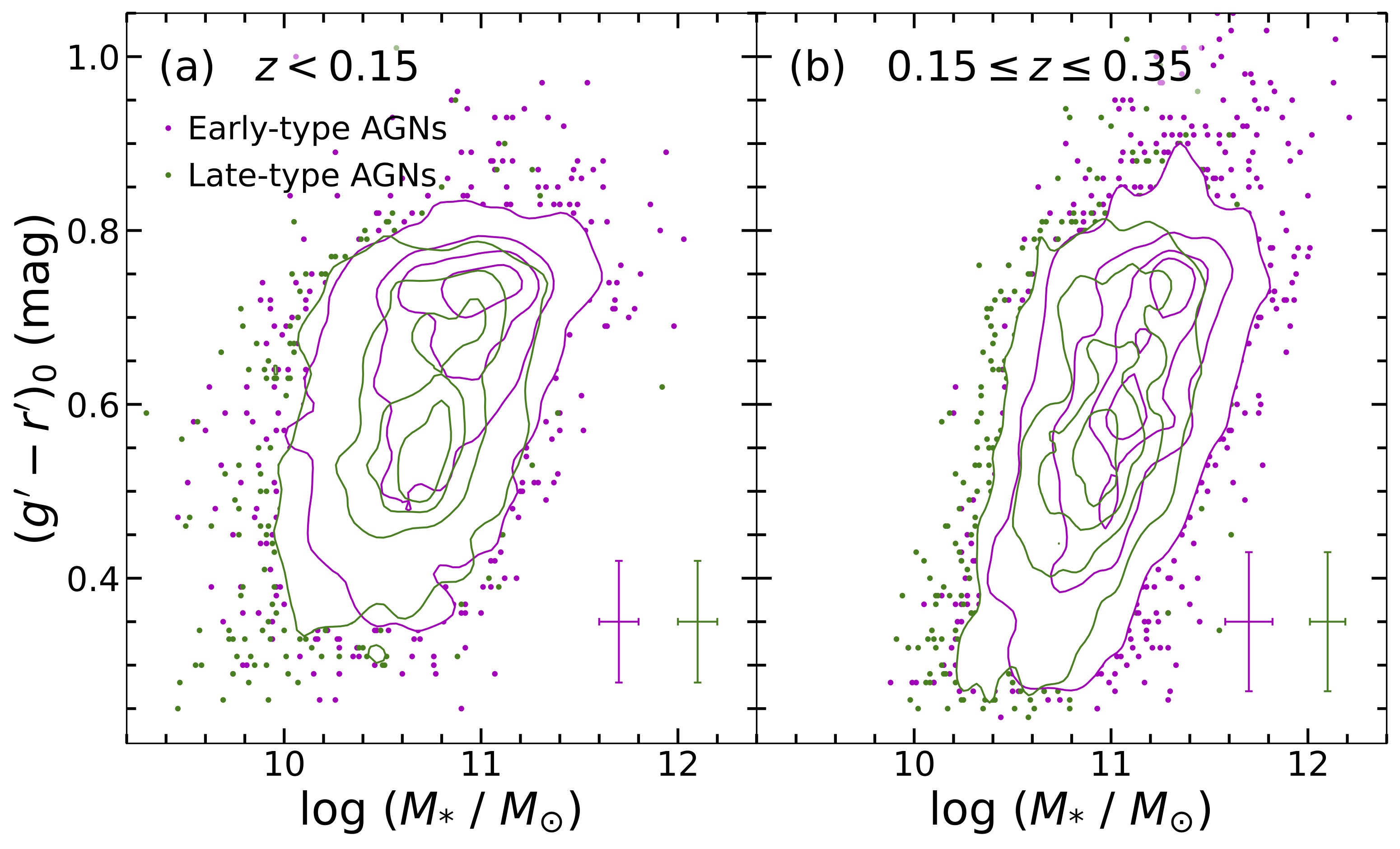}
\caption{Rest-frame SDSS $g^{\prime}-r^{\prime}$ color [$(g^{\prime}-r^{\prime})_0$] versus stellar mass ($M_*$) for AGNs. AGNs with early-type and late-type morphologies are shown in magenta and green, respectively. Objects with $z<0.15$ are shown in panel (a), and objects with $0.15\leq z \leq 0.35$ are shown in panel (b). Contours indicate the distribution of 10\%, 30\%, 60\%, and 90\% of the entire sample, respectively. Dots are individual objects located outside the 90\% contour. The lower-right corner of each panel shows the typical uncertainties.}
\end{figure*}

\begin{figure*}[t!]
\renewcommand{\figurename}{Extended Fig.}
\centering
\includegraphics[width=0.7\textwidth]{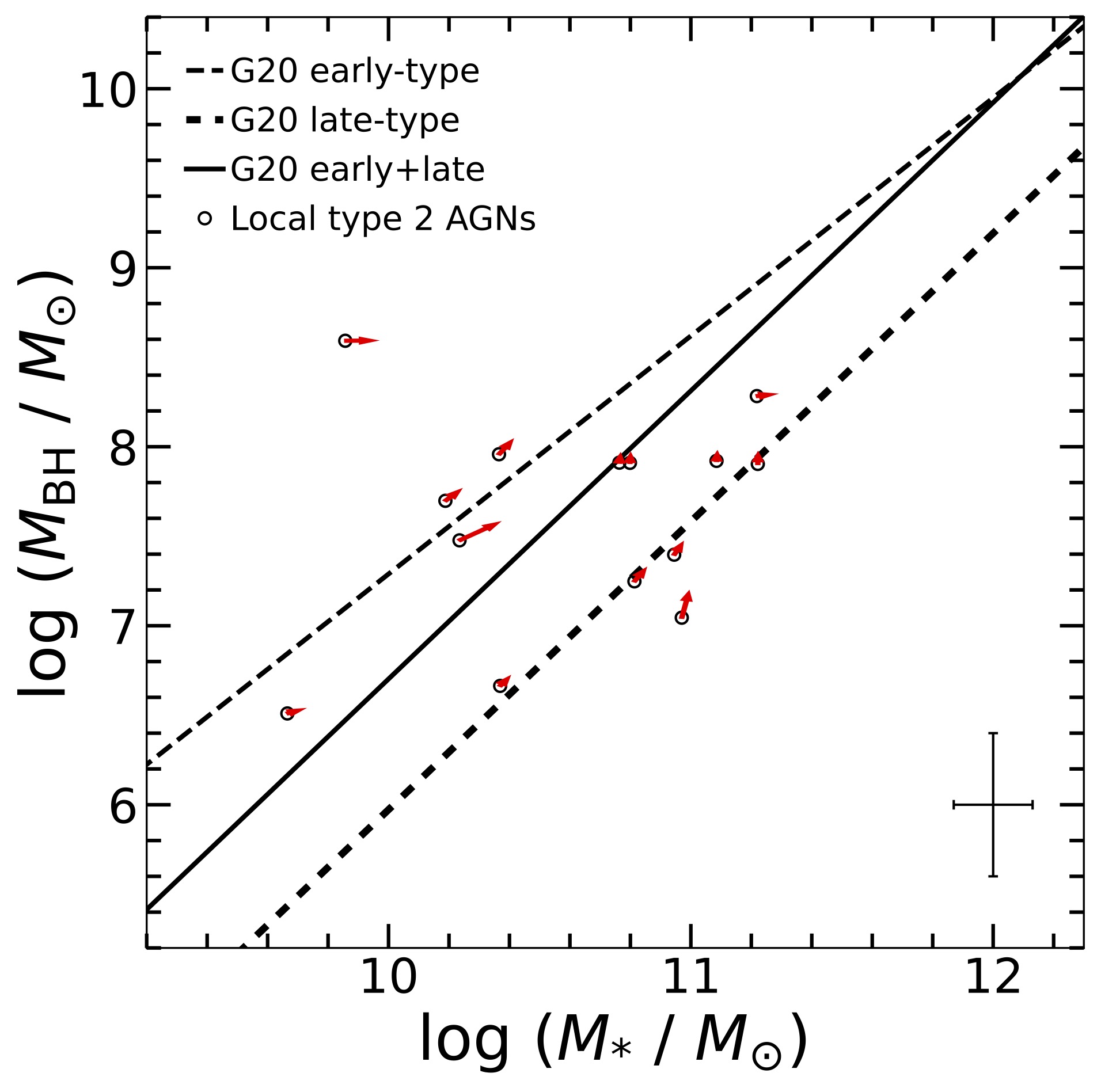}
\caption{The relation between $M_{\rm BH}$ and $M_*$ for local ($z<0.1$) type~2 AGNs. BH masses are derived from broad emission lines in the near-infrared (see Methods). Their evolution to $z=0$ is indicated by the direction and length of each arrow. Best-fit relations from G20 for local early-type galaxies (black dashed line), late-type galaxies (black dotted line), and both types combined (black solid line; intrinsic scatter of 0.81~dex indicated as shaded gray region). Typical uncertainties are shown in the lower-right corner.}
\end{figure*}

\begin{figure*}[t!]
\renewcommand{\figurename}{Extended Fig.}
\centering
\includegraphics[width=0.7\textwidth]{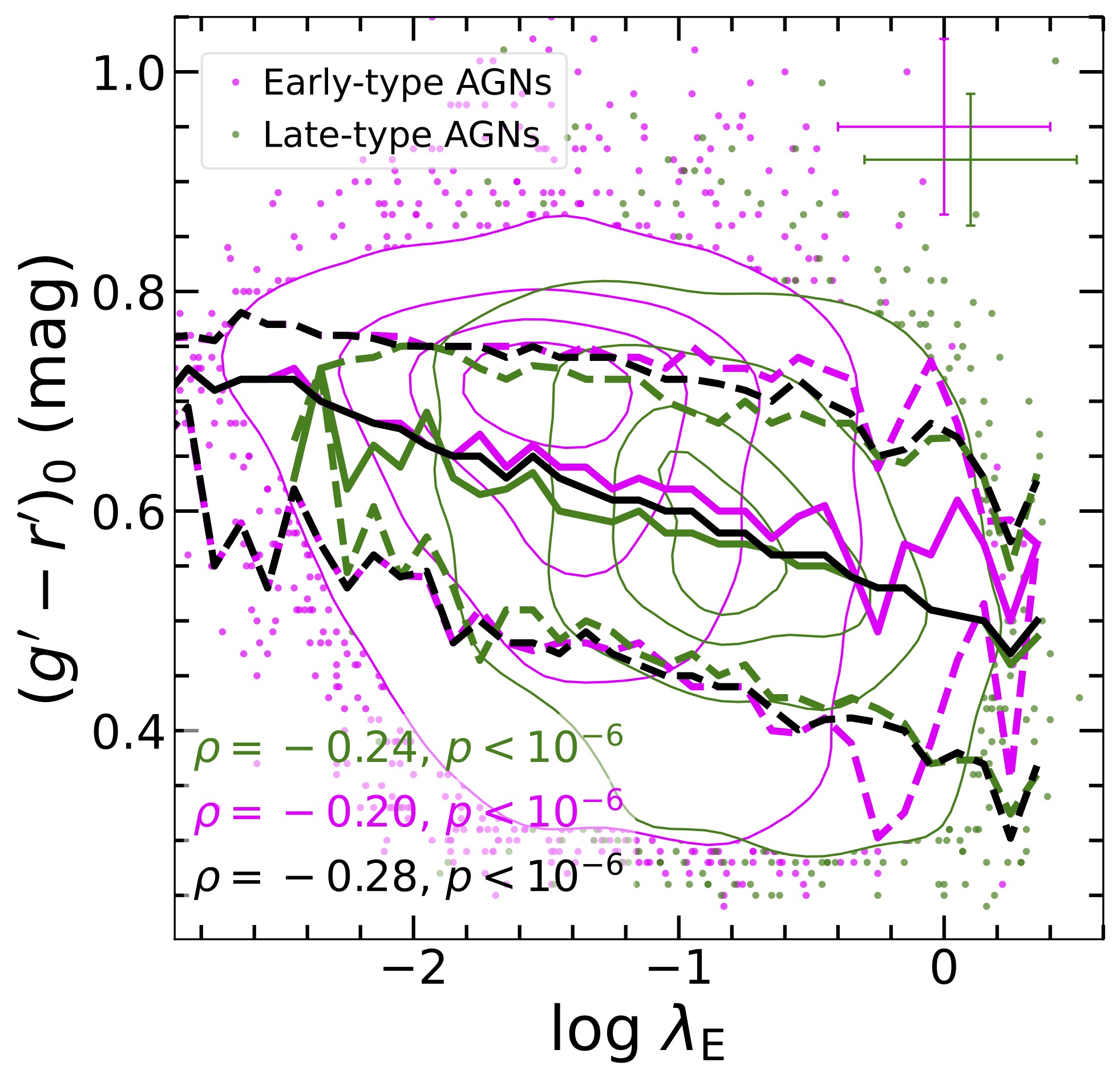}
\caption{Rest-frame optical color $(g^{\prime}-r^{\prime})_0$ versus $\lambda_{\rm E}$ for AGNs. AGNs with early-type, late-type, and both types combined morphologies are shown in magenta, green, and black, respectively. Contours indicate the distribution of 10\%, 30\%, 60\%, and 90\% of the entire sample, respectively. Dots are individual objects located outside the 90\% contour. Thick solid and dashed lines represent 50th, 16th, and 84th percentiles of the objects. The Spearman correlation coefficient $\rho$ and $p$-value are shown in the lower-left corner. Typical uncertainties are shown in the upper-right corner.}
\end{figure*}

\begin{figure*}[t!]
\renewcommand{\figurename}{Extended Fig.}
\centering
\includegraphics[width=0.7\textwidth]{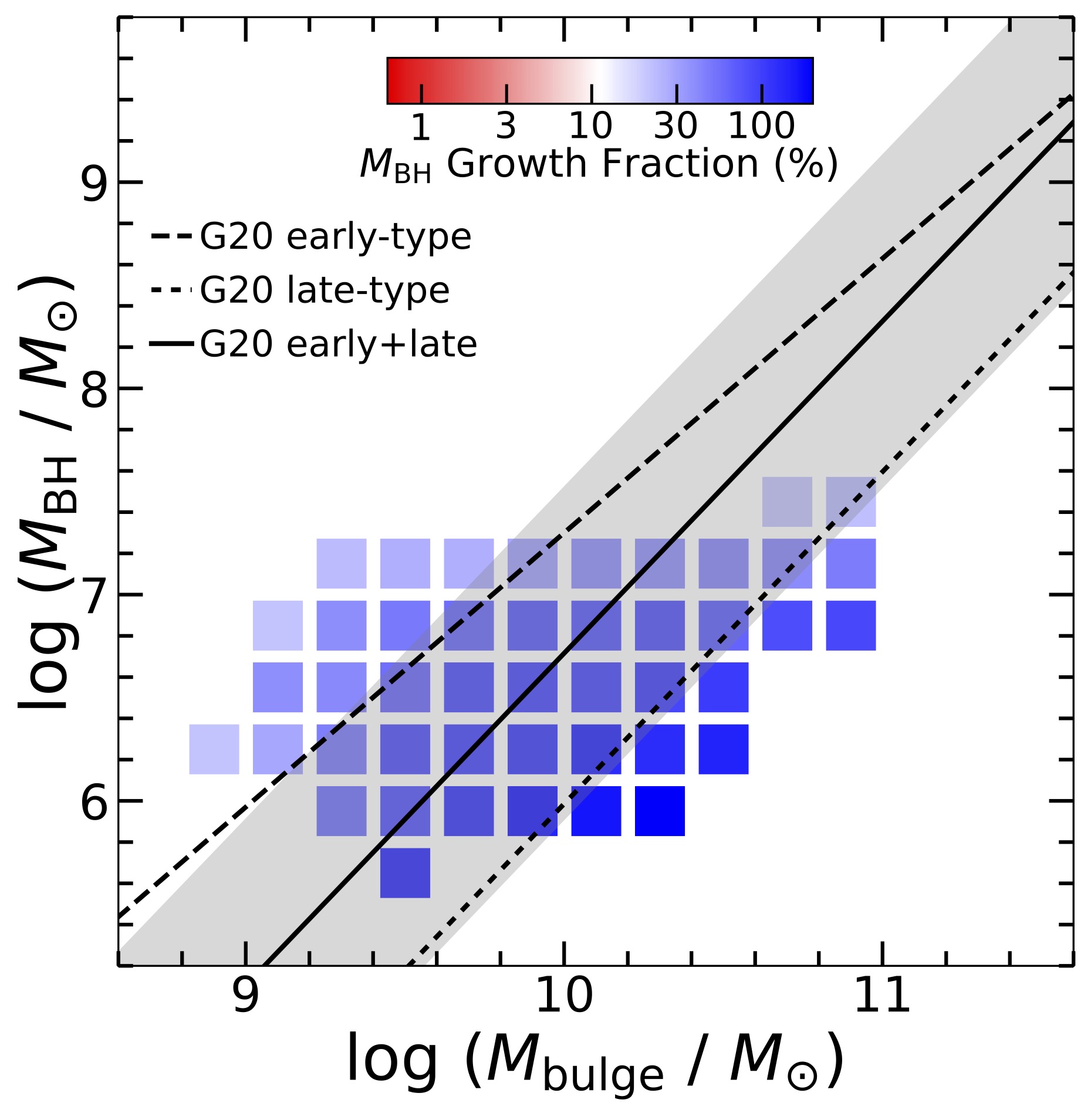}
\caption{The relation between $M_{\rm BH}$ and $M_{\rm bulge}$ for $z\leq 0.35$ AGNs with $M_{\rm BH}$ growth fraction larger than $20\%$. Colors represent growth fraction of $M_{\rm BH}$. $M_{\rm bulge}$ is predicted from $M_*$ with a bulge-to-total ratio of 0.2. We show the best-fit relations from G20 for local early-type galaxies (black dashed line), late-type galaxies (black dotted line), and both types combined (black solid line; intrinsic scatter of 0.81~dex indicated as shaded gray region).}
\end{figure*}



\section*{Supplementary material}

**Supplementary Table 1** Image Properties of the Low-$z$ Type 1 AGNs

| Band | Exposure time (s) | PSF FWHM (arcsec) | $3\sigma$ surface brightness depth (mag arcsec$^{-2}$) | $5\sigma$ point source depth (mag) |
|---|---|---|---|---|
| (1) | (2) | (3) | (4) | (5) |
| $g_{\rm P1}$ | [645, 810, 1000] | [1.04, 1.14, 1.25] | [26.54, 26.75, 26.94] | [23.13, 23.34, 23.53] |
| $r_{\rm P1}$ | [712, 880, 1140] | [0.96, 1.05, 1.15] | [26.41, 26.63, 26.82] | [23.06, 23.29, 23.50] |
| $i_{\rm P1}$ | [1028, 1440, 2025] | [0.90, 0.98, 1.07] | [26.25, 26.48, 26.69] | [23.02, 23.26, 23.48] |
| $z_{\rm P1}$ | [600, 720, 870] | [0.86, 0.94, 1.02] | [25.61, 25.81, 26.00] | [22.32, 22.53, 22.71] |
| $y_{\rm P1}$ | [660, 790, 960] | [0.83, 0.90, 0.97] | [24.51, 24.74, 24.96] | [21.43, 21.62, 21.81] |

Statistics indicate 16th, 50th, and 84th percentiles.

**Supplementary Table 2** Models and Parameters for SED Fitting

| Parameter | Value |
|---|---|
| Star formation history: delayed | |
| e-folding time of the main stellar population: $\tau_{\rm main}$ (Gyr) | 0.001, 0.05, 0.1, 0.25, 0.5, 0.75, 1.0, 1.25, 1.5, 2.0, 2.5, 3.0, 3.5, 4.0, 4.5, 5.0, 6.0, 7.0, 8.0, 9.0, 10, 12, 15, 20 |
| Age of the main stellar population: $t_{\rm main}$ (Gyr) | 9.0, 9.5, 10.0, 10.5, 11.0, 11.5, 12.0, 12.5, 13.0 |
| Mass fraction of the latest burst population: $f_{\rm burst}$ | 0 |
| Single stellar population: Bruzual & Charlot (2003) | |
| Initial mass function | Chabrier (2003) |
| Metallicity | 0.02 (solar) |
| Nebular emission | |
| Ionization parameter: $\log U$ | $-2.0$ |
| Dust attenuation: modified starburst attenuation law | |
| The color excess of the nebular lines: $E(B-V)$ (mag) | 0.0, 0.001, 0.005, 0.01, 0.03, 0.05 0.1, 0.2, 0.3, 0.4, 0.5, 0.6, 0.8 |
| Ratio of $E(B-V)$ between stellar continuum and emission line | 0.44 |

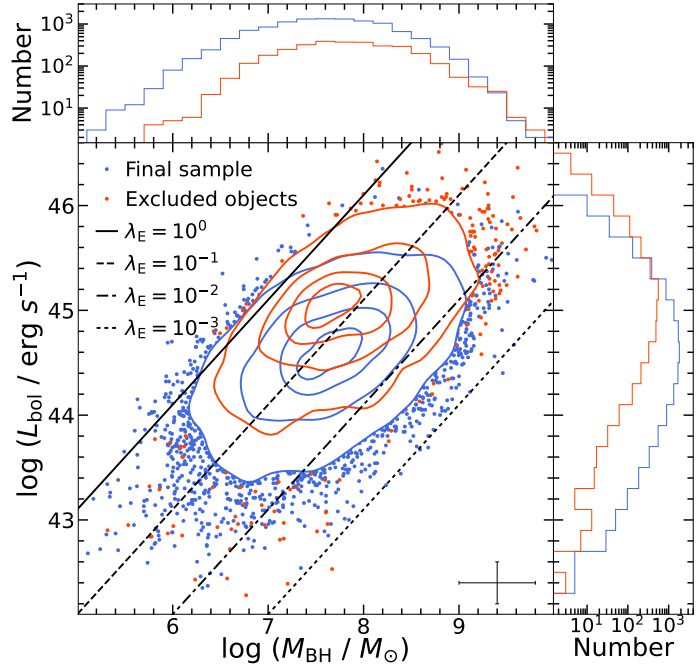

**Supplementary Fig. 1** Distribution of AGNs in the final sample (blue) and those that are excluded (red) on the $M_{\rm BH}$ versus $L_{\rm bol}$ plane. Contours indicate the distribution of 10%, 30%, 60%, and 90% of the entire sample, respectively. Dots are individual objects located outside the 90% contour. Solid, dashed, dash-dotted, and dotted lines represent Eddington ratios of $\lambda_{\rm E} = 10^0, 10^{-1}, 10^{-2}$, and $10^{-3}$, respectively. Typical uncertainties in the lower-right corner indicate a conservative $1\sigma$ uncertainty of 0.4 dex for $M_{\rm BH}$.

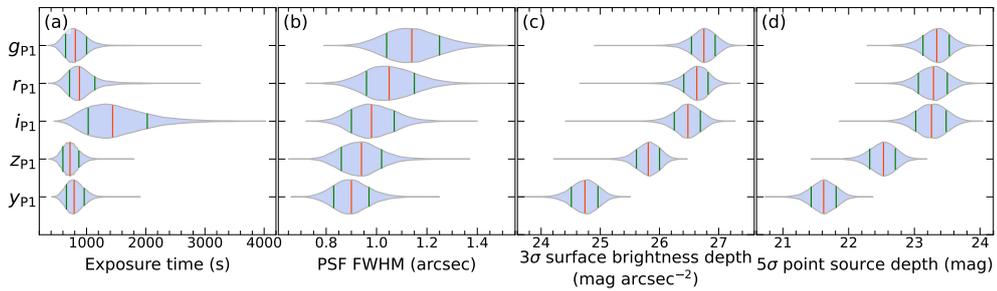

**Supplementary Fig. 2** Distributions of image properties of the AGN sample. Panels (a–d) show the distributions of exposure time, PSF FWHM, $3\sigma$ surface brightness depth, and $5\sigma$ point source depth, respectively. The red and two adjacent green vertical lines indicate the 50th, 16th, and 84th percentiles of the population. $\sigma$ is the standard deviation of 100 flux measurements made using a circular aperture of diameter $15''$ for extended sources and 2 times the PSF FWHM for point sources, randomly placed in the source-masked background of the image. The statistics are shown in Supplementary Table 1.

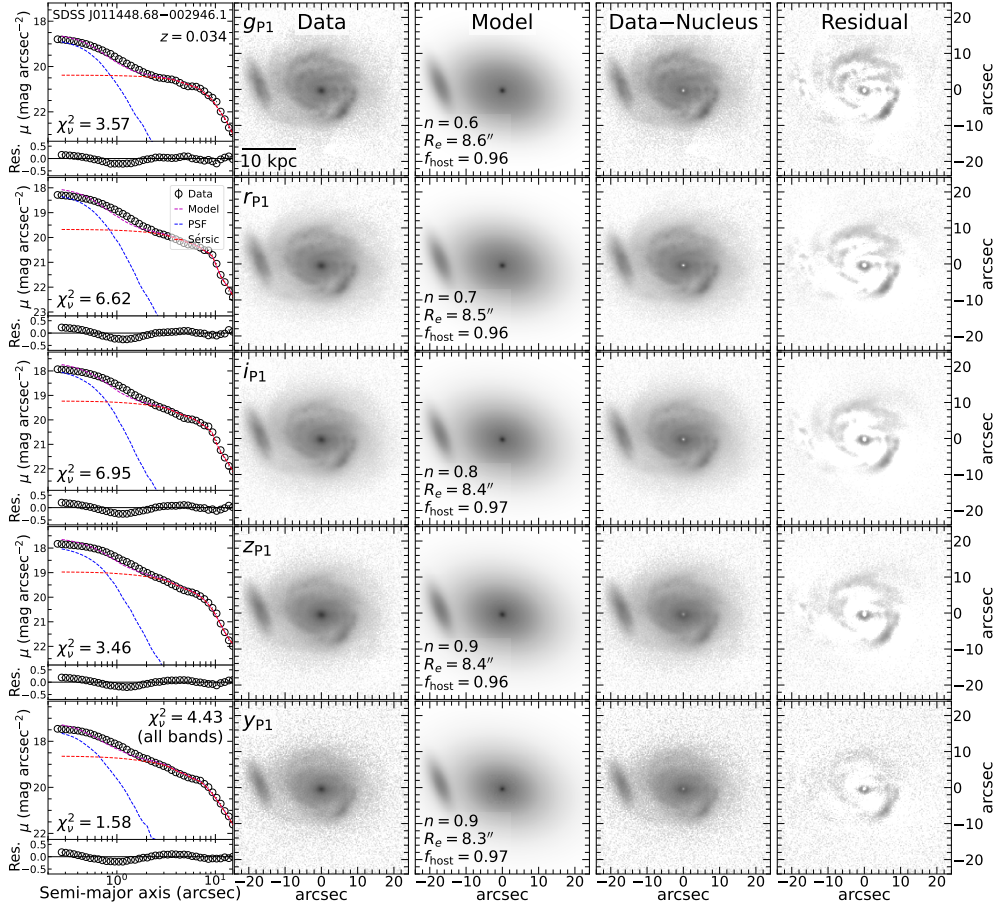

**Supplementary Fig. 3** An example of our simultaneous multiwavelength decomposition of SDSS J011448.68−002946.1, a host galaxy with a late-type morphology at $z = 0.034$. Rows from top to bottom are the results for filters $grizy_{\rm P1}$. The upper panel of the left column shows the radial profile of the surface brightness (open circles with error bars), PSF component (blue), Sérsic component (red), and total (PSF + Sérsic) model (purple). The $\chi_\nu^2$ from GALFITM for each band is given in the lower-left corner, while that for all five bands is shown in the upper-right corner of the first panel in the bottom row. The lower sub-panel gives the residuals between the data and the model (data−model). The images illustrate, from left to right, the original data, best-fit total model, data with the nucleus component modeled by the PSF subtracted, and residuals of the total model. The best-fit Sérsic index ($n$), half-light radius ($R_e$), and host fraction $f_{\rm host} \equiv$ Sérsic / (PSF + Sérsic) are shown in the third column.

3

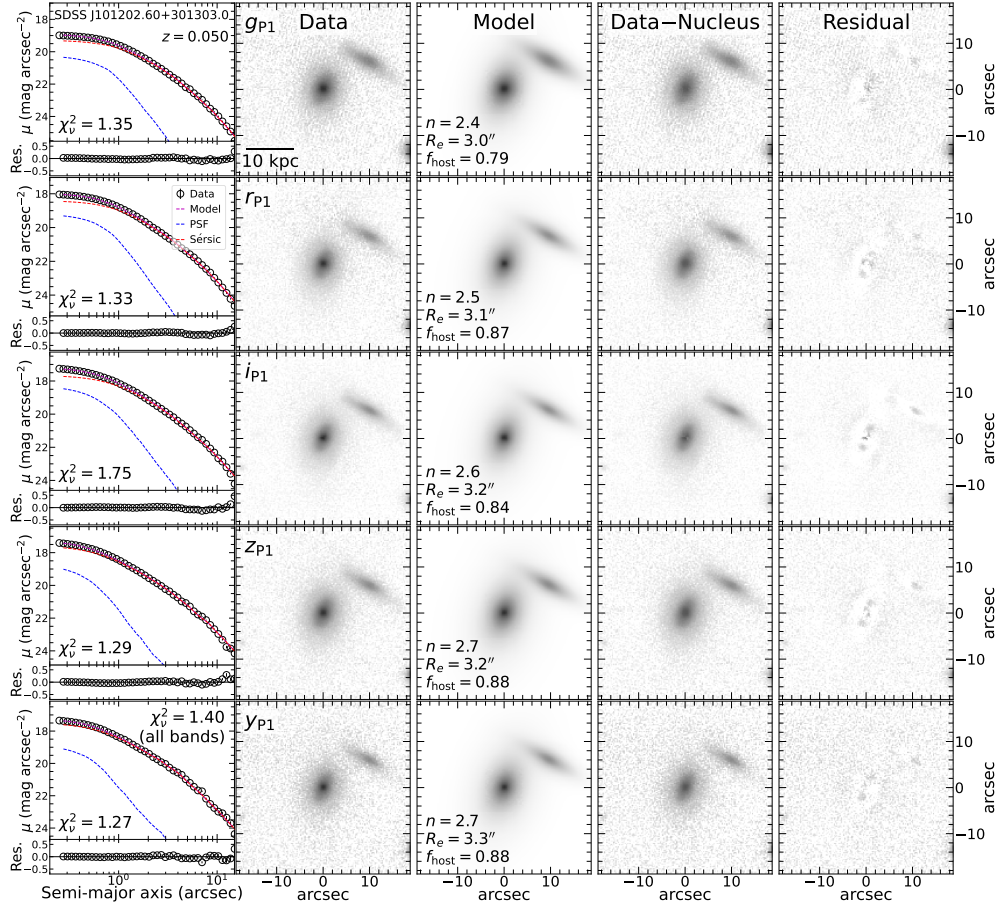

**Supplementary Fig. 4** An example of our simultaneous multiwavelength decomposition of SDSS J101202.60+301303.0, a host galaxy with a early-type morphology at $z = 0.050$. Rows from top to bottom are the results for filters $grizy_{\rm P1}$. The upper panel of the left column shows the radial profile of the surface brightness (open circles with error bars), PSF component (blue), Sérsic component (red), and total (PSF + Sérsic) model (purple). The $\chi^2_\nu$ from `GALFITM` for each band is given in the lower-left corner, while that for all five bands is shown in the upper-right corner of the first panel in the bottom row. The lower sub-panel gives the residuals between the data and the model (data−model). The images illustrate, from left to right, the original data, best-fit total model, data with the nucleus component modeled by the PSF subtracted, and residuals of the total model. The best-fit Sérsic index ($n$), half-light radius ($R_e$), and host fraction $f_{\rm host} \equiv$ Sérsic / (PSF + Sérsic) are shown in the third column.

4

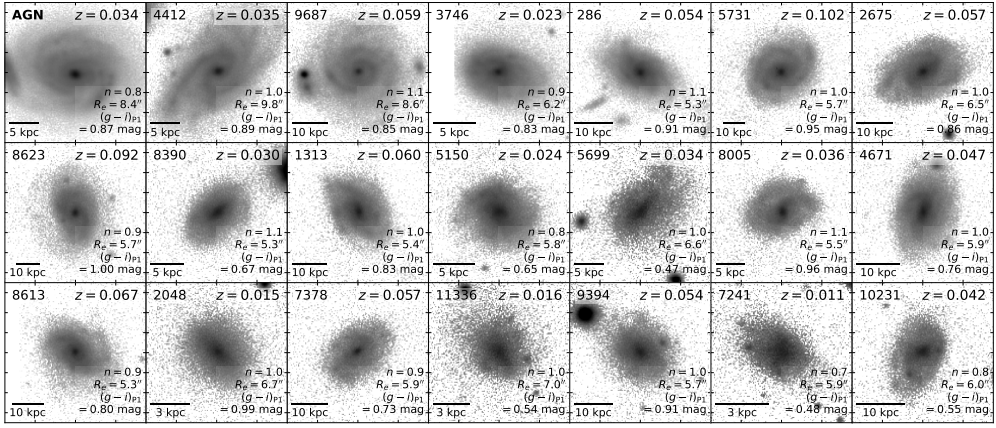

**Supplementary Fig. 5** An example of a collection of $i_{P1}$ band images of 20 inactive galaxies that have similar structural properties and color as the target AGN (top left panel). The target AGN is the same object in Supplementary Figure 3. Galaxy index and redshift are shown at the top, while the best-fit $(g-i)_{P1}$ color, half-light radius ($R_e$), and Sérsic index ($n$) in the $i_{P1}$ band are shown at the bottom right of each panel.

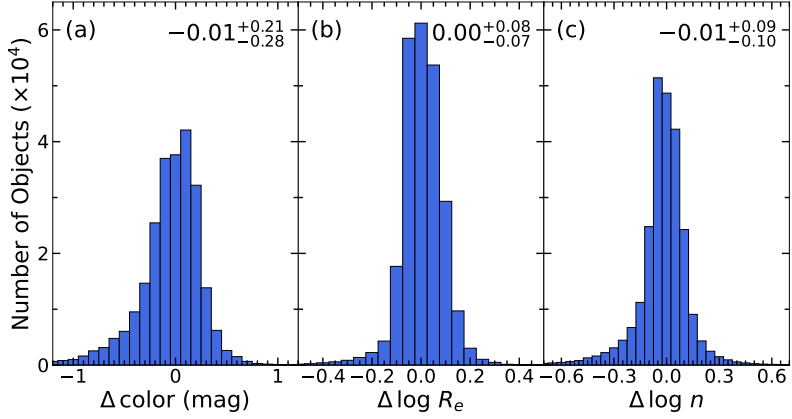

**Supplementary Fig. 6** Distributions of the differences ($\Delta$ = AGN−galaxy) between AGNs and inactive galaxies matched in structure and color. Panels (a–c) show the distributions of the differences in observed-frame $(g-i)_{P1}$ color, $R_e$, and $n$ in the $i_{P1}$ band, respectively. The median and its differences with 16th and 84th percentiles are shown in the top-right corner of each panel.

5

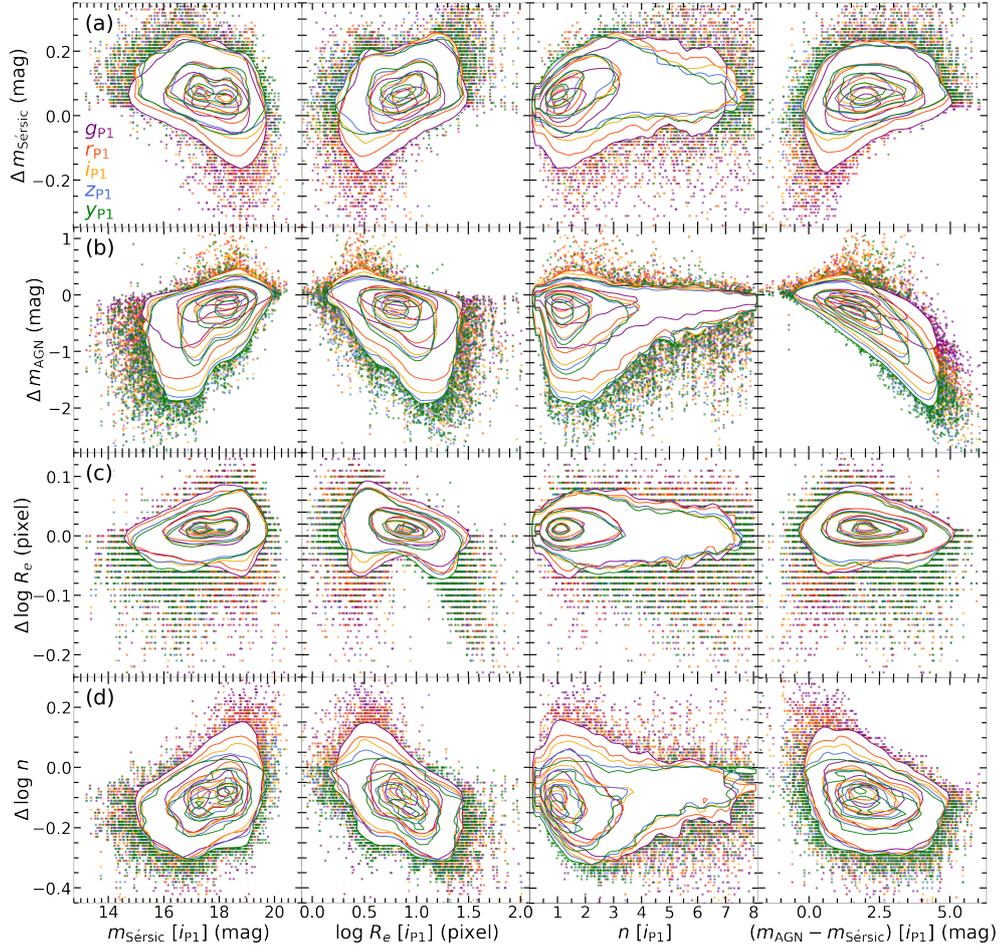

**Supplementary Fig. 7** Differences (Δ = AGN−galaxy) between the results measured in mock AGNs and those measured in the inactive galaxies used to generate the mock AGNs versus AGN properties. Panels (a–d) show differences on $m_{\text{Sérsic}}$, $m_{\text{AGN}}$, $R_e$, and $n$ versus $m_{\text{Sérsic}}$, $R_e$, $n$, and $m_{\text{AGN}} - m_{\text{Sérsic}}$ of the AGN sample, respectively. Purple, red, orange, blue, and green colors represent parameters measured in the $grizy_{\text{P1}}$ bands, respectively. Contours indicate the distribution of 10%, 30%, 60%, and 90% of the entire sample, respectively. Individual dots are the median values of 20 mock AGNs for each AGN target located outside the 90% contour.

6

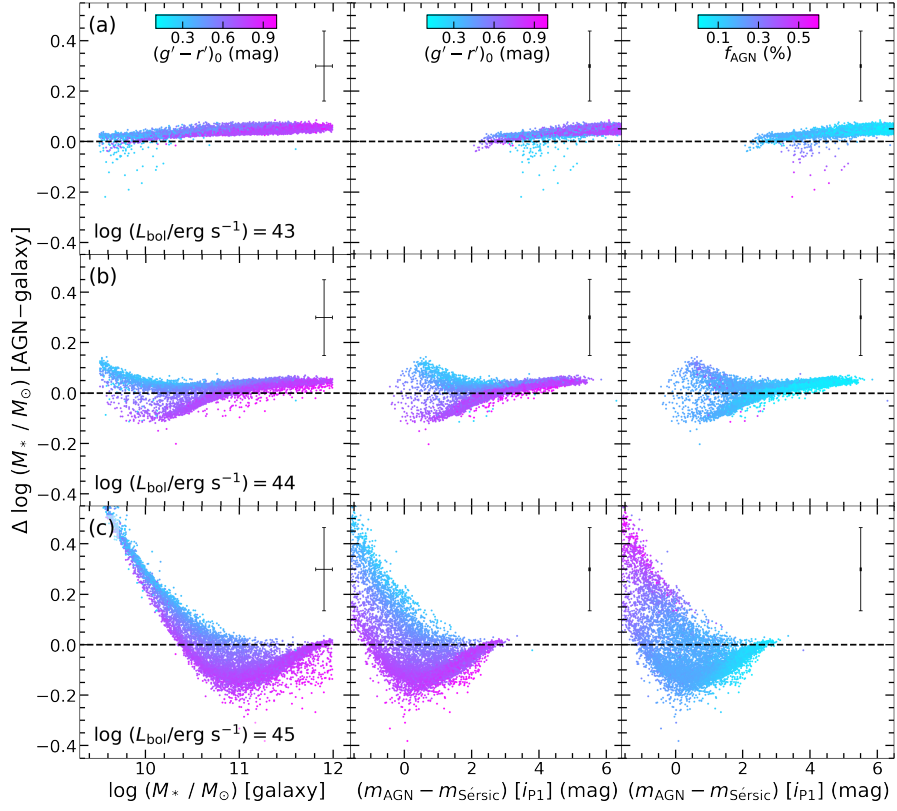

**Supplementary Fig. 8** Differences ($\Delta$ = AGN−galaxy) between the stellar masses measured in mock AGNs and those measured in the inactive galaxies used to generate the mock AGNs versus AGN properties. Left column shows stellar mass differences versus $M_*$ of inactive galaxies, while middle and right columns show stellar mass differences versus $m_{\rm AGN} - m_{\rm Sérsic}$ in the $i_{\rm P1}$ band. Mock are shown for AGNs with (a) $\log(L_{\rm bol}/{\rm erg~s^{-1}}) = 43$, (b) $\log(L_{\rm bol}/{\rm erg~s^{-1}}) = 44$, and (c) $\log(L_{\rm bol}/{\rm erg~s^{-1}}) = 45$. Colors in the left and middle columns represent the rest-frame $(g'-r')_0$ color of the galaxies, and the colors in right column represent $f_{\rm AGN}$ derived from SED-fitting to mock AGNs. Typical uncertainties are shown in the upper-right corner.

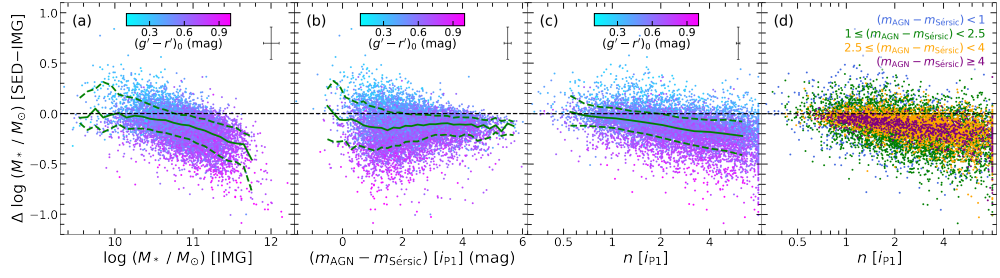

**Supplementary Fig. 9** Differences ($\Delta = $ SED$-$IMG) between stellar masses of AGNs derived from SED-fitting to AGN+host SEDs (SED) and those from SED-fitting to AGN contribution-subtracted host SEDs from image decomposition (IMG) versus AGN properties. Panel (a) shows stellar mass differences versus $M_*$ from IMG method, panel (b) shows stellar mass differences versus $m_{\rm AGN} - m_{\rm Sérsic}$ in the $i_{\rm P1}$ band, while panels (c) and (d) show stellar mass differences versus Sérsic index $n$ in the $i_{\rm P1}$ band. Colors in panels (a), (b), and (c) represent rest-frame optical color $(g'-r')_0$ of galaxies, while colors in panel (d) represent objects with different AGN-to-host contrast ($m_{\rm AGN} - m_{\rm Sérsic}$) in the $i_{\rm P1}$ band. Green solid and dashed curves indicate running medians and 16th/84th percentiles of the objects. Typical uncertainties are shown in the upper-right corner.

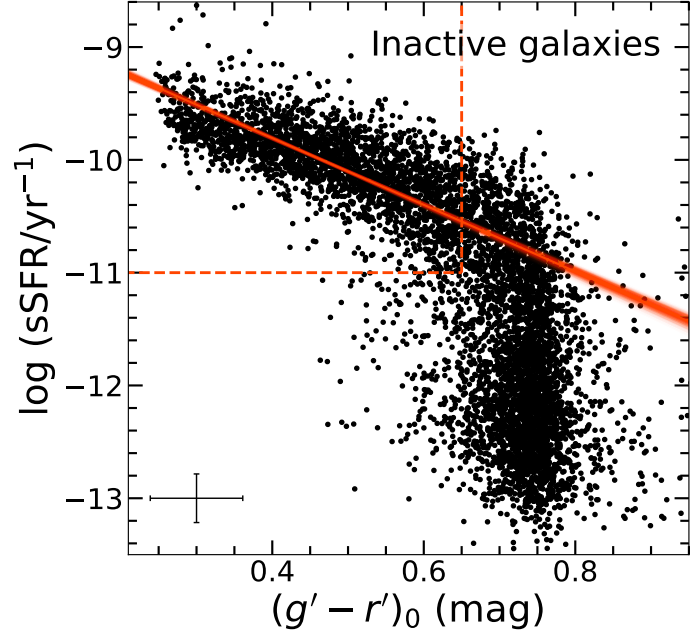

**Supplementary Fig. 10** Specific star formation rate (sSFR) versus rest-frame SDSS $g'-r'$ color $[(g'-r')_0]$ for the inactive galaxy sample. The points represent individual objects, with typical errorbars indicating $1\sigma$ uncertainty shown in the lower-left corner. The red solid line shows the best-fit linear relation to objects with $(g'-r')_0 < 0.65$ mag and sSFR $> 10^{-11}$ yr$^{-1}$ (above and to the left of the red dashed lines) using `linmix`; the faint lines demarcate uncertainties of the fit.

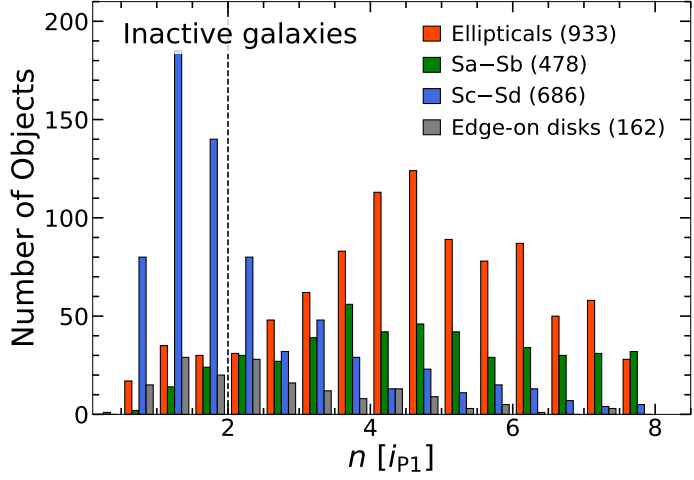

**Supplementary Fig. 11** Distribution of Sérsic index $n$ in the $i_{P1}$ band for the inactive galaxy sample. Visual morphological classifications (red: elliptical galaxies; green: Sa−Sb type galaxies; blue: Sc−Sd type galaxies; grey: edge-on disk galaxies) are from the Galaxy Zoo 2 project. The black vertical dashed line at $n = 2$ indicates our criterion to distinguish between late-type, disk-dominated ($n \leq 2$) and early-type, bulge-dominated ($n > 2$) galaxies.

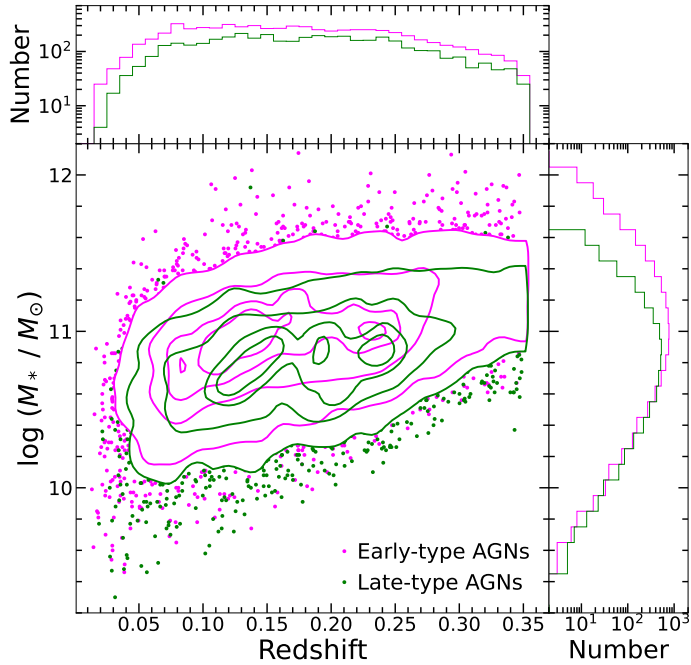

**Supplementary Fig. 12** Comparison of redshifts and stellar masses for AGNs with early-type (magenta) and late-type (green) morphologies. Contours indicate the distribution of 10%, 30%, 60%, and 90% of the entire sample, respectively. Dots are individual objects located outside the 90% contour.

9




\begin{thebibliography}{100}
\expandafter\ifx\csname url\endcsname\relax
  \def\url#1{\burl{#1}}\fi
\expandafter\ifx\csname urlprefix\endcsname\relax\def\urlprefix{URL }\fi
\providecommand{\bibinfo}[2]{#2}
\providecommand{\eprint}[2][]{\url{#2}}
\providecommand{\doi}[1]{\url{https://doi.org/#1}}
\bibcommenthead

\bibitem{Kormendy&Ho2013ARA&A}
\bibinfo{author}{{Kormendy}, J.} \& \bibinfo{author}{{Ho}, L.~C.}
\newblock \bibinfo{title}{{Coevolution (Or Not) of Supermassive Black Holes and
  Host Galaxies}}.
\newblock \emph{\bibinfo{journal}{\araa}} \textbf{\bibinfo{volume}{51}},
  \bibinfo{pages}{511--653} (\bibinfo{year}{2013}).

\bibitem{Di_Matteo+2005Natur}
\bibinfo{author}{{Di Matteo}, T.}, \bibinfo{author}{{Springel}, V.} \&
  \bibinfo{author}{{Hernquist}, L.}
\newblock \bibinfo{title}{{Energy input from quasars regulates the growth and
  activity of black holes and their host galaxies}}.
\newblock \emph{\bibinfo{journal}{\nat}} \textbf{\bibinfo{volume}{433}},
  \bibinfo{pages}{604--607} (\bibinfo{year}{2005}).

\bibitem{Weinberger+2018MNRAS}
\bibinfo{author}{{Weinberger}, R.} \emph{et~al.}
\newblock \bibinfo{title}{{Supermassive black holes and their feedback effects
  in the IllustrisTNG simulation}}.
\newblock \emph{\bibinfo{journal}{\mnras}} \textbf{\bibinfo{volume}{479}},
  \bibinfo{pages}{4056--4072} (\bibinfo{year}{2018}).

\bibitem{Peng2007ApJ}
\bibinfo{author}{{Peng}, C.~Y.}
\newblock \bibinfo{title}{{How Mergers May Affect the Mass Scaling Relation
  between Gravitationally Bound Systems}}.
\newblock \emph{\bibinfo{journal}{\apj}} \textbf{\bibinfo{volume}{671}},
  \bibinfo{pages}{1098--1107} (\bibinfo{year}{2007}).

\bibitem{Greene+2020ARA&A}
\bibinfo{author}{{Greene}, J.~E.}, \bibinfo{author}{{Strader}, J.} \&
  \bibinfo{author}{{Ho}, L.~C.}
\newblock \bibinfo{title}{{Intermediate-Mass Black Holes}}.
\newblock \emph{\bibinfo{journal}{\araa}} \textbf{\bibinfo{volume}{58}},
  \bibinfo{pages}{257--312} (\bibinfo{year}{2020}).

\bibitem{Vestergaard&Peterson2006ApJ}
\bibinfo{author}{{Vestergaard}, M.} \& \bibinfo{author}{{Peterson}, B.~M.}
\newblock \bibinfo{title}{{Determining Central Black Hole Masses in Distant
  Active Galaxies and Quasars. II. Improved Optical and UV Scaling
  Relationships}}.
\newblock \emph{\bibinfo{journal}{\apj}} \textbf{\bibinfo{volume}{641}},
  \bibinfo{pages}{689--709} (\bibinfo{year}{2006}).

\bibitem{Wang+2009ApJ}
\bibinfo{author}{{Wang}, J.-G.} \emph{et~al.}
\newblock \bibinfo{title}{{Estimating Black Hole Masses in Active Galactic
  Nuclei Using the Mg II {\ensuremath{\lambda}}2800 Emission Line}}.
\newblock \emph{\bibinfo{journal}{\apj}} \textbf{\bibinfo{volume}{707}},
  \bibinfo{pages}{1334--1346} (\bibinfo{year}{2009}).

\bibitem{Peng+2006bApJ}
\bibinfo{author}{{Peng}, C.~Y.} \emph{et~al.}
\newblock \bibinfo{title}{{Probing the Coevolution of Supermassive Black Holes
  and Galaxies Using Gravitationally Lensed Quasar Hosts}}.
\newblock \emph{\bibinfo{journal}{\apj}} \textbf{\bibinfo{volume}{649}},
  \bibinfo{pages}{616--634} (\bibinfo{year}{2006}).

\bibitem{Alexander+2008AJ}
\bibinfo{author}{{Alexander}, D.~M.} \emph{et~al.}
\newblock \bibinfo{title}{{Weighing the Black Holes in z {\ensuremath{\approx}}
  2 Submillimeter-Emitting Galaxies Hosting Active Galactic Nuclei}}.
\newblock \emph{\bibinfo{journal}{\aj}} \textbf{\bibinfo{volume}{135}},
  \bibinfo{pages}{1968--1981} (\bibinfo{year}{2008}).

\bibitem{Schulze&Wisotzki2014MNRAS}
\bibinfo{author}{{Schulze}, A.} \& \bibinfo{author}{{Wisotzki}, L.}
\newblock \bibinfo{title}{{Accounting for selection effects in the BH-bulge
  relations: no evidence for cosmological evolution}}.
\newblock \emph{\bibinfo{journal}{\mnras}} \textbf{\bibinfo{volume}{438}},
  \bibinfo{pages}{3422--3433} (\bibinfo{year}{2014}).

\bibitem{Neeleman+2021ApJ}
\bibinfo{author}{{Neeleman}, M.} \emph{et~al.}
\newblock \bibinfo{title}{{The Kinematics of z {\ensuremath{\gtrsim}} 6 Quasar
  Host Galaxies}}.
\newblock \emph{\bibinfo{journal}{\apj}} \textbf{\bibinfo{volume}{911}},
  \bibinfo{pages}{141} (\bibinfo{year}{2021}).

\bibitem{Kelly2007ApJ}
\bibinfo{author}{{Kelly}, B.~C.}
\newblock \bibinfo{title}{{Some Aspects of Measurement Error in Linear
  Regression of Astronomical Data}}.
\newblock \emph{\bibinfo{journal}{\apj}} \textbf{\bibinfo{volume}{665}},
  \bibinfo{pages}{1489--1506} (\bibinfo{year}{2007}).

\bibitem{Bennert+2021ApJ}
\bibinfo{author}{{Bennert}, V.~N.} \emph{et~al.}
\newblock \bibinfo{title}{{A Local Baseline of the Black Hole Mass Scaling
  Relations for Active Galaxies. IV. Correlations Between M$_{BH}$ and Host
  Galaxy {\ensuremath{\sigma}}, Stellar Mass, and Luminosity}}.
\newblock \emph{\bibinfo{journal}{\apj}} \textbf{\bibinfo{volume}{921}},
  \bibinfo{pages}{36} (\bibinfo{year}{2021}).

\bibitem{Reines&Volonteri2015ApJ}
\bibinfo{author}{{Reines}, A.~E.} \& \bibinfo{author}{{Volonteri}, M.}
\newblock \bibinfo{title}{{Relations between Central Black Hole Mass and Total
  Galaxy Stellar Mass in the Local Universe}}.
\newblock \emph{\bibinfo{journal}{\apj}} \textbf{\bibinfo{volume}{813}},
  \bibinfo{pages}{82} (\bibinfo{year}{2015}).

\bibitem{Kim&Ho2019ApJ}
\bibinfo{author}{{Kim}, M.} \& \bibinfo{author}{{Ho}, L.~C.}
\newblock \bibinfo{title}{{Evidence for a Young Stellar Population in Nearby
  Type 1 Active Galaxies}}.
\newblock \emph{\bibinfo{journal}{\apj}} \textbf{\bibinfo{volume}{876}},
  \bibinfo{pages}{35} (\bibinfo{year}{2019}).

\bibitem{Ho2008ARA&A}
\bibinfo{author}{{Ho}, L.~C.}
\newblock \bibinfo{title}{{Nuclear Activity in Nearby Galaxies}}.
\newblock \emph{\bibinfo{journal}{\araa}} \textbf{\bibinfo{volume}{46}},
  \bibinfo{pages}{475--539} (\bibinfo{year}{2008}).

\bibitem{Boroson&Green1992ApJS}
\bibinfo{author}{{Boroson}, T.~A.} \& \bibinfo{author}{{Green}, R.~F.}
\newblock \bibinfo{title}{{The Emission-Line Properties of Low-Redshift
  Quasi-stellar Objects}}.
\newblock \emph{\bibinfo{journal}{\apjs}} \textbf{\bibinfo{volume}{80}},
  \bibinfo{pages}{109} (\bibinfo{year}{1992}).

\bibitem{Urrutia+2012ApJ}
\bibinfo{author}{{Urrutia}, T.} \emph{et~al.}
\newblock \bibinfo{title}{{Spitzer Observations of Young Red Quasars}}.
\newblock \emph{\bibinfo{journal}{\apj}} \textbf{\bibinfo{volume}{757}},
  \bibinfo{pages}{125} (\bibinfo{year}{2012}).

\bibitem{Sun+2015ApJ}
\bibinfo{author}{{Sun}, M.} \emph{et~al.}
\newblock \bibinfo{title}{{Evolution in the Black Hole{\textemdash}Galaxy
  Scaling Relations and the Duty Cycle of Nuclear Activity in Star-forming
  Galaxies}}.
\newblock \emph{\bibinfo{journal}{\apj}} \textbf{\bibinfo{volume}{802}},
  \bibinfo{pages}{14} (\bibinfo{year}{2015}).

\bibitem{Zhao+2021ApJ}
\bibinfo{author}{{Zhao}, Y.} \emph{et~al.}
\newblock \bibinfo{title}{{The Diverse Morphology, Stellar Population, and
  Black Hole Scaling Relations of the Host Galaxies of Nearby Quasars}}.
\newblock \emph{\bibinfo{journal}{\apj}} \textbf{\bibinfo{volume}{911}},
  \bibinfo{pages}{94} (\bibinfo{year}{2021}).

\bibitem{Mathur+2012ApJ}
\bibinfo{author}{{Mathur}, S.}, \bibinfo{author}{{Fields}, D.},
  \bibinfo{author}{{Peterson}, B.~M.} \& \bibinfo{author}{{Grupe}, D.}
\newblock \bibinfo{title}{{Supermassive Black Holes, Pseudobulges, and the
  Narrow-line Seyfert 1 Galaxies}}.
\newblock \emph{\bibinfo{journal}{\apj}} \textbf{\bibinfo{volume}{754}},
  \bibinfo{pages}{146} (\bibinfo{year}{2012}).

\bibitem{Greene&Ho2006ApJ}
\bibinfo{author}{{Greene}, J.~E.} \& \bibinfo{author}{{Ho}, L.~C.}
\newblock \bibinfo{title}{{The M$_{BH}$-{\ensuremath{\sigma}}$_{*}$ Relation in
  Local Active Galaxies}}.
\newblock \emph{\bibinfo{journal}{\apjl}} \textbf{\bibinfo{volume}{641}},
  \bibinfo{pages}{L21--L24} (\bibinfo{year}{2006}).

\bibitem{Madau&Dickinson2014ARA&A}
\bibinfo{author}{{Madau}, P.} \& \bibinfo{author}{{Dickinson}, M.}
\newblock \bibinfo{title}{{Cosmic Star-Formation History}}.
\newblock \emph{\bibinfo{journal}{\araa}} \textbf{\bibinfo{volume}{52}},
  \bibinfo{pages}{415--486} (\bibinfo{year}{2014}).

\bibitem{Graham&Sahu2023}
\bibinfo{author}{{Graham}, A.~W.} \& \bibinfo{author}{{Sahu}, N.}
\newblock \bibinfo{title}{{Appreciating mergers for understanding the
  non-linear M$_{bh}$-M$_{*,spheroid}$ and M$_{bh}$-M$_{*, galaxy}$ relations,
  updated herein, and the implications for the (reduced) role of AGN
  feedback}}.
\newblock \emph{\bibinfo{journal}{\mnras}} \textbf{\bibinfo{volume}{518}},
  \bibinfo{pages}{2177--2200} (\bibinfo{year}{2023}).

\bibitem{Steinborn+2018MNRAS}
\bibinfo{author}{{Steinborn}, L.~K.} \emph{et~al.}
\newblock \bibinfo{title}{{Cosmological simulations of black hole growth II:
  how (in)significant are merger events for fuelling nuclear activity?}}
\newblock \emph{\bibinfo{journal}{\mnras}} \textbf{\bibinfo{volume}{481}},
  \bibinfo{pages}{341--360} (\bibinfo{year}{2018}).

\bibitem{Zhuang&Ho2020ApJ}
\bibinfo{author}{{Zhuang}, M.-Y.} \& \bibinfo{author}{{Ho}, L.~C.}
\newblock \bibinfo{title}{{The Interplay between Star Formation and Black Hole
  Accretion in Nearby Active Galaxies}}.
\newblock \emph{\bibinfo{journal}{\apj}} \textbf{\bibinfo{volume}{896}},
  \bibinfo{pages}{108} (\bibinfo{year}{2020}).

\bibitem{Zhuang&Ho2022ApJ}
\bibinfo{author}{{Zhuang}, M.-Y.} \& \bibinfo{author}{{Ho}, L.~C.}
\newblock \bibinfo{title}{{The Star-forming Main Sequence of the Host Galaxies
  of Low-redshift Quasars}}.
\newblock \emph{\bibinfo{journal}{\apj}} \textbf{\bibinfo{volume}{934}},
  \bibinfo{pages}{130} (\bibinfo{year}{2022}).

\bibitem{Farrah+2022MNRAS}
\bibinfo{author}{{Farrah}, D.} \emph{et~al.}
\newblock \bibinfo{title}{{Stellar and black hole assembly in $z < 0.3$
  infrared-luminous mergers: intermittent starbursts versus super-Eddington
  accretion}}.
\newblock \emph{\bibinfo{journal}{\mnras}} \textbf{\bibinfo{volume}{513}},
  \bibinfo{pages}{4770--4786} (\bibinfo{year}{2022}).

\bibitem{Borys+2005ApJ}
\bibinfo{author}{{Borys}, C.} \emph{et~al.}
\newblock \bibinfo{title}{{The Relationship between Stellar and Black Hole Mass
  in Submillimeter Galaxies}}.
\newblock \emph{\bibinfo{journal}{\apj}} \textbf{\bibinfo{volume}{635}},
  \bibinfo{pages}{853--863} (\bibinfo{year}{2005}).

\bibitem{Veilleux+2006ApJ}
\bibinfo{author}{{Veilleux}, S.} \emph{et~al.}
\newblock \bibinfo{title}{{A Deep Hubble Space Telescope H-Band Imaging Survey
  of Massive Gas-rich Mergers}}.
\newblock \emph{\bibinfo{journal}{\apj}} \textbf{\bibinfo{volume}{643}},
  \bibinfo{pages}{707--723} (\bibinfo{year}{2006}).

\bibitem{Chen+2022ApJ}
\bibinfo{author}{{Chen}, C.-C.} \emph{et~al.}
\newblock \bibinfo{title}{{JWST Sneaks a Peek at the Stellar Morphology of
  $z\sim2$ Submillimeter Galaxies: Bulge Formation at Cosmic Noon}}.
\newblock \emph{\bibinfo{journal}{\apjl}} \textbf{\bibinfo{volume}{939}},
  \bibinfo{pages}{L7} (\bibinfo{year}{2022}).

\bibitem{Greve+2005}
\bibinfo{author}{{Greve}, T.~R.} \emph{et~al.}
\newblock \bibinfo{title}{{An interferometric CO survey of luminous
  submillimetre galaxies}}.
\newblock \emph{\bibinfo{journal}{\mnras}} \textbf{\bibinfo{volume}{359}},
  \bibinfo{pages}{1165--1183} (\bibinfo{year}{2005}).

\bibitem{Shangguan+2019}
\bibinfo{author}{{Shangguan}, J.} \emph{et~al.}
\newblock \bibinfo{title}{{Interstellar Medium and Star Formation of Starburst
  Galaxies on the Merger Sequence}}.
\newblock \emph{\bibinfo{journal}{\apj}} \textbf{\bibinfo{volume}{870}},
  \bibinfo{pages}{104} (\bibinfo{year}{2019}).

\bibitem{Orban+2011MNRAS}
\bibinfo{author}{{Orban de Xivry}, G.} \emph{et~al.}
\newblock \bibinfo{title}{{The role of secular evolution in the black hole
  growth of narrow-line Seyfert 1 galaxies}}.
\newblock \emph{\bibinfo{journal}{\mnras}} \textbf{\bibinfo{volume}{417}},
  \bibinfo{pages}{2721--2736} (\bibinfo{year}{2011}).

\bibitem{Shangguan+2020ApJ}
\bibinfo{author}{{Shangguan}, J.}, \bibinfo{author}{{Ho}, L.~C.},
  \bibinfo{author}{{Bauer}, F.~E.}, \bibinfo{author}{{Wang}, R.} \&
  \bibinfo{author}{{Treister}, E.}
\newblock \bibinfo{title}{{AGN Feedback and Star Formation of Quasar Host
  Galaxies: Insights from the Molecular Gas}}.
\newblock \emph{\bibinfo{journal}{\apj}} \textbf{\bibinfo{volume}{899}},
  \bibinfo{pages}{112} (\bibinfo{year}{2020}).

\bibitem{Xie+2021ApJ}
\bibinfo{author}{{Xie}, Y.}, \bibinfo{author}{{Ho}, L.~C.},
  \bibinfo{author}{{Zhuang}, M.-Y.} \& \bibinfo{author}{{Shangguan}, J.}
\newblock \bibinfo{title}{{The Infrared Emission and Vigorous Star Formation of
  Low-redshift Quasars}}.
\newblock \emph{\bibinfo{journal}{\apj}} \textbf{\bibinfo{volume}{910}},
  \bibinfo{pages}{124} (\bibinfo{year}{2021}).

\bibitem{Kennicutt1998ApJ}
\bibinfo{author}{{Kennicutt}, J., Robert~C.}
\newblock \bibinfo{title}{{The Global Schmidt Law in Star-forming Galaxies}}.
\newblock \emph{\bibinfo{journal}{\apj}} \textbf{\bibinfo{volume}{498}},
  \bibinfo{pages}{541--552} (\bibinfo{year}{1998}).

\bibitem{Martini2004}
\bibinfo{author}{{Martini}, P.}
\newblock \bibinfo{editor}{{Ho}, L.~C.} (ed.) \emph{\bibinfo{title}{{QSO
  Lifetimes}}}.
\newblock (ed.\bibinfo{editor}{{Ho}, L.~C.})
  \emph{\bibinfo{booktitle}{Coevolution of Black Holes and Galaxies}},
  \bibinfo{pages}{169} (\bibinfo{year}{2004}).
\newblock \eprint{astro-ph/0304009}.

\bibitem{Habouzit+2017MNRAS}
\bibinfo{author}{{Habouzit}, M.}, \bibinfo{author}{{Volonteri}, M.} \&
  \bibinfo{author}{{Dubois}, Y.}
\newblock \bibinfo{title}{{Blossoms from black hole seeds: properties and early
  growth regulated by supernova feedback}}.
\newblock \emph{\bibinfo{journal}{\mnras}} \textbf{\bibinfo{volume}{468}},
  \bibinfo{pages}{3935--3948} (\bibinfo{year}{2017}).

\bibitem{Catmabacak+2022MNRAS}
\bibinfo{author}{{{\c{C}}atmabacak}, O.} \emph{et~al.}
\newblock \bibinfo{title}{{Black hole-galaxy scaling relations in FIRE: the
  importance of black hole location and mergers}}.
\newblock \emph{\bibinfo{journal}{\mnras}} \textbf{\bibinfo{volume}{511}},
  \bibinfo{pages}{506--535} (\bibinfo{year}{2022}).

\bibitem{Koss+2021ApJS}
\bibinfo{author}{{Koss}, M.~J.} \emph{et~al.}
\newblock \bibinfo{title}{{BAT AGN Spectroscopic Survey. XX. Molecular Gas in
  Nearby Hard-X-Ray-selected AGN Galaxies}}.
\newblock \emph{\bibinfo{journal}{\apjs}} \textbf{\bibinfo{volume}{252}},
  \bibinfo{pages}{29} (\bibinfo{year}{2021}).

\bibitem{Molina+2021ApJ}
\bibinfo{author}{{Molina}, J.} \emph{et~al.}
\newblock \bibinfo{title}{{Compact Molecular Gas Distribution in Quasar Host
  Galaxies}}.
\newblock \emph{\bibinfo{journal}{\apj}} \textbf{\bibinfo{volume}{908}},
  \bibinfo{pages}{231} (\bibinfo{year}{2021}).

\bibitem{Aird+2019MNRAS}
\bibinfo{author}{{Aird}, J.}, \bibinfo{author}{{Coil}, A.~L.} \&
  \bibinfo{author}{{Georgakakis}, A.}
\newblock \bibinfo{title}{{X-rays across the galaxy population - III. The
  incidence of AGN as a function of star formation rate}}.
\newblock \emph{\bibinfo{journal}{\mnras}} \textbf{\bibinfo{volume}{484}},
  \bibinfo{pages}{4360--4378} (\bibinfo{year}{2019}).

\bibitem{Decarli+2018ApJ}
\bibinfo{author}{{Decarli}, R.} \emph{et~al.}
\newblock \bibinfo{title}{{An ALMA [C II] Survey of 27 Quasars at $z > 5.94$}}.
\newblock \emph{\bibinfo{journal}{\apj}} \textbf{\bibinfo{volume}{854}},
  \bibinfo{pages}{97} (\bibinfo{year}{2018}).

\bibitem{Inayoshi+2020ARA&A}
\bibinfo{author}{{Inayoshi}, K.}, \bibinfo{author}{{Visbal}, E.} \&
  \bibinfo{author}{{Haiman}, Z.}
\newblock \bibinfo{title}{{The Assembly of the First Massive Black Holes}}.
\newblock \emph{\bibinfo{journal}{\araa}} \textbf{\bibinfo{volume}{58}},
  \bibinfo{pages}{27--97} (\bibinfo{year}{2020}).

\bibitem{Daddi+2005ApJ}
\bibinfo{author}{{Daddi}, E.} \emph{et~al.}
\newblock \bibinfo{title}{{Passively Evolving Early-Type Galaxies at $1.4
  \lesssim z \lesssim 2.5$ in the Hubble Ultra Deep Field}}.
\newblock \emph{\bibinfo{journal}{\apj}} \textbf{\bibinfo{volume}{626}},
  \bibinfo{pages}{680--697} (\bibinfo{year}{2005}).

\bibitem{van_Dokkum+2010ApJ}
\bibinfo{author}{{van Dokkum}, P.~G.} \emph{et~al.}
\newblock \bibinfo{title}{{The Growth of Massive Galaxies Since z = 2}}.
\newblock \emph{\bibinfo{journal}{\apj}} \textbf{\bibinfo{volume}{709}},
  \bibinfo{pages}{1018--1041} (\bibinfo{year}{2010}).

\bibitem{Oser+2010ApJ}
\bibinfo{author}{{Oser}, L.}, \bibinfo{author}{{Ostriker}, J.~P.},
  \bibinfo{author}{{Naab}, T.}, \bibinfo{author}{{Johansson}, P.~H.} \&
  \bibinfo{author}{{Burkert}, A.}
\newblock \bibinfo{title}{{The Two Phases of Galaxy Formation}}.
\newblock \emph{\bibinfo{journal}{\apj}} \textbf{\bibinfo{volume}{725}},
  \bibinfo{pages}{2312--2323} (\bibinfo{year}{2010}).

\bibitem{Davari+2017}
\bibinfo{author}{{Davari}, R.~H.}, \bibinfo{author}{{Ho}, L.~C.},
  \bibinfo{author}{{Mobasher}, B.} \& \bibinfo{author}{{Canalizo}, G.}
\newblock \bibinfo{title}{{Detection of Prominent Stellar Disks in the
  Progenitors of Present-day Massive Elliptical Galaxies}}.
\newblock \emph{\bibinfo{journal}{\apj}} \textbf{\bibinfo{volume}{836}},
  \bibinfo{pages}{75} (\bibinfo{year}{2017}).

\bibitem{Zhuang+2021ApJ}
\bibinfo{author}{{Zhuang}, M.-Y.}, \bibinfo{author}{{Ho}, L.~C.} \&
  \bibinfo{author}{{Shangguan}, J.}
\newblock \bibinfo{title}{{Black Hole Accretion Correlates with Star Formation
  Rate and Star Formation Efficiency in Nearby Luminous Type 1 Active
  Galaxies}}.
\newblock \emph{\bibinfo{journal}{\apj}} \textbf{\bibinfo{volume}{906}},
  \bibinfo{pages}{38} (\bibinfo{year}{2021}).

\bibitem{Ward+2022MNRAS}
\bibinfo{author}{{Ward}, S.~R.}, \bibinfo{author}{{Harrison}, C.~M.},
  \bibinfo{author}{{Costa}, T.} \& \bibinfo{author}{{Mainieri}, V.}
\newblock \bibinfo{title}{{Cosmological simulations predict that AGN
  preferentially live in gas-rich, star-forming galaxies despite effective
  feedback}}.
\newblock \emph{\bibinfo{journal}{\mnras}} \textbf{\bibinfo{volume}{514}},
  \bibinfo{pages}{2936--2957} (\bibinfo{year}{2022}).

\bibitem{King&Pounds2015ARA&A}
\bibinfo{author}{{King}, A.} \& \bibinfo{author}{{Pounds}, K.}
\newblock \bibinfo{title}{{Powerful Outflows and Feedback from Active Galactic
  Nuclei}}.
\newblock \emph{\bibinfo{journal}{\araa}} \textbf{\bibinfo{volume}{53}},
  \bibinfo{pages}{115--154} (\bibinfo{year}{2015}).

\bibitem{Tremaine+2002ApJ}
\bibinfo{author}{{Tremaine}, S.} \emph{et~al.}
\newblock \bibinfo{title}{{The Slope of the Black Hole Mass versus Velocity
  Dispersion Correlation}}.
\newblock \emph{\bibinfo{journal}{\apj}} \textbf{\bibinfo{volume}{574}},
  \bibinfo{pages}{740--753} (\bibinfo{year}{2002}).

\bibitem{Gao+2019}
\bibinfo{author}{{Gao}, H.}, \bibinfo{author}{{Ho}, L.~C.},
  \bibinfo{author}{{Barth}, A.~J.} \& \bibinfo{author}{{Li}, Z.-Y.}
\newblock \bibinfo{title}{{The Carnegie-Irvine Galaxy Survey. VIII.
  Demographics of Bulges along the Hubble Sequence}}.
\newblock \emph{\bibinfo{journal}{\apjs}} \textbf{\bibinfo{volume}{244}},
  \bibinfo{pages}{34} (\bibinfo{year}{2019}).

\bibitem{Ding_YZ+2022ApJ}
\bibinfo{author}{{Ding}, Y.}, \bibinfo{author}{{Li}, R.},
  \bibinfo{author}{{Ho}, L.~C.} \& \bibinfo{author}{{Ricci}, C.}
\newblock \bibinfo{title}{{Accretion Disk Outflow during the X-Ray Flare of the
  Super-Eddington Active Nucleus of I Zwicky 1}}.
\newblock \emph{\bibinfo{journal}{\apj}} \textbf{\bibinfo{volume}{931}},
  \bibinfo{pages}{77} (\bibinfo{year}{2022}).

\bibitem{Dave+2019MNRAS}
\bibinfo{author}{{Dav{\'e}}, R.} \emph{et~al.}
\newblock \bibinfo{title}{{SIMBA: Cosmological simulations with black hole
  growth and feedback}}.
\newblock \emph{\bibinfo{journal}{\mnras}} \textbf{\bibinfo{volume}{486}},
  \bibinfo{pages}{2827--2849} (\bibinfo{year}{2019}).

\bibitem{Liu+2019ApJS}
\bibinfo{author}{{Liu}, H.-Y.} \emph{et~al.}
\newblock \bibinfo{title}{{A Comprehensive and Uniform Sample of Broad-line
  Active Galactic Nuclei from the SDSS DR7}}.
\newblock \emph{\bibinfo{journal}{\apjs}} \textbf{\bibinfo{volume}{243}},
  \bibinfo{pages}{21} (\bibinfo{year}{2019}).

\bibitem{Abazajian+2009ApJS}
\bibinfo{author}{{Abazajian}, K.~N.} \emph{et~al.}
\newblock \bibinfo{title}{{The Seventh Data Release of the Sloan Digital Sky
  Survey}}.
\newblock \emph{\bibinfo{journal}{\apjs}} \textbf{\bibinfo{volume}{182}},
  \bibinfo{pages}{543--558} (\bibinfo{year}{2009}).

\bibitem{York+2000AJ}
\bibinfo{author}{{York}, D.~G.} \emph{et~al.}
\newblock \bibinfo{title}{{The Sloan Digital Sky Survey: Technical Summary}}.
\newblock \emph{\bibinfo{journal}{\aj}} \textbf{\bibinfo{volume}{120}},
  \bibinfo{pages}{1579--1587} (\bibinfo{year}{2000}).

\bibitem{Shen+2011ApJS}
\bibinfo{author}{{Shen}, Y.} \emph{et~al.}
\newblock \bibinfo{title}{{A Catalog of Quasar Properties from Sloan Digital
  Sky Survey Data Release 7}}.
\newblock \emph{\bibinfo{journal}{\apjs}} \textbf{\bibinfo{volume}{194}},
  \bibinfo{pages}{45} (\bibinfo{year}{2011}).

\bibitem{Oh+2015ApJS}
\bibinfo{author}{{Oh}, K.} \emph{et~al.}
\newblock \bibinfo{title}{{A New Catalog of Type 1 AGNs and its Implications on
  the AGN Unified Model}}.
\newblock \emph{\bibinfo{journal}{\apjs}} \textbf{\bibinfo{volume}{219}},
  \bibinfo{pages}{1} (\bibinfo{year}{2015}).

\bibitem{Lyke+2020ApJS}
\bibinfo{author}{{Lyke}, B.~W.} \emph{et~al.}
\newblock \bibinfo{title}{{The Sloan Digital Sky Survey Quasar Catalog:
  Sixteenth Data Release}}.
\newblock \emph{\bibinfo{journal}{\apjs}} \textbf{\bibinfo{volume}{250}},
  \bibinfo{pages}{8} (\bibinfo{year}{2020}).

\bibitem{Rakshit+2020ApJS}
\bibinfo{author}{{Rakshit}, S.}, \bibinfo{author}{{Stalin}, C.~S.} \&
  \bibinfo{author}{{Kotilainen}, J.}
\newblock \bibinfo{title}{{Spectral Properties of Quasars from Sloan Digital
  Sky Survey Data Release 14: The Catalog}}.
\newblock \emph{\bibinfo{journal}{\apjs}} \textbf{\bibinfo{volume}{249}},
  \bibinfo{pages}{17} (\bibinfo{year}{2020}).

\bibitem{McLure&Dunlop2004MNRAS}
\bibinfo{author}{{McLure}, R.~J.} \& \bibinfo{author}{{Dunlop}, J.~S.}
\newblock \bibinfo{title}{{The cosmological evolution of quasar black hole
  masses}}.
\newblock \emph{\bibinfo{journal}{\mnras}} \textbf{\bibinfo{volume}{352}},
  \bibinfo{pages}{1390--1404} (\bibinfo{year}{2004}).

\bibitem{Ho&Kim2015ApJ}
\bibinfo{author}{{Ho}, L.~C.} \& \bibinfo{author}{{Kim}, M.}
\newblock \bibinfo{title}{{A Revised Calibration of the Virial Mass Estimator
  for Black Holes in Active Galaxies Based on Single-epoch
  H{\ensuremath{\beta}} Spectra}}.
\newblock \emph{\bibinfo{journal}{\apj}} \textbf{\bibinfo{volume}{809}},
  \bibinfo{pages}{123} (\bibinfo{year}{2015}).

\bibitem{Ho&Kim2014}
\bibinfo{author}{{Ho}, L.~C.} \& \bibinfo{author}{{Kim}, M.}
\newblock \bibinfo{title}{{The Black Hole Mass Scale of Classical and Pseudo
  Bulges in Active Galaxies}}.
\newblock \emph{\bibinfo{journal}{\apj}} \textbf{\bibinfo{volume}{789}},
  \bibinfo{pages}{17} (\bibinfo{year}{2014}).

\bibitem{Greene&Ho2005ApJ}
\bibinfo{author}{{Greene}, J.~E.} \& \bibinfo{author}{{Ho}, L.~C.}
\newblock \bibinfo{title}{{Estimating Black Hole Masses in Active Galaxies
  Using the H{\ensuremath{\alpha}} Emission Line}}.
\newblock \emph{\bibinfo{journal}{\apj}} \textbf{\bibinfo{volume}{630}},
  \bibinfo{pages}{122--129} (\bibinfo{year}{2005}).

\bibitem{Woo+2015ApJ}
\bibinfo{author}{{Woo}, J.-H.}, \bibinfo{author}{{Yoon}, Y.},
  \bibinfo{author}{{Park}, S.}, \bibinfo{author}{{Park}, D.} \&
  \bibinfo{author}{{Kim}, S.~C.}
\newblock \bibinfo{title}{{The Black Hole Mass-Stellar Velocity Dispersion
  Relation of Narrow-line Seyfert 1 Galaxies}}.
\newblock \emph{\bibinfo{journal}{\apj}} \textbf{\bibinfo{volume}{801}},
  \bibinfo{pages}{38} (\bibinfo{year}{2015}).

\bibitem{Du+2016ApJ}
\bibinfo{author}{{Du}, P.} \emph{et~al.}
\newblock \bibinfo{title}{{Supermassive Black Holes with High Accretion Rates
  in Active Galactic Nuclei. V. A New Size-Luminosity Scaling Relation for the
  Broad-line Region}}.
\newblock \emph{\bibinfo{journal}{\apj}} \textbf{\bibinfo{volume}{825}},
  \bibinfo{pages}{126} (\bibinfo{year}{2016}).

\bibitem{Du&Wang2019ApJ}
\bibinfo{author}{{Du}, P.} \& \bibinfo{author}{{Wang}, J.-M.}
\newblock \bibinfo{title}{{The Radius-Luminosity Relationship Depends on
  Optical Spectra in Active Galactic Nuclei}}.
\newblock \emph{\bibinfo{journal}{\apj}} \textbf{\bibinfo{volume}{886}},
  \bibinfo{pages}{42} (\bibinfo{year}{2019}).

\bibitem{Dalla_Bonta+2020ApJ}
\bibinfo{author}{{Dalla Bont{\`a}}, E.} \emph{et~al.}
\newblock \bibinfo{title}{{The Sloan Digital Sky Survey Reverberation Mapping
  Project: Estimating Masses of Black Holes in Quasars with Single-epoch
  Spectroscopy}}.
\newblock \emph{\bibinfo{journal}{\apj}} \textbf{\bibinfo{volume}{903}},
  \bibinfo{pages}{112} (\bibinfo{year}{2020}).

\bibitem{Fonseca2020}
\bibinfo{author}{{Fonseca Alvarez}, G.} \emph{et~al.}
\newblock \bibinfo{title}{{The Sloan Digital Sky Survey Reverberation Mapping
  Project: The H{\ensuremath{\beta}} Radius-Luminosity Relation}}.
\newblock \emph{\bibinfo{journal}{\apj}} \textbf{\bibinfo{volume}{899}},
  \bibinfo{pages}{73} (\bibinfo{year}{2020}).

\bibitem{Treu+2007ApJ}
\bibinfo{author}{{Treu}, T.}, \bibinfo{author}{{Woo}, J.-H.},
  \bibinfo{author}{{Malkan}, M.~A.} \& \bibinfo{author}{{Blandford}, R.~D.}
\newblock \bibinfo{title}{{Cosmic Evolution of Black Holes and Spheroids. II.
  Scaling Relations at z=0.36}}.
\newblock \emph{\bibinfo{journal}{\apj}} \textbf{\bibinfo{volume}{667}},
  \bibinfo{pages}{117--130} (\bibinfo{year}{2007}).

\bibitem{Bennert+2010ApJ}
\bibinfo{author}{{Bennert}, V.~N.} \emph{et~al.}
\newblock \bibinfo{title}{{Cosmic Evolution of Black Holes and Spheroids. IV.
  The M $_{BH}$-L $_{sph}$ Relation}}.
\newblock \emph{\bibinfo{journal}{\apj}} \textbf{\bibinfo{volume}{708}},
  \bibinfo{pages}{1507--1527} (\bibinfo{year}{2010}).

\bibitem{Falomo+2014MNRAS}
\bibinfo{author}{{Falomo}, R.}, \bibinfo{author}{{Bettoni}, D.},
  \bibinfo{author}{{Karhunen}, K.}, \bibinfo{author}{{Kotilainen}, J.~K.} \&
  \bibinfo{author}{{Uslenghi}, M.}
\newblock \bibinfo{title}{{Low-redshift quasars in the Sloan Digital Sky Survey
  Stripe 82. The host galaxies}}.
\newblock \emph{\bibinfo{journal}{\mnras}} \textbf{\bibinfo{volume}{440}},
  \bibinfo{pages}{476--493} (\bibinfo{year}{2014}).

\bibitem{Matsuoka+2014ApJ}
\bibinfo{author}{{Matsuoka}, Y.}, \bibinfo{author}{{Strauss}, M.~A.},
  \bibinfo{author}{{Price}, I., Ted~N.} \& \bibinfo{author}{{DiDonato}, M.~S.}
\newblock \bibinfo{title}{{Massive Star-forming Host Galaxies of Quasars on
  Sloan Digital Sky Survey Stripe 82}}.
\newblock \emph{\bibinfo{journal}{\apj}} \textbf{\bibinfo{volume}{780}},
  \bibinfo{pages}{162} (\bibinfo{year}{2014}).

\bibitem{Bentz&Manne-Nicholas2018ApJ}
\bibinfo{author}{{Bentz}, M.~C.} \& \bibinfo{author}{{Manne-Nicholas}, E.}
\newblock \bibinfo{title}{{Black Hole-Galaxy Scaling Relationships for Active
  Galactic Nuclei with Reverberation Masses}}.
\newblock \emph{\bibinfo{journal}{\apj}} \textbf{\bibinfo{volume}{864}},
  \bibinfo{pages}{146} (\bibinfo{year}{2018}).

\bibitem{Ishino+2020PASJ}
\bibinfo{author}{{Ishino}, T.} \emph{et~al.}
\newblock \bibinfo{title}{{Subaru Hyper Suprime-Cam view of quasar host
  galaxies at $z < 1$}}.
\newblock \emph{\bibinfo{journal}{\pasj}} \textbf{\bibinfo{volume}{72}},
  \bibinfo{pages}{83} (\bibinfo{year}{2020}).

\bibitem{Li_Junyao+2021aApJ}
\bibinfo{author}{{Li}, J.} \emph{et~al.}
\newblock \bibinfo{title}{{The Sizes of Quasar Host Galaxies in the Hyper
  Suprime-Cam Subaru Strategic Program}}.
\newblock \emph{\bibinfo{journal}{\apj}} \textbf{\bibinfo{volume}{918}},
  \bibinfo{pages}{22} (\bibinfo{year}{2021}).

\bibitem{Li_Jennifer+2021ApJ}
\bibinfo{author}{{Li}, J. I.~H.} \emph{et~al.}
\newblock \bibinfo{title}{{The Sloan Digital Sky Survey Reverberation Mapping
  Project: The M$_{BH}$-Host Relations at $0.2 \lesssim z \lesssim 0.6$ from
  Reverberation Mapping and Hubble Space Telescope Imaging}}.
\newblock \emph{\bibinfo{journal}{\apj}} \textbf{\bibinfo{volume}{906}},
  \bibinfo{pages}{103} (\bibinfo{year}{2021}).

\bibitem{Son+2022JKAS}
\bibinfo{author}{{Son}, S.}, \bibinfo{author}{{Kim}, M.},
  \bibinfo{author}{{Barth}, A.~J.} \& \bibinfo{author}{{Ho}, L.~C.}
\newblock \bibinfo{title}{{Relation between Black Hole Mass and Bulge
  Luminosity in Hard X-Ray Selected Type 1 AGNs}}.
\newblock \emph{\bibinfo{journal}{Journal of Korean Astronomical Society}}
  \textbf{\bibinfo{volume}{55}}, \bibinfo{pages}{37--57}
  (\bibinfo{year}{2022}).

\bibitem{Heckman1980}
\bibinfo{author}{{Heckman}, T.~M.}
\newblock \bibinfo{title}{{An Optical and Radio Survey of the Nuclei of Bright
  Galaxies - Activity in the Normal Galactic Nuclei}}.
\newblock \emph{\bibinfo{journal}{\aap}} \textbf{\bibinfo{volume}{87}},
  \bibinfo{pages}{152} (\bibinfo{year}{1980}).

\bibitem{Baldwin+1981PASP}
\bibinfo{author}{{Baldwin}, J.~A.}, \bibinfo{author}{{Phillips}, M.~M.} \&
  \bibinfo{author}{{Terlevich}, R.}
\newblock \bibinfo{title}{{Classification parameters for the emission-line
  spectra of extragalactic objects.}}
\newblock \emph{\bibinfo{journal}{\pasp}} \textbf{\bibinfo{volume}{93}},
  \bibinfo{pages}{5--19} (\bibinfo{year}{1981}).

\bibitem{Chambers+2016arXiv}
\bibinfo{author}{{Chambers}, K.~C.} \emph{et~al.}
\newblock \bibinfo{title}{{The Pan-STARRS1 Surveys}}.
\newblock \emph{\bibinfo{journal}{arXiv e-prints}}
  \bibinfo{pages}{arXiv:1612.05560} (\bibinfo{year}{2016}).

\bibitem{Kron1980}
\bibinfo{author}{{Kron}, R.~G.}
\newblock \bibinfo{title}{{Photometry of a complete sample of faint galaxies.}}
\newblock \emph{\bibinfo{journal}{\apjs}} \textbf{\bibinfo{volume}{43}},
  \bibinfo{pages}{305--325} (\bibinfo{year}{1980}).

\bibitem{Flewelling+2020ApJS}
\bibinfo{author}{{Flewelling}, H.~A.} \emph{et~al.}
\newblock \bibinfo{title}{{The Pan-STARRS1 Database and Data Products}}.
\newblock \emph{\bibinfo{journal}{\apjs}} \textbf{\bibinfo{volume}{251}},
  \bibinfo{pages}{7} (\bibinfo{year}{2020}).

\bibitem{1996A&AS..117..393B}
\bibinfo{author}{{Bertin}, E.} \& \bibinfo{author}{{Arnouts}, S.}
\newblock \bibinfo{title}{{SExtractor: Software for source extraction.}}
\newblock \emph{\bibinfo{journal}{\aaps}} \textbf{\bibinfo{volume}{117}},
  \bibinfo{pages}{393--404} (\bibinfo{year}{1996}).

\bibitem{Bertin+2002ASPC}
\bibinfo{author}{{Bertin}, E.} \emph{et~al.}
\newblock \bibinfo{editor}{{Bohlender}, D.~A.}, \bibinfo{editor}{{Durand}, D.}
  \& \bibinfo{editor}{{Handley}, T.~H.} (eds) \emph{\bibinfo{title}{{The
  TERAPIX Pipeline}}}.
\newblock (eds \bibinfo{editor}{{Bohlender}, D.~A.}, \bibinfo{editor}{{Durand},
  D.} \& \bibinfo{editor}{{Handley}, T.~H.})
  \emph{\bibinfo{booktitle}{Astronomical Data Analysis Software and Systems
  XI}}, Vol. \bibinfo{volume}{281} of \emph{\bibinfo{series}{ASP Conf. Ser.}},
  \bibinfo{pages}{228} (\bibinfo{publisher}{San Francisco, CA: ASP},
  \bibinfo{year}{2002}).

\bibitem{Farrow+2014MNRAS}
\bibinfo{author}{{Farrow}, D.~J.} \emph{et~al.}
\newblock \bibinfo{title}{{Pan-STARRS1: Galaxy clustering in the Small Area
  Survey 2}}.
\newblock \emph{\bibinfo{journal}{\mnras}} \textbf{\bibinfo{volume}{437}},
  \bibinfo{pages}{748--770} (\bibinfo{year}{2014}).

\bibitem{Moffat1969}
\bibinfo{author}{{Moffat}, A.~F.~J.}
\newblock \bibinfo{title}{{A Theoretical Investigation of Focal Stellar Images
  in the Photographic Emulsion and Application to Photographic Photometry}}.
\newblock \emph{\bibinfo{journal}{\aap}} \textbf{\bibinfo{volume}{3}},
  \bibinfo{pages}{455} (\bibinfo{year}{1969}).

\bibitem{Haussler+2013MNRAS}
\bibinfo{author}{{H{\"a}u{\ss}ler}, B.} \emph{et~al.}
\newblock \bibinfo{title}{{MegaMorph - multiwavelength measurement of galaxy
  structure: complete S{\'e}rsic profile information from modern surveys}}.
\newblock \emph{\bibinfo{journal}{\mnras}} \textbf{\bibinfo{volume}{430}},
  \bibinfo{pages}{330--369} (\bibinfo{year}{2013}).

\bibitem{Vika+2013MNRAS}
\bibinfo{author}{{Vika}, M.} \emph{et~al.}
\newblock \bibinfo{title}{{MegaMorph - multiwavelength measurement of galaxy
  structure. S{\'e}rsic profile fits to galaxies near and far}}.
\newblock \emph{\bibinfo{journal}{\mnras}} \textbf{\bibinfo{volume}{435}},
  \bibinfo{pages}{623--649} (\bibinfo{year}{2013}).

\bibitem{Sersic1968}
\bibinfo{author}{{Sersic}, J.~L.}
\newblock \emph{\bibinfo{title}{{Atlas de Galaxias Australes}}}
  (\bibinfo{year}{1968}).

\bibitem{Kelvin+2012MNRAS}
\bibinfo{author}{{Kelvin}, L.~S.} \emph{et~al.}
\newblock \bibinfo{title}{{Galaxy And Mass Assembly (GAMA): Structural
  Investigation of Galaxies via Model Analysis}}.
\newblock \emph{\bibinfo{journal}{\mnras}} \textbf{\bibinfo{volume}{421}},
  \bibinfo{pages}{1007--1039} (\bibinfo{year}{2012}).

\bibitem{Peng+2002AJ}
\bibinfo{author}{{Peng}, C.~Y.}, \bibinfo{author}{{Ho}, L.~C.},
  \bibinfo{author}{{Impey}, C.~D.} \& \bibinfo{author}{{Rix}, H.-W.}
\newblock \bibinfo{title}{{Detailed Structural Decomposition of Galaxy
  Images}}.
\newblock \emph{\bibinfo{journal}{\aj}} \textbf{\bibinfo{volume}{124}},
  \bibinfo{pages}{266--293} (\bibinfo{year}{2002}).

\bibitem{Peng+2010AJ}
\bibinfo{author}{{Peng}, C.~Y.}, \bibinfo{author}{{Ho}, L.~C.},
  \bibinfo{author}{{Impey}, C.~D.} \& \bibinfo{author}{{Rix}, H.-W.}
\newblock \bibinfo{title}{{Detailed Decomposition of Galaxy Images. II. Beyond
  Axisymmetric Models}}.
\newblock \emph{\bibinfo{journal}{\aj}} \textbf{\bibinfo{volume}{139}},
  \bibinfo{pages}{2097--2129} (\bibinfo{year}{2010}).

\bibitem{Haussler+2022A&A}
\bibinfo{author}{{H{\"a}u{\ss}ler}, B.} \emph{et~al.}
\newblock \bibinfo{title}{{GALAPAGOS-2/GALFITM/GAMA - Multi-wavelength
  measurement of galaxy structure: Separating the properties of spheroid and
  disk components in modern surveys}}.
\newblock \emph{\bibinfo{journal}{\aap}} \textbf{\bibinfo{volume}{664}},
  \bibinfo{pages}{A92} (\bibinfo{year}{2022}).

\bibitem{Zhuang&Shen2023}
\bibinfo{author}{{Zhuang}, M.-Y.} \& \bibinfo{author}{{Shen}, Y.}
\newblock \bibinfo{title}{{Characterization of JWST NIRCam PSFs and
  Implications for AGN+Host Image Decomposition}}.
\newblock \emph{\bibinfo{journal}{arXiv e-prints}}
  \bibinfo{pages}{arXiv:2304.13776} (\bibinfo{year}{2023}).

\bibitem{Casura+2022MNRAS}
\bibinfo{author}{{Casura}, S.} \emph{et~al.}
\newblock \bibinfo{title}{{Galaxy And Mass Assembly (GAMA): bulge-disc
  decomposition of KiDS data in the nearby Universe}}.
\newblock \emph{\bibinfo{journal}{\mnras}} \textbf{\bibinfo{volume}{516}},
  \bibinfo{pages}{942--974} (\bibinfo{year}{2022}).

\bibitem{Haussler+2007ApJS}
\bibinfo{author}{{H{\"a}ussler}, B.} \emph{et~al.}
\newblock \bibinfo{title}{{GEMS: Galaxy Fitting Catalogs and Testing Parametric
  Galaxy Fitting Codes: GALFIT and GIM2D}}.
\newblock \emph{\bibinfo{journal}{\apjs}} \textbf{\bibinfo{volume}{172}},
  \bibinfo{pages}{615--633} (\bibinfo{year}{2007}).

\bibitem{Gao2017}
\bibinfo{author}{{Gao}, H.} \& \bibinfo{author}{{Ho}, L.~C.}
\newblock \bibinfo{title}{{An Optimal Strategy for Accurate Bulge-to-disk
  Decomposition of Disk Galaxies}}.
\newblock \emph{\bibinfo{journal}{\apj}} \textbf{\bibinfo{volume}{845}},
  \bibinfo{pages}{114} (\bibinfo{year}{2017}).

\bibitem{Fan+2014ApJ}
\bibinfo{author}{{Fan}, L.} \emph{et~al.}
\newblock \bibinfo{title}{{Structure and Morphology of X-Ray-selected Active
  Galactic Nucleus Hosts at $1 < z < 3$ in the CANDELS-COSMOS Field}}.
\newblock \emph{\bibinfo{journal}{\apjl}} \textbf{\bibinfo{volume}{784}},
  \bibinfo{pages}{L9} (\bibinfo{year}{2014}).

\bibitem{Simard+2011ApJS}
\bibinfo{author}{{Simard}, L.}, \bibinfo{author}{{Mendel}, J.~T.},
  \bibinfo{author}{{Patton}, D.~R.}, \bibinfo{author}{{Ellison}, S.~L.} \&
  \bibinfo{author}{{McConnachie}, A.~W.}
\newblock \bibinfo{title}{{A Catalog of Bulge+disk Decompositions and Updated
  Photometry for 1.12 Million Galaxies in the Sloan Digital Sky Survey}}.
\newblock \emph{\bibinfo{journal}{\apjs}} \textbf{\bibinfo{volume}{196}},
  \bibinfo{pages}{11} (\bibinfo{year}{2011}).

\bibitem{2006ApJS..166..470Richards+}
\bibinfo{author}{{Richards}, G.~T.} \emph{et~al.}
\newblock \bibinfo{title}{{Spectral Energy Distributions and Multiwavelength
  Selection of Type 1 Quasars}}.
\newblock \emph{\bibinfo{journal}{\apjs}} \textbf{\bibinfo{volume}{166}},
  \bibinfo{pages}{470--497} (\bibinfo{year}{2006}).

\bibitem{2019A&A...622A.103Boquien+}
\bibinfo{author}{{Boquien}, M.} \emph{et~al.}
\newblock \bibinfo{title}{{CIGALE: a python Code Investigating GALaxy
  Emission}}.
\newblock \emph{\bibinfo{journal}{\aap}} \textbf{\bibinfo{volume}{622}},
  \bibinfo{pages}{A103} (\bibinfo{year}{2019}).

\bibitem{Yang+2022ApJ}
\bibinfo{author}{{Yang}, G.} \emph{et~al.}
\newblock \bibinfo{title}{{Fitting AGN/Galaxy X-Ray-to-radio SEDs with CIGALE
  and Improvement of the Code}}.
\newblock \emph{\bibinfo{journal}{\apj}} \textbf{\bibinfo{volume}{927}},
  \bibinfo{pages}{192} (\bibinfo{year}{2022}).

\bibitem{2018JOSS....3..695Green}
\bibinfo{author}{{Green}, G.~M.}
\newblock \bibinfo{title}{{dustmaps: A Python interface for maps of
  interstellar dust}}.
\newblock \emph{\bibinfo{journal}{The Journal of Open Source Software}}
  \textbf{\bibinfo{volume}{3}}, \bibinfo{pages}{695} (\bibinfo{year}{2018}).

\bibitem{1998ApJ...500..525Schlegel+}
\bibinfo{author}{{Schlegel}, D.~J.}, \bibinfo{author}{{Finkbeiner}, D.~P.} \&
  \bibinfo{author}{{Davis}, M.}
\newblock \bibinfo{title}{{Maps of Dust Infrared Emission for Use in Estimation
  of Reddening and Cosmic Microwave Background Radiation Foregrounds}}.
\newblock \emph{\bibinfo{journal}{\apj}} \textbf{\bibinfo{volume}{500}},
  \bibinfo{pages}{525--553} (\bibinfo{year}{1998}).

\bibitem{2011ApJ...737..103Schlafly&Finkbeiner}
\bibinfo{author}{{Schlafly}, E.~F.} \& \bibinfo{author}{{Finkbeiner}, D.~P.}
\newblock \bibinfo{title}{{Measuring Reddening with Sloan Digital Sky Survey
  Stellar Spectra and Recalibrating SFD}}.
\newblock \emph{\bibinfo{journal}{\apj}} \textbf{\bibinfo{volume}{737}},
  \bibinfo{pages}{103} (\bibinfo{year}{2011}).

\bibitem{BC03}
\bibinfo{author}{{Bruzual}, G.} \& \bibinfo{author}{{Charlot}, S.}
\newblock \bibinfo{title}{{Stellar population synthesis at the resolution of
  2003}}.
\newblock \emph{\bibinfo{journal}{\mnras}} \textbf{\bibinfo{volume}{344}},
  \bibinfo{pages}{1000--1028} (\bibinfo{year}{2003}).

\bibitem{Inoue2011MNRAS}
\bibinfo{author}{{Inoue}, A.~K.}
\newblock \bibinfo{title}{{Rest-frame ultraviolet-to-optical spectral
  characteristics of extremely metal-poor and metal-free galaxies}}.
\newblock \emph{\bibinfo{journal}{\mnras}} \textbf{\bibinfo{volume}{415}},
  \bibinfo{pages}{2920--2931} (\bibinfo{year}{2011}).

\bibitem{Calzetti+2000ApJ}
\bibinfo{author}{{Calzetti}, D.} \emph{et~al.}
\newblock \bibinfo{title}{{The Dust Content and Opacity of Actively
  Star-forming Galaxies}}.
\newblock \emph{\bibinfo{journal}{\apj}} \textbf{\bibinfo{volume}{533}},
  \bibinfo{pages}{682--695} (\bibinfo{year}{2000}).

\bibitem{Mitchell+2013MNRAS}
\bibinfo{author}{{Mitchell}, P.~D.}, \bibinfo{author}{{Lacey}, C.~G.},
  \bibinfo{author}{{Baugh}, C.~M.} \& \bibinfo{author}{{Cole}, S.}
\newblock \bibinfo{title}{{How well can we really estimate the stellar masses
  of galaxies from broad-band photometry?}}
\newblock \emph{\bibinfo{journal}{\mnras}} \textbf{\bibinfo{volume}{435}},
  \bibinfo{pages}{87--114} (\bibinfo{year}{2013}).

\bibitem{Ciesla+2015A&A}
\bibinfo{author}{{Ciesla}, L.} \emph{et~al.}
\newblock \bibinfo{title}{{Constraining the properties of AGN host galaxies
  with spectral energy distribution modelling}}.
\newblock \emph{\bibinfo{journal}{\aap}} \textbf{\bibinfo{volume}{576}},
  \bibinfo{pages}{A10} (\bibinfo{year}{2015}).

\bibitem{Chabrier2003PASP}
\bibinfo{author}{{Chabrier}, G.}
\newblock \bibinfo{title}{{Galactic Stellar and Substellar Initial Mass
  Function}}.
\newblock \emph{\bibinfo{journal}{\pasp}} \textbf{\bibinfo{volume}{115}},
  \bibinfo{pages}{763--795} (\bibinfo{year}{2003}).

\bibitem{Kroupa2001MNRAS}
\bibinfo{author}{{Kroupa}, P.}
\newblock \bibinfo{title}{{On the variation of the initial mass function}}.
\newblock \emph{\bibinfo{journal}{\mnras}} \textbf{\bibinfo{volume}{322}},
  \bibinfo{pages}{231--246} (\bibinfo{year}{2001}).

\bibitem{Bell+2003ApJS}
\bibinfo{author}{{Bell}, E.~F.}, \bibinfo{author}{{McIntosh}, D.~H.},
  \bibinfo{author}{{Katz}, N.} \& \bibinfo{author}{{Weinberg}, M.~D.}
\newblock \bibinfo{title}{{The Optical and Near-Infrared Properties of
  Galaxies. I. Luminosity and Stellar Mass Functions}}.
\newblock \emph{\bibinfo{journal}{\apjs}} \textbf{\bibinfo{volume}{149}},
  \bibinfo{pages}{289--312} (\bibinfo{year}{2003}).

\bibitem{Stalevski+2012MNRAS}
\bibinfo{author}{{Stalevski}, M.}, \bibinfo{author}{{Fritz}, J.},
  \bibinfo{author}{{Baes}, M.}, \bibinfo{author}{{Nakos}, T.} \&
  \bibinfo{author}{{Popovi{\'c}}, L.~{\v{C}}.}
\newblock \bibinfo{title}{{3D radiative transfer modelling of the dusty tori
  around active galactic nuclei as a clumpy two-phase medium}}.
\newblock \emph{\bibinfo{journal}{\mnras}} \textbf{\bibinfo{volume}{420}},
  \bibinfo{pages}{2756--2772} (\bibinfo{year}{2012}).

\bibitem{Stalevski+2016MNRAS}
\bibinfo{author}{{Stalevski}, M.} \emph{et~al.}
\newblock \bibinfo{title}{{The dust covering factor in active galactic
  nuclei}}.
\newblock \emph{\bibinfo{journal}{\mnras}} \textbf{\bibinfo{volume}{458}},
  \bibinfo{pages}{2288--2302} (\bibinfo{year}{2016}).

\bibitem{Baldry+2004ApJ}
\bibinfo{author}{{Baldry}, I.~K.} \emph{et~al.}
\newblock \bibinfo{title}{{Quantifying the Bimodal Color-Magnitude Distribution
  of Galaxies}}.
\newblock \emph{\bibinfo{journal}{\apj}} \textbf{\bibinfo{volume}{600}},
  \bibinfo{pages}{681--694} (\bibinfo{year}{2004}).

\bibitem{Blanton&Moustakas2009ARA&A}
\bibinfo{author}{{Blanton}, M.~R.} \& \bibinfo{author}{{Moustakas}, J.}
\newblock \bibinfo{title}{{Physical Properties and Environments of Nearby
  Galaxies}}.
\newblock \emph{\bibinfo{journal}{\araa}} \textbf{\bibinfo{volume}{47}},
  \bibinfo{pages}{159--210} (\bibinfo{year}{2009}).

\bibitem{Blanton+2003ApJ}
\bibinfo{author}{{Blanton}, M.~R.} \emph{et~al.}
\newblock \bibinfo{title}{{The Broadband Optical Properties of Galaxies with
  Redshifts $0.02<z<0.22$}}.
\newblock \emph{\bibinfo{journal}{\apj}} \textbf{\bibinfo{volume}{594}},
  \bibinfo{pages}{186--207} (\bibinfo{year}{2003}).

\bibitem{Magorrian+1998AJ}
\bibinfo{author}{{Magorrian}, J.} \emph{et~al.}
\newblock \bibinfo{title}{{The Demography of Massive Dark Objects in Galaxy
  Centers}}.
\newblock \emph{\bibinfo{journal}{\aj}} \textbf{\bibinfo{volume}{115}},
  \bibinfo{pages}{2285--2305} (\bibinfo{year}{1998}).

\bibitem{Trump+2013ApJ}
\bibinfo{author}{{Trump}, J.~R.} \emph{et~al.}
\newblock \bibinfo{title}{{A Census of Broad-line Active Galactic Nuclei in
  Nearby Galaxies: Coeval Star Formation and Rapid Black Hole Growth}}.
\newblock \emph{\bibinfo{journal}{\apj}} \textbf{\bibinfo{volume}{763}},
  \bibinfo{pages}{133} (\bibinfo{year}{2013}).

\bibitem{Salim+2016ApJS}
\bibinfo{author}{{Salim}, S.} \emph{et~al.}
\newblock \bibinfo{title}{{GALEX-SDSS-WISE Legacy Catalog (GSWLC): Star
  Formation Rates, Stellar Masses, and Dust Attenuations of 700,000
  Low-redshift Galaxies}}.
\newblock \emph{\bibinfo{journal}{\apjs}} \textbf{\bibinfo{volume}{227}},
  \bibinfo{pages}{2} (\bibinfo{year}{2016}).

\bibitem{Salim+2018ApJ}
\bibinfo{author}{{Salim}, S.}, \bibinfo{author}{{Boquien}, M.} \&
  \bibinfo{author}{{Lee}, J.~C.}
\newblock \bibinfo{title}{{Dust Attenuation Curves in the Local Universe:
  Demographics and New Laws for Star-forming Galaxies and High-redshift
  Analogs}}.
\newblock \emph{\bibinfo{journal}{\apj}} \textbf{\bibinfo{volume}{859}},
  \bibinfo{pages}{11} (\bibinfo{year}{2018}).

\bibitem{Weinmann+2006MNRAS}
\bibinfo{author}{{Weinmann}, S.~M.}, \bibinfo{author}{{van den Bosch}, F.~C.},
  \bibinfo{author}{{Yang}, X.} \& \bibinfo{author}{{Mo}, H.~J.}
\newblock \bibinfo{title}{{Properties of galaxy groups in the Sloan Digital Sky
  Survey - I. The dependence of colour, star formation and morphology on halo
  mass}}.
\newblock \emph{\bibinfo{journal}{\mnras}} \textbf{\bibinfo{volume}{366}},
  \bibinfo{pages}{2--28} (\bibinfo{year}{2006}).

\bibitem{Huang+2012ApJ}
\bibinfo{author}{{Huang}, S.}, \bibinfo{author}{{Haynes}, M.~P.},
  \bibinfo{author}{{Giovanelli}, R.} \& \bibinfo{author}{{Brinchmann}, J.}
\newblock \bibinfo{title}{{The Arecibo Legacy Fast ALFA Survey: The Galaxy
  Population Detected by ALFALFA}}.
\newblock \emph{\bibinfo{journal}{\apj}} \textbf{\bibinfo{volume}{756}},
  \bibinfo{pages}{113} (\bibinfo{year}{2012}).

\bibitem{Brinchmann+2004MNRAS}
\bibinfo{author}{{Brinchmann}, J.} \emph{et~al.}
\newblock \bibinfo{title}{{The physical properties of star-forming galaxies in
  the low-redshift Universe}}.
\newblock \emph{\bibinfo{journal}{\mnras}} \textbf{\bibinfo{volume}{351}},
  \bibinfo{pages}{1151--1179} (\bibinfo{year}{2004}).

\bibitem{Ravindranath+2004ApJ}
\bibinfo{author}{{Ravindranath}, S.} \emph{et~al.}
\newblock \bibinfo{title}{{The Evolution of Disk Galaxies in the GOODS-South
  Field: Number Densities and Size Distribution}}.
\newblock \emph{\bibinfo{journal}{\apjl}} \textbf{\bibinfo{volume}{604}},
  \bibinfo{pages}{L9--L12} (\bibinfo{year}{2004}).

\bibitem{Gabor+2009ApJ}
\bibinfo{author}{{Gabor}, J.~M.} \emph{et~al.}
\newblock \bibinfo{title}{{Active Galactic Nucleus Host Galaxy Morphologies in
  COSMOS}}.
\newblock \emph{\bibinfo{journal}{\apj}} \textbf{\bibinfo{volume}{691}},
  \bibinfo{pages}{705--722} (\bibinfo{year}{2009}).

\bibitem{Dewsnap+2022arXiv}
\bibinfo{author}{{Dewsnap}, C.} \emph{et~al.}
\newblock \bibinfo{title}{{GALFIT-ing AGN Host Galaxies in COSMOS: HST vs.
  Subaru}}.
\newblock \emph{\bibinfo{journal}{\apj}} \textbf{\bibinfo{volume}{944}},
  \bibinfo{pages}{137} (\bibinfo{year}{2023}).

\bibitem{Willett+2013MNRAS}
\bibinfo{author}{{Willett}, K.~W.} \emph{et~al.}
\newblock \bibinfo{title}{{Galaxy Zoo 2: detailed morphological classifications
  for 304 122 galaxies from the Sloan Digital Sky Survey}}.
\newblock \emph{\bibinfo{journal}{\mnras}} \textbf{\bibinfo{volume}{435}},
  \bibinfo{pages}{2835--2860} (\bibinfo{year}{2013}).

\bibitem{Hart+2016MNRAS}
\bibinfo{author}{{Hart}, R.~E.} \emph{et~al.}
\newblock \bibinfo{title}{{Galaxy Zoo: comparing the demographics of spiral arm
  number and a new method for correcting redshift bias}}.
\newblock \emph{\bibinfo{journal}{\mnras}} \textbf{\bibinfo{volume}{461}},
  \bibinfo{pages}{3663--3682} (\bibinfo{year}{2016}).

\bibitem{Kormendy&Kennicutt2004ARA&A}
\bibinfo{author}{{Kormendy}, J.} \& \bibinfo{author}{{Kennicutt}, J.,
  Robert~C.}
\newblock \bibinfo{title}{{Secular Evolution and the Formation of Pseudobulges
  in Disk Galaxies}}.
\newblock \emph{\bibinfo{journal}{\araa}} \textbf{\bibinfo{volume}{42}},
  \bibinfo{pages}{603--683} (\bibinfo{year}{2004}).

\bibitem{Gao+2020ApJS}
\bibinfo{author}{{Gao}, H.}, \bibinfo{author}{{Ho}, L.~C.},
  \bibinfo{author}{{Barth}, A.~J.} \& \bibinfo{author}{{Li}, Z.-Y.}
\newblock \bibinfo{title}{{The Carnegie-Irvine Galaxy Survey. IX.
  Classification of Bulge Types and Statistical Properties of Pseudo Bulges}}.
\newblock \emph{\bibinfo{journal}{\apjs}} \textbf{\bibinfo{volume}{247}},
  \bibinfo{pages}{20} (\bibinfo{year}{2020}).

\bibitem{Kim+2017ApJS}
\bibinfo{author}{{Kim}, M.}, \bibinfo{author}{{Ho}, L.~C.},
  \bibinfo{author}{{Peng}, C.~Y.}, \bibinfo{author}{{Barth}, A.~J.} \&
  \bibinfo{author}{{Im}, M.}
\newblock \bibinfo{title}{{Stellar Photometric Structures of the Host Galaxies
  of Nearby Type 1 Active Galactic Nuclei}}.
\newblock \emph{\bibinfo{journal}{\apjs}} \textbf{\bibinfo{volume}{232}},
  \bibinfo{pages}{21} (\bibinfo{year}{2017}).

\bibitem{Kim+2021ApJS}
\bibinfo{author}{{Kim}, M.}, \bibinfo{author}{{Barth}, A.~J.},
  \bibinfo{author}{{Ho}, L.~C.} \& \bibinfo{author}{{Son}, S.}
\newblock \bibinfo{title}{{A Hubble Space Telescope Imaging Survey of
  Low-redshift Swift-BAT Active Galaxies}}.
\newblock \emph{\bibinfo{journal}{\apjs}} \textbf{\bibinfo{volume}{256}},
  \bibinfo{pages}{40} (\bibinfo{year}{2021}).

\bibitem{Zhao+2019ApJ}
\bibinfo{author}{{Zhao}, D.}, \bibinfo{author}{{Ho}, L.~C.},
  \bibinfo{author}{{Zhao}, Y.}, \bibinfo{author}{{Shangguan}, J.} \&
  \bibinfo{author}{{Kim}, M.}
\newblock \bibinfo{title}{{The Role of Major Mergers and Nuclear Star Formation
  in Nearby Obscured Quasars}}.
\newblock \emph{\bibinfo{journal}{\apj}} \textbf{\bibinfo{volume}{877}},
  \bibinfo{pages}{52} (\bibinfo{year}{2019}).

\bibitem{Martin-Navarro+2022MNRAS}
\bibinfo{author}{{Mart{\'\i}n-Navarro}, I.}, \bibinfo{author}{{Shankar}, F.} \&
  \bibinfo{author}{{Mezcua}, M.}
\newblock \bibinfo{title}{{Rejuvenation triggers nuclear activity in nearby
  galaxies}}.
\newblock \emph{\bibinfo{journal}{\mnras}} \textbf{\bibinfo{volume}{513}},
  \bibinfo{pages}{L10--L14} (\bibinfo{year}{2022}).

\bibitem{Frank+1992apa..book}
\bibinfo{author}{{Frank}, J.}, \bibinfo{author}{{King}, A.} \&
  \bibinfo{author}{{Raine}, D.}
\newblock \emph{\bibinfo{title}{{Accretion power in astrophysics.}}}
  (\bibinfo{publisher}{Cambridge: Cambridge Univ. Press},
  \bibinfo{year}{1992}).

\bibitem{Aversa+2015ApJ}
\bibinfo{author}{{Aversa}, R.}, \bibinfo{author}{{Lapi}, A.},
  \bibinfo{author}{{de Zotti}, G.}, \bibinfo{author}{{Shankar}, F.} \&
  \bibinfo{author}{{Danese}, L.}
\newblock \bibinfo{title}{{Black Hole and Galaxy Coevolution from Continuity
  Equation and Abundance Matching}}.
\newblock \emph{\bibinfo{journal}{\apj}} \textbf{\bibinfo{volume}{810}},
  \bibinfo{pages}{74} (\bibinfo{year}{2015}).

\bibitem{Delvecchio+2020ApJ}
\bibinfo{author}{{Delvecchio}, I.} \emph{et~al.}
\newblock \bibinfo{title}{{The Evolving AGN Duty Cycle in Galaxies Since z
  {\ensuremath{\sim}} 3 as Encoded in the X-Ray Luminosity Function}}.
\newblock \emph{\bibinfo{journal}{\apj}} \textbf{\bibinfo{volume}{892}},
  \bibinfo{pages}{17} (\bibinfo{year}{2020}).

\bibitem{Ananna+2022ApJS}
\bibinfo{author}{{Ananna}, T.~T.} \emph{et~al.}
\newblock \bibinfo{title}{{BASS. XXX. Distribution Functions of DR2 Eddington
  Ratios, Black Hole Masses, and X-Ray Luminosities}}.
\newblock \emph{\bibinfo{journal}{\apjs}} \textbf{\bibinfo{volume}{261}},
  \bibinfo{pages}{9} (\bibinfo{year}{2022}).

\bibitem{Lamperti+2017MNRAS}
\bibinfo{author}{{Lamperti}, I.} \emph{et~al.}
\newblock \bibinfo{title}{{BAT AGN Spectroscopic Survey - IV: Near-Infrared
  Coronal Lines, Hidden Broad Lines, and Correlation with Hard X-ray
  Emission}}.
\newblock \emph{\bibinfo{journal}{\mnras}} \textbf{\bibinfo{volume}{467}},
  \bibinfo{pages}{540--572} (\bibinfo{year}{2017}).

\bibitem{Ricci+2022ApJS}
\bibinfo{author}{{Ricci}, F.} \emph{et~al.}
\newblock \bibinfo{title}{{BASS. XXIX. The Near-infrared View of the Broad-line
  Region (BLR): The Effects of Obscuration in BLR Characterization}}.
\newblock \emph{\bibinfo{journal}{\apjs}} \textbf{\bibinfo{volume}{261}},
  \bibinfo{pages}{8} (\bibinfo{year}{2022}).

\bibitem{Abbott+2021ApJS}
\bibinfo{author}{{Abbott}, T.~M.~C.} \emph{et~al.}
\newblock \bibinfo{title}{{The Dark Energy Survey Data Release 2}}.
\newblock \emph{\bibinfo{journal}{\apjs}} \textbf{\bibinfo{volume}{255}},
  \bibinfo{pages}{20} (\bibinfo{year}{2021}).

\bibitem{Cappellari+2013MNRAS}
\bibinfo{author}{{Cappellari}, M.} \emph{et~al.}
\newblock \bibinfo{title}{{The ATLAS$^{3D}$ project - XX. Mass-size and
  mass-{\ensuremath{\sigma}} distributions of early-type galaxies: bulge
  fraction drives kinematics, mass-to-light ratio, molecular gas fraction and
  stellar initial mass function}}.
\newblock \emph{\bibinfo{journal}{\mnras}} \textbf{\bibinfo{volume}{432}},
  \bibinfo{pages}{1862--1893} (\bibinfo{year}{2013}).

\bibitem{dataproduct}
\bibinfo{author}{{Zhuang}, M.-Y.} \& \bibinfo{author}{{Ho}, L. C.}
\newblock \bibinfo{title}{{Basic Information and Derived Properties of the Final AGN Sample}}.
\newblock \emph{\bibinfo{publisher}{National Astronomical Data Center of China}}
  \bibinfo{identifier}{https://doi.org/10.12149/101278} (\bibinfo{year}{2023}).

\end{thebibliography}
\end{document}